\documentstyle[aaspp4,psfig]{article}
\begin{document}
\title { Beam Instabilities in  Magnetized Pair Plasma}
\author{ Maxim Lyutikov 
\footnote{Currently at the Canadian Institute
for Theoretical Astrophysics, 60 St. George, Toronto, Ont,  M5S 3H8, Canada}}
\affil{Theoretical Astrophysics, California Institute of Technology,
Pasadena, California 91125}








\begin{abstract} 
Beam instabilities in the strongly magnetized
electron-positron plasma of pulsar magnetospheres
are considered. We analyze the resonance conditions
and  estimate the  growth rates of the Cherenkov and cyclotron
instabilities
of  the ordinary (O), extraordinary (X)  and Alfv\'{e}n modes
in two limiting regimes: kinetic and hydrodynamic.
As a preliminary step,
we reconsider wave dispersion  and  polarization
properties in a one dimensional pair plasma
 taking into
account relativistic  thermal effects. 
We then find the location of the Cherenkov and cyclotron resonance of the 
X, O and Alfv\'{e}n waves with the particles from the primary
beam.
The importance of the different instabilities as a source of the
coherent pulsar radiation generation is then estimated taking into account
the angular dependence of the
growth rates and the  limitations on the  length of the coherent wave-particle
interaction imposed by the  curvature of the magnetic field lines.
We conclude, that in the pulsar magnetosphere the Cherenkov-type
instabilities occur in the hydrodynamic regimes, while 
 the cyclotron-type
instabilities occur in the kinetic regime. 
We argue, that  electromagnetic cyclotron-type  instabilities
on the X, O and probably Alfv\'{e}n waves are more likely to
develop in the pulsar magnetosphere. 
\end{abstract}

\section{Introduction}

At the moment the most promising theories of the pulsar radio 
emission generation are based on the plasma emission model, in 
which the high brightness radio emission is generated by plasma
instabilities developing in the outflowing plasma
 (\cite{Melrose-DB}).
To find the  instability that can be responsible for the
generation of pulsar radio emission it is essential to know
the dispersion relations of the normal modes of the medium
and take into account the evolution of the modes as
they propagate outward in the pulsar magnetosphere.

In a standard model of pulsar magnetospheres 
(\cite{GoldreichJulian}, \cite{Sturrock}) rotating, strongly magnetized
 neutron stars induce strong electric fields that pull
the charges from their surfaces. Inside the closed  field
lines of the neutron star magnetosphere, a steady charge
distribution established,  compensating the induced electric
field. On the open field lines,  the neutron star
 generates a dense flow of relativistic
electron-positron pairs penetrated by a highly relativistic
 electron or positron beam.
This relativistic flow generates the observed pulsar radio emission.

First, we  present 
a  consideration of the properties of
 linear waves in a strongly magnetized
electron-positron plasma with similar distributions
of electrons and positron. We  take into account  possible relativistic temperatures
of plasma component. The particles are assumed
to be in their ground gyrational state so that the plasma is one-dimensional. 
Properties of  a 
one-dimensional  pair plasma are considerably different
from the properties of  a well studied electron-ion plasma.
 Previously,  \cite{SuvorovChugunov} investigated the
dispersion relations in a one-dimensional plasma with a power law distribution
of the particles,  \cite{arons1} 
 considered waves in a cold plasma  and  in a warm 
plasma using a  water bag distribution.
For the
propagation of the coupled O-Alfven modes,
 \cite{arons1}
 considered  the case of infinitely
strong magnetic field and found that the O branch is always superluminous
and thus cannot be excited by the Cherenkov mechanism. On the other hand,
 \cite{Volokitin} took into account finite magnetic field
and found that the O wave becomes subluminous for large wave numbers
allowing Cherenkov excitation.
\cite{LyutikovDisp}
considered  the waves in pair plasma taking into account the relative
motion of electrons and positrons.
 \cite{ZankGreaves} used a fluid theory to
consider waves in a {\it  nonrelativistic}  three-dimensional  pair plasma. 
In our approach we use
 relativistic kinetic theory assuming that the 
possible  resonant and nonresonant contributions
from the beam particles may be considered as perturbations to the
initial plasma state.

We resort to similar distributions of the 
pair plasma  with equal densities and  
neglect the curvature of the magnetic field
lines. These approximations require some justifications. First,
the pulsar plasma is  nonneutral due to the
presence of the primary beam.
We neglect the nonneutrality of plasma since
the density of plasma $n_p$  is thought to be much larger than the 
Goldreich-Julian density 
 $n_{GJ}={\bf \Omega \cdot B}/(2\,
\pi\, e\, c)$,
$n _p = \lambda\, n_{GJ} =10^3-10^6 n_{GJ} $
($\Omega$ is a rotational
frequency of the neutron star,
 $q$ is a charge of a particle, $B$ is magnetic field,
 $c$ is a speed of light, $\lambda$ is the multiplicity factor).
 Secondly,
the inhomogeneity of the magnetic field results in a curvature 
drift of the particles perpendicular to the osculating plane. 
It can be shown that for the typical parameters in the
pulsar magnetosphere the drift velocity of plasma particles
due to the curvature of magnetic field lines
could be neglected in the calculations of the real part of the
dielectric tensor unless the curvature of field lines $R_B$ satisfies the
condition
$R_B \ll \gamma_p^2 r_{L}$ (here $\gamma_p$ is the average
streaming energy of plasma particles in the pulsar frame
and $ r_{L}=c/\omega_B$ ia a Larmor radius). 
This follows from the assumption of
a nonrelativistic transverse motion and 
 the transformation of the radius of curvature
seen
in the center of gyration frame $R_{B,{ \rm cg}} = R_B/\gamma^2$
($R_{B}$ is the radius of curvature in the pulsar frame,
$\gamma$ is the Lorentz factor of the particle). This
 condition is well satisfied inside the
pulsar magnetosphere for the plasma streaming energy $\gamma_p < 10^4$.
By contrast, the
drift velocity may be  very  important for the high energy resonant particles.

In this paper we consider wave excitation in a strongly magnetized
pair plasma in the approximation of {\it straight} magnetic field lines, thus
omitting an important Cherenkov-drift resonance 
(\cite{LyutikovBlandfordMachabeli},
 \cite{LyutikovMachabeliBlandford2}).
This is an important mechanism that may be responsible for the 
generation of the cone type emission in pulsars. The
 electromagnetic  Cherenkov-drift
instability occurs in the kinetic regime on the 
high frequency vacuum-like O and X  waves.
It has the same advantages as the electromagnetic  cyclotron instabilities
considered in this paper.  

In presenting our results 
 we were trying to find a balance between
 providing an exact general relation and finding a
simple  useful analytical approximation, which 
gives an idea of how modes behave in  different regimes. 
Several approximations will be extensively used.
For the case of radio waves propagating in the pulsar magnetosphere
there is naturally a small parameter, $\omega/\omega_B$ 
($\omega$ is a frequency of a wave), so that
in many cases complicated dispersion relations and
polarization properties  can be simplified 
by expanding in this parameter. When making such expansion
one should be especially careful
  near the 
points where dispersion curves almost intersect (Eq. \ref{tte}).
A formal expansion in $\omega/\omega_B$ diverges near the intersection
point. In this case, since the intersection formally occurs only for the
parallel propagation, the relevant small parameter is the angle of propagation
with respect to magnetic field.  
Another simplification that we will often use, is the expansion
in a ratio of the  plasma frequency  $\omega_p$ to the cyclotron frequency
$\omega_B = e B/mc$ ($m$ is a mass of a particle) 
$\omega_p \ll \omega_B$. This is a good approximation for the
most parts of the pulsar magnetosphere.

The main conlusions of our work are the following.
For the chosen parameters of the 
magnetosphere plasma, the Cherenkov-type electrostatic
beam instabilities develop in a hydrodynamic regime, while 
cyclotron-type  electromagnetic instabilities  develop in a kinetic
regime.
Electrostatic beam instabilities in the pulsar plasma are
generally weaker than the
 electromagnetic    instabilities.
In addition,
Cherenkov instabilities have  largest growth rate near the stellar
surface, where the Cherenkov resonance can occurs {\it only}
 on  the Alfv\'{e}n  mode. However,
 this mode 
{\it cannot } escape to infinity, even though it has
some electromagnetic component.
 Another factor that limits the
development of the 
Cherenkov-type instabilities is that they grow within a much 
 narrower angles than cyclotron 
instabilities. In a curved magnetic field this 
results in a shorter length of the coherent 
 wave-particle interaction. 
 
The relative weakness of   electrostatic instabilities as compared
 to electromagnetic instabilities is an unusual characteristics
of the strongly relativistic beams. The reason is that
for the particles in the primary beam, which contribute to the
development of the instability, the effective parallel mass is
$m_{{\rm eff}\parallel} \,= \, \gamma_b^3  m \approx 10^{21}\, m$.
This
suppresses the development of the electrostatic instabilities.
In contrast,
the effective transverse mass , $m_{{\rm eff}\perp} = \gamma_b  m$,
is less  affected by the large parallel
momentum. The electromagnetic instabilities
are less  suppressed by the large streaming momenta.
Thus, the relativistic velocities and one-dimensionality
of the distribution function result in a strong suppression
of the electrostatic instabilities as compared to
electromagnetic instabilities.

The calculations presented here provided a basis for the 
model of pulsar radio emission presented in 
\cite{LyutikovBlandfordMachabeli}.

\section{Plasma Parameters}
\label{PlasmaParameters}

\begin{figure}
\psfig{file=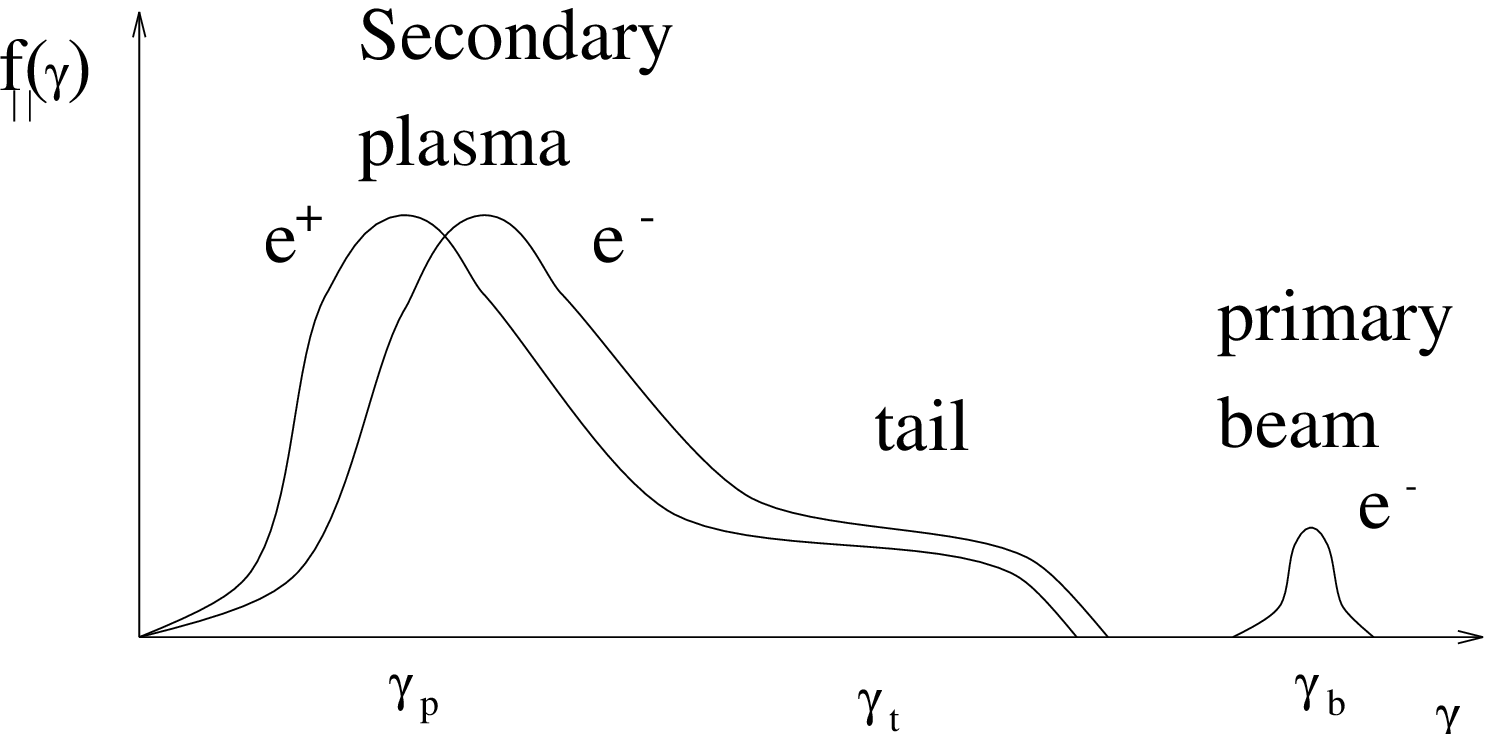,width=12.0cm}
\caption{
 Distribution function for a one-dimensional electron-positron
plasma of pulsar magnetosphere.
\label{Distributionfunction}}
\end{figure}

To a large extent a possible mechanism for the generation of pulsar radio
emission is predicated on the choice of parameters of the plasma flow that
is generated by  a rotating neutron star.
At this point we know
only the general features of the
distribution function of the particles in a pulsar magnetosphere
(\cite{Tademaru},\cite{Arons1981}, \cite{DaughertyHarding1983}).
It is believed to comprise (see Fig. \ref{Distributionfunction})
(i) a highly relativistic primary beam
with the Lorentz factor $\,  \gamma_b \approx 10^7$ and
density equal to the Goldreich-Julian density $n_{GJ}$,
  (ii)   a secondary
electron positron plasma with a bulk streaming Lorentz factor
$\,  \gamma_p \approx 10 -1000$, a similar  scatter in energy
$T_p \approx \,  \gamma_p$ and a density much larger
than  the beam density $n_p \approx \, \lambda\, n_{GJ} =10^3-10^6 n_{GJ}$,
 (iii) a tail of plasma
distribution with the energy up to $\,  \gamma_t =10^4- 10^5$.

We will normalize the  density of the pair plasma  to the
 Goldreich-Julian density.
\begin{equation}
n_{\alpha}
= \lambda\, n_{GJ} =10^3-10^6 n_{GJ},
\hskip .3 truein
\omega_p^2 = \lambda \omega_b^2 = 2  \lambda \omega_B \Omega
\label{qpq}
\end{equation}
(subscript ${\alpha}$ in Eq. (\ref{qpq}) refers to the
electrons and positrons of the bulk plasma).
Secondary pairs are born with almost the same energy
 in the avalanche-like
process above in the polar cap  (\cite{Arons83}).
The combination of the pair plasma and primary beam is expected to screen
the rotationally induced electric field so that the flow is force-free.

Another relation between the parameters of the plasma and the beam
comes from the energy argument that the primary particles stop producing
the pairs when the energy in the pair plasma becomes equal to the energy in the
primary beam:
\begin{equation}
2 \,
<\,\gamma\,>_{\pm}\, n_{\pm}=\gamma_b^{\prime} \, n_{GJ}, \hskip .6 truein
\mbox{at the pair formation front}
\label{vg}
\end{equation}
where $<\,\gamma\,>_{\pm}$ and $ n_{\pm}$  are the initial
 average energies and densities of pair plasma, $\gamma_b^{\prime}$
is the energy of the beam in the pulsar frame and $ n_{GJ}$ is the density
of the beam.
It is assumed that the initial densities, temperatures
 and velocities of the plasma
components are equal. For cold components $<\,\gamma\,>_{\pm}\,=\,
\gamma_p$, ($\gamma_p$ is the stream $\gamma$-factor
of the bulk plasma  with respect to the pulsar frame), 
 while for the relativistically hot components with a temperature
$T_p$ the average energy is
 $<\,\gamma\,>^{(0)}_{\pm}\,=\,   \gamma_p \, T_p$, where 
$ T_p/2 $ is the average energy of particles in the plasma frame.

In what follows, the quantities measured in the pulsar frame will be denoted
with a prime. The relations between plasma parameters measured in the 
pulsar and plasma frames are
\begin{equation}
\gamma_b ={\gamma_b^{\prime} \over 2 \gamma_p},\hskip .2 truein 
\omega_p = {\omega_p ^{\prime} \over \sqrt{\gamma_p}} ,\hskip .2 truein
\omega_B= \omega_B  ^{\prime}  
\label{jjj1}
\end{equation}

In this paper we neglect the difference of energies of secondary
plasma components which arises as the flow propagates outward in 
curved magnetic field lines. 
For the consideration of the effects of relative velocity on the 
wave dispersion see  \cite{LyutikovDisp}.

The uncertainty in the physics of the pair formation front 
forces
us to allow for  a broad range of plasma parameters. 
Accordingly, the growth rates of the particular instabilities
can vary considerably depending on the assumed parameters.
The numerical estimates will be given for a typical pulsar with the 
period P= 0.5 s (light cylinder radius $R_{ll}= \,2.4 \times 10^9$ cm),
and the  surface magnetic field $ B= 10^{12}\, G$ and the primary
 beam Lorentz factor
$\gamma_b =2 \times 10^7$ (e.g., \cite{Arons83}). These assumptions and the
equation (\ref{vg}) reduce the number of free parameters to two:
plasma temperature and the bulk streaming energy $\gamma_p$ 
(or temperature and the multiplicity factor $\lambda$).
Consequently,
we will consider two separate cases of cold and relativistically
hot plasma. For numerical estimates we will use the following
fiducial numbers: $\gamma_p=100$, $\lambda 10^5$,
 $T_p \ll 1$ for the cold plasma, and
 $\gamma_p=100$, $\lambda = 10^4$,
 $T_p  \approx 10$ for the relativistically hot plasma
( $T_p$ is the invariant temperature of plasma in units of $m c^2$).

The radial dependence of the parameters is assumed to follow 
 the dipole geometry of the
magnetic field:
\begin{eqnarray}
&&
{\omega_B}(r)\,=\, {\omega_B}(R_{NS}) \left( { R_{NS}\over y} \right)^3,
\mbox{} \nonumber \\ \mbox{}
&&
\omega_p(r)\,=\, \omega_p(R_{NS}) \left( { R_{NS}\over y} \right)^{3/2}.
\label{vg21}
\end{eqnarray}

\section{Response Tensor for a One Dimensional Plasma In Staright Magnetic Field}
\label{DielectricTensor}

In the limits of applicability
 of our simplifying assumptions,
 the dielectric tensor is (\cite{LyutikovMachabeliBlandford2})
 \begin{eqnarray}
\epsilon_{xx}&&=1-{1\over 2} \sum _{\alpha}
 \,{ \omega_{p \alpha}^2\over\omega^2 }\,
\int {d p_{z}\over \gamma}  \left( (\omega- k_{z} v_{z} )
A^+_{\alpha}
 f_{\alpha}  
 \right) = \epsilon_{yy}
\mbox{} \nonumber \\ \mbox{}
\epsilon_{{z} {z}}&&=1-\sum _{\alpha} \,{ \omega_{p \alpha}^2 }\,
 \int {d p_{z}\over \gamma^3}  {f_{\alpha} \over 
\Omega^{o \,2}_{\alpha}  } 
-\sum _{\alpha} \,
{ \omega_{p \alpha}^2\over \omega^2}\,
\int {d p_{z}\over \gamma}  f_{\alpha}   \,
{ (k_x^2+ k_y^2) \, v_{z}^2 \, \over 
\Omega _{\alpha}^+ \Omega_{\alpha}^- }
\mbox{} \nonumber \\ \mbox{}
\epsilon_{xy}&&=-{i\over 2} \sum _{\alpha}
\,{ \omega_{p \alpha}^2\over \omega^2}
 \int {d p_{z}\over \gamma} \left( (\omega- k_{z} v_{z} ) A^-_{\alpha} \right)
 f_{\alpha} = - \epsilon_{yx}
\mbox{} \nonumber \\ \mbox{}
\epsilon_{xz}&&={1\over 2} \sum _{\alpha}
\,{ \omega_{p \alpha}^2\over \omega^2 }
 \int {d p_{z}\over \gamma} v_{z} 
  \left( k_x   A^+_{\alpha}   +  i k_y A^-_{\alpha} \right )   f_{\alpha} 
\mbox{} \nonumber \\ \mbox{}
\epsilon_{z x} &&= {1\over 2} \sum _{\alpha}
\,{ \omega_{p \alpha}^2\over \omega^2 }
 \int {d p_{z}\over \gamma} v_{z} 
 \left(  k_x  A^+_{\alpha} -
 i k_y A^-_{\alpha} \right )  f_{\alpha} 
\mbox{} \nonumber \\ \mbox{}
\epsilon_{yz}&&= - {1\over 2} \sum _{\alpha}
\,{ \omega_{p \alpha}^2\over \omega^2}
 \int {d p_{z}\over \gamma}{ v_{z} \over c}  \left(
k_y A^+_{\alpha} - i  k_x c   A^-_{\alpha}
\right) f_{\alpha}
\mbox{} \nonumber \\ \mbox{}
\epsilon_{z y} &&= - {1\over 2} \sum _{\alpha}
\,{ \omega_{p \alpha}^2\over \omega^2}
 \int {d p_{z}\over \gamma}{ v_{z} \over c}  \left(
k_y A^+_{\alpha} +  i  k_x  A^-_{\alpha} \right)
 f_{\alpha}
\label{epsilon}
\end{eqnarray}
Here
\begin{eqnarray}
&&
A^+_{\alpha}=\left({1\over \Omega^+ _{\alpha} }+{1\over \Omega^-
 _{\alpha} } \right),
\hskip .4 truein
A^-_{\alpha}=\left({1\over \Omega^-_{\alpha} }-{1\over \Omega^+
_{\alpha} } \right),
\mbox{} \nonumber \\ \mbox{}
&&
\Omega^{\pm}_{\alpha}
=\omega - k_{z} v_{z}    \pm \omega_{B } \gamma^{-1},
\hskip .1truein
\Omega^o _{\alpha} =\omega - k_{z} v_{z} ,
\label{As}
\end{eqnarray}
 where $f_{\alpha}$ are one dimensional distribution functions of
the components ${\alpha}$,
$v_{z} $ is a velocity along the local magnetic field,  $\gamma$ is a 
Lorentz factor of a particle,  $k_x$, $k_{z}$ and $k_y$ are
the corresponding components of the  wave vector and magnetic field
is directed along the $z$ axis.

For  stationary and spatially uniform plasma we can use  Fourier analysis
which  reduces the
problem to the following system of equations for the perturbations
in the electric field:
\begin{equation}
\Lambda _{\alpha \beta} E_{\beta}(\omega,{\bf k})=0
\label{a1a1}
\end{equation}
where $\epsilon_{\alpha \beta}(\omega,{\bf k})$ is the dielectric tensor
of the medium and
\begin{equation}
\Lambda _{\alpha \beta} =  k_{\alpha} k_{\beta} -k^2 c^2 \delta_{\alpha \beta}
+
\omega^2 \epsilon_{\alpha \beta}(\omega,{\bf k})
\label{dett1}
\end{equation}
The normal modes satisfy a dispersion relation
\begin{equation}
Det\left| \Lambda _{\alpha \beta}
 \right|=0
\label{dettt}
\end{equation}
whose roots determine the time behavior of the perturbations.

\section{Waves in Cold  Pair Plasma in Rest Frame}
\label{WavesCold}

\subsection{Dielectric Tensor}

In this section we consider waves   in a cold,
strongly magnetized, electron-positron plasma in its rest frame.
If the average velocities of the electrons  and positrons of the secondary plasma
are the same, then  
retaining only nonresonant terms we find from Eq. (\ref{epsilon})
the dielectric tensor for cold pair plasma with coincident
distribution functions:
\begin{eqnarray}
\epsilon_{xx}&&=   1 + {2\, \omega_p^2\over \omega_B^2
       -\omega ^2  } 
\, =\epsilon_{yy}
\mbox{} \nonumber \\ \mbox{}
\epsilon_{z z}&&=  1 - {{2\,{{{\it \omega_p}}^2}}\over {{{\omega }^2}}} 
\mbox{} \nonumber \\ \mbox{}
\epsilon_{xy}&&= 
 \epsilon_{yx}=
\epsilon_{xz}=
\epsilon_{z x}=
\epsilon_{y z}=
\epsilon_{z y}=0
\label{epsilon21}
\end{eqnarray}
where $ \omega_p^2 =\, 4\, \pi n_p e^2/m$ is the plasma frequency,
$ {\omega_B} = |e| B/ m c$ is the nonrelativistic 
positively defined cyclotron frequency.

\subsection{Dispersion of the Normal Modes}
\label{Coldwaves}

\begin{figure}
\psfig{file=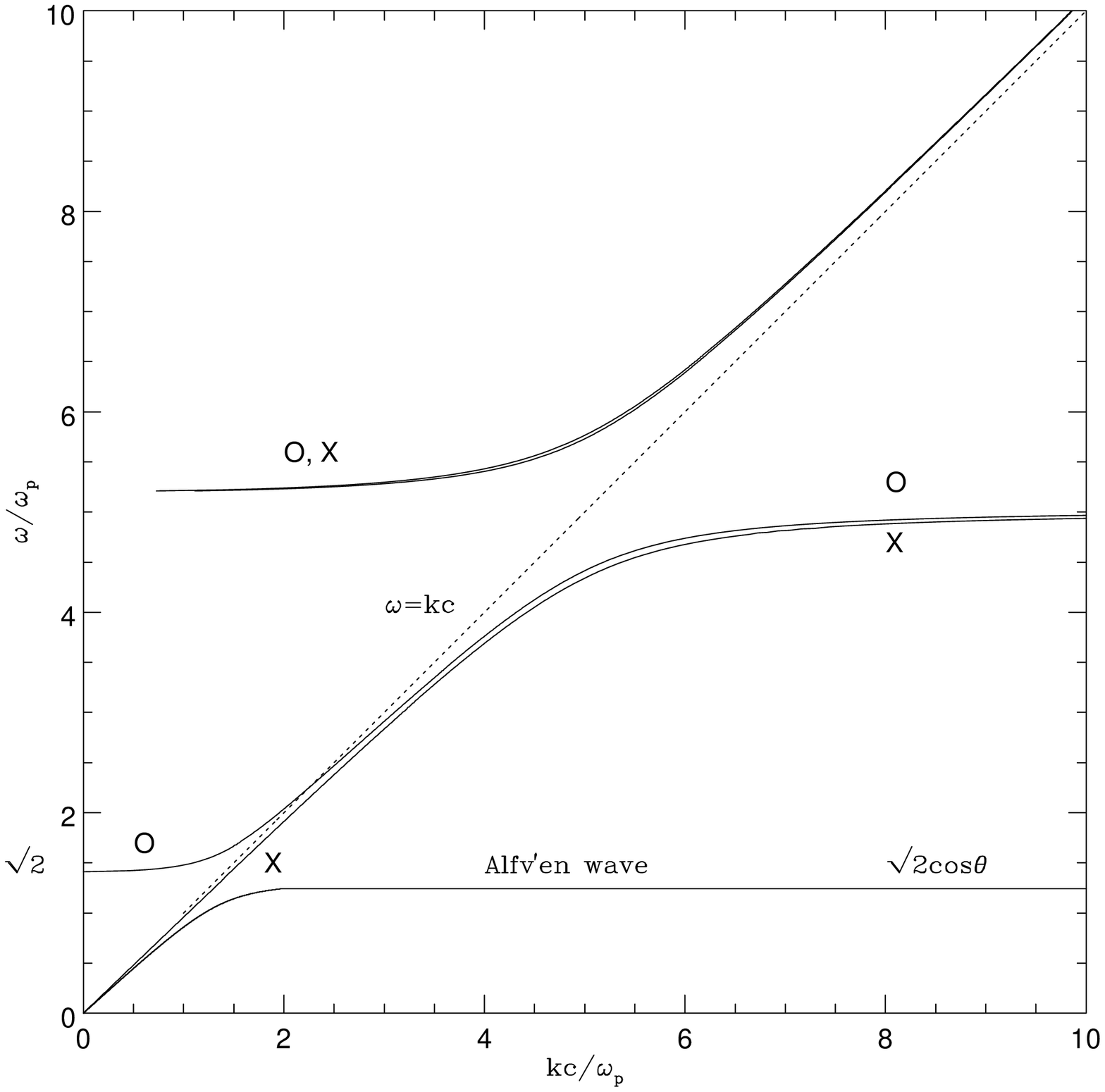,width=12.0cm}
\caption[ Dispersion curves for the waves in a cold electron-positron
plasma in the plasma frame]{}
 {
Dispersion curves for the waves in a cold electron-positron
plasma in the plasma frame for oblique propagation ($\theta=0.5$). 
There are three
modes: Ordinary (O), Extraordinary (X) and  Alfv\'{e}n.
For graphic  purposes  the gyrofrequency was chosen to be $\omega_B=5 \omega_p$.
In the high frequency regime $\omega \gg \omega_B$ there are two
subluminous waves with the dispersion relation
$ \omega^2 \approx  k^2 c^2 + 2 \omega_p^2 $ for the X and O modes.
Both X and O modes have resonances at $ \omega = \omega_B$ and
Alfv\'{e}n has a  resonance  at $ \omega = \sqrt{2} \cos \theta \omega_p$.
O mode has a cutoff at $ \omega = \sqrt{2} \omega_p$.
O mode crosses the vacuum dispersion relation at
$ \omega^2 = 2 \omega_p^2 + \omega_B^2 \sin \theta^2$.}
\label{WaveCold}
\end{figure}

\begin{figure}
\psfig{file=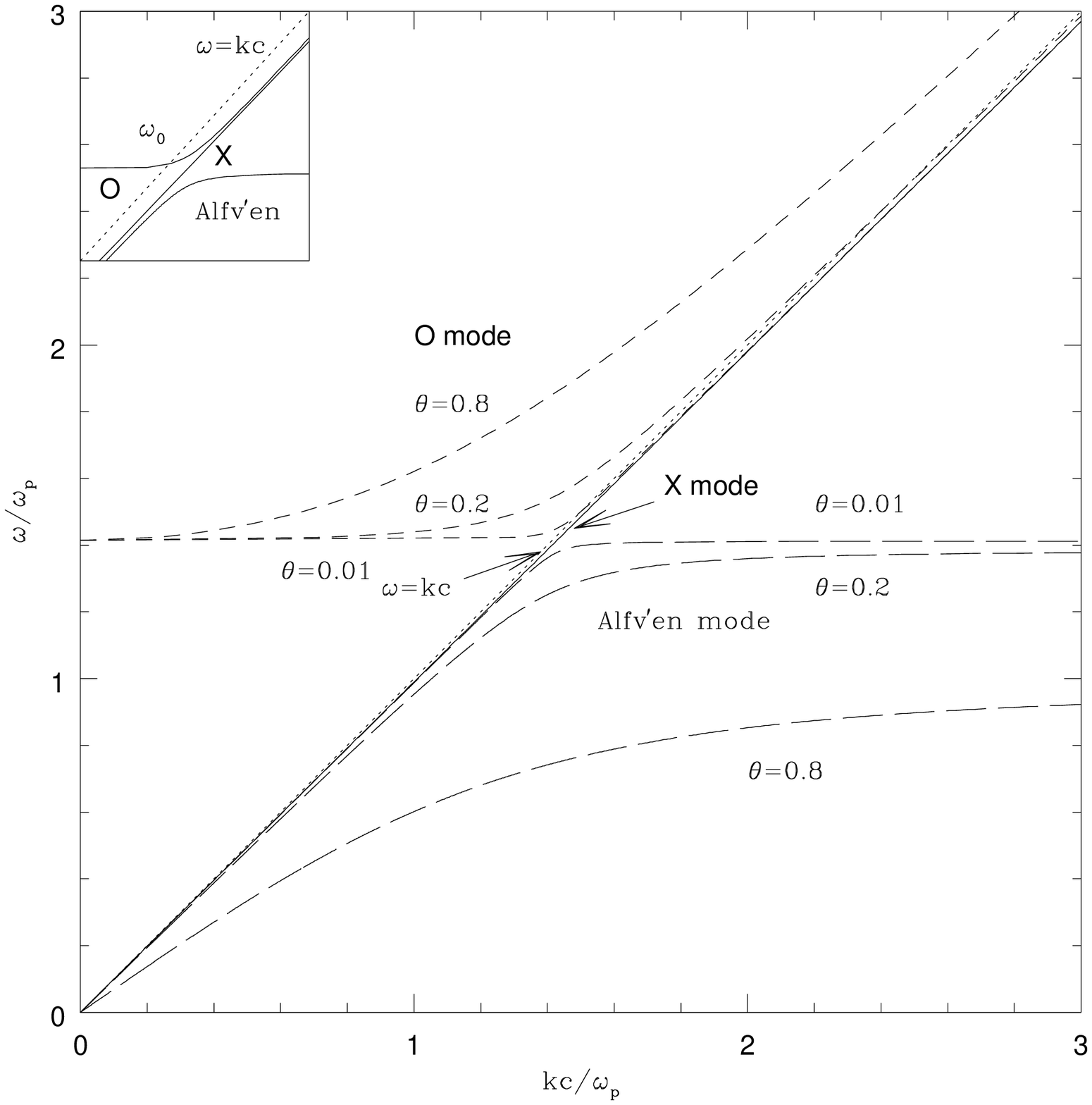,width=12.0cm}
\caption[ Dispersion curves for the waves in a cold electron-positron
plasma in the plasma frame in the limit $\omega_p \ll \omega_B$]{
Dispersion curves for the waves in a cold electron-positron
plasma in the plasma frame in the limit $\omega_p \ll \omega_B$. There are three modes
represented by the  dashed (O mode), solid
(X mode) and long dashed (Alfv\'{e}n mode). The dotted line
represents the vacuum dispersion relation. 
For the exact parallel
propagation,  the dispersion curves for the O mode and
Alfv\'{e}n mode intersect.
The insert in the upper left corner shows the region near the cross-over
point $\omega_0$.
\label{fig1}
}
\end{figure}

Equation (\ref{dettt}) with the dielectric tensor (\ref{epsilon21})
   factorizes giving 
the three wave branches: X  and two coupled
 O
and Alfv\'{e}n branches with the index of refraction $n$ given by 
\begin{eqnarray}
&  n^2 = 1 -
\mbox{{\Large $
 {{2\,{{{\it  \omega_p}}^2}}\over
       {{{\omega }^2} - {{{\it  {\omega_B}}}^2}}}
$}}
 \hskip 1.5 truein
 & \mbox{ X mode}
\label{fjhi1}
\mbox{} \\ \mbox{}
&  n^2 =
\mbox{{\Large $
  {{\left( {{\omega }^2} - 2\,{{{\it  \omega_p}}^2} \right) \,
       \left( {{\omega }^2} - {{{\it  {\omega_B}}}^2} - 2\,{{{\it  \omega_p}}^2}
          \right) }\over
 {{{\omega }^4} - {{\omega }^2}\,{{{\it  {\omega_B}}}^2} -
       2\,{{\omega }^2}\,{{{\it  \omega_p}}^2} +
       2\,{{{\it  {\omega_B}}}^2}\,{{{\it  \omega_p}}^2}\,{{\cos  ^2 \theta}}}}
$}}
 & \mbox{ Alfv\'{e}n and O  mode}
\label{fjhi}
\end{eqnarray}
where $\theta$ is the angle of propagtion with respect to the magnetic field
(see Figures  \ref{WaveCold} and  \ref{fig1}).

Equation (\ref{fjhi1}) describes  the transverse X wave with the
electric vector perpendicular to the {\bf k-B} plane and  
equation (\ref{fjhi}) describes the
coupled longitudinal-transverse wave which  has two branches:
 O quasi-transverse wave with
 the electric vector in the {\bf k-B} plane and quasi-longitudinal
Alfv\'{e}n wave with the electric vector along {\bf  B}.
 
\subsection{Parallel Propagation}

The normal modes of the plasma for the parallel propagation are
given by
\begin{eqnarray}
\hskip -.3 truein
&& \omega_l\,=\, \sqrt{2} \, \omega_p
\label{fhdu1334}
\mbox{}  \\ \mbox{}
&& n_t^2=  1 +
{2\, \omega_p^2\over \omega_B^2 -\omega_t ^2}
\label{fhdu133}
\end{eqnarray}
(subscripts $_l$ and $_t$ refer to longitudinal and transverse polarizations of waves).

 For exactly  parallel
propagation,  the dispersion curves for the O mode and
Alfv\'{e}n mode intersect at
\begin{equation}
\omega^{\ast}= \sqrt{2} \omega_p  
\hskip .3 truein 
k^{\ast} c
 \approx \sqrt{2} \omega_p \left(
1+ \omega_p^2/{\omega_B}^2\right)
 , \mbox{ for ${\omega_p \over \omega_B }\ll 1$ }
\label{tte}
\end{equation}
This intersection occurs only
for the parallel propagation,  while for oblique propagation
the dispersion
curves for O, X and  Alfv\'{e}n modes are well
separated. It follows that equation (\ref{fhdu1334})
describes the  O mode for   $k \,< \, k^{\ast} $ and the
Alfv\'{e}n mode for $k \,>\, k^{\ast} $, while equation (\ref{fhdu133})
describes the X mode for all frequencies and the 
Alfv\'{e}n mode for  $k \,< \, k^{\ast} $ and the O mode for
 $k \,>\, k^{\ast} $.

As the wave propagates in a curved magnetic field of the pulsar
magnetosphere, it's path on the CMA diagram (see Sections \ref{CMADiagram})
depends on the branch that the wave belongs to. In the linear regime in wave amplitude
and with the adiabatically changing  parameters of a medium, a wave always stays
on a given branch. For example, original electrostatic wave
emitted along  the magnetic field can acquire 
electromagnetic components as it propagates
in  the presence of a  curved magnetic field.
The linear evolution of the longitudinal plasma wave, emitted originally
along the field line, will drastically depend on which branch (O
or  Alfv\'{e}n) the wave actually belongs to.
The
propagation of O and Alfv\'{e}n  waves in the inhomogeneous
plasma of a pulsar magnetosphere differs considerably
(\cite{arons1}, \cite{BarnardArons}).
If the plasma wave is emitted with the wave vector (or frequency) above the 
intersection  point $k \,>\, k^{\ast} $, then the linear
transformation of the longitudinal plasma wave will follow the
 Alfv\'{e}n branch, which cannot escape magnetosphere.
On the other hand, if the plasma wave is emitted with the wave vector 
below the intersection point $k \,<, k^{\ast} $, then the linear
transformation of the longitudinal plasma wave will follow the
O branch, which may escape from plasma.

\subsection{Oblique Propagation}

In the pulsar magnetosphere
 the waves that may be important for the generation of the
observed radio emission have frequencies much less than the 
gyrofrequency. In what follows, we will often use the low frequency 
approximation when all the relevant frequencies 
are much less than the gyrofrequency. In the cold plasma in its rest frame this 
implies: $\omega \ll \omega_B$. 
The solution of equation (\ref{fjhi1}) in the low frequency limit
describes a {\it subluminous}  transverse electromagnetic wave:
\begin{equation}
\omega^2= k^2 c^2 \left( 1- { 2 \omega_p^2 \over \omega_B^2} \right)=k^2 v_A^2 ,
\hskip .2 truein \mbox{ $\omega \ll \omega_B$},
\hskip .2 truein \mbox{ X mode}
\label{X}
\end{equation}
where $v_A$ is the Alfv\'{e}n velocity in a strongly magnetized plasma.

Solutions of equation (\ref{fjhi})  are more complicated. The simple form
for the dispersion relation may be obtained near the  cross-over point,
where the dispersion relation of the  O mode crosses the  vacuum
dispersion relation or in the asymptotic regimes far from the  cross-over point.

Solving (\ref{fjhi}) with the refractive index set to 
unity we find the cross-over point for the O wave.
\begin{equation}
\omega_0^2 =\, k_0^2\, c^2  =\,
  2\,{{{\it  \omega_p}}^2} + {{{\it  {\omega_B}}}^2}\,{{\sin  ^2 \theta}}
\label{fjhi2}
\end{equation}

Near the cross-over point, the approximate 
dispersion relation for the O mode
 may be found using the relation
\begin{equation}
\omega- \omega_0 = - \left. \left( { \partial K( \omega,{\bf k})
 \over \partial {\bf k} }\right)
 \right/ 
 \left( { \partial K ( \omega,{\bf k})  \over \partial \omega } \right)
d {\bf k}
\label{fjhi02}
\end{equation}
where $ K ( \omega,{\bf k}) = 0 $ is the dispersion 
equation for $\omega ( {\bf k})$.

From Eq. (\ref{fjhi}) we find
\begin{equation}
\omega = k_0\, c + \kappa ( k-  k_0) c
\label{fjhi3}
\end{equation}
where 
\begin{eqnarray}
&&
\kappa = {1\over c} \left. {\partial \omega \over \partial k } \right| _
{ k = k_0} = \,
 {{{{{\it  {\omega_B}}}^4}\,{{\cos  ^2 \theta}}\,{{\sin  ^2 \theta}}}\over
 {4\,{{{\it  \omega_p}}^4} + 2\,{{{\it  {\omega_B}}}^2}\,{{{\it  \omega_p}}^2}\,
       {{\sin  ^2 \theta}} + {{{\it  {\omega_B}}}^4}\,{{\cos  ^2 \theta}}\,
       {{\sin  ^2 \theta}}}}
\mbox{} \nonumber \\ \mbox{}
&&
\, \approx \,
 1 - {{4\,{{{\it  \omega_p}}^4}}\over
     {4\,{{{\it  \omega_p}}^4} + {{{\it  {\omega_B}}}^4}\,{{\cos  ^2 \theta}}\,
        {{\sin  ^2 \theta}}}}
\label{fjhi4}
\end{eqnarray}
where we used the assumption $  {\omega_B} \, \gg  \omega_p$.
From (\ref{fjhi4}) it follows, that 
the behavior  of the dispersion relation of the O wave 
 near the cross-over point shows a very sensitive dependence 
on the angle of propagation.
There exist a critical angle $\theta _c = {2  \omega_p^2 /\omega_B^2}$ 
at which the dispersion relation changes:
\begin{eqnarray}
\hskip -.7 truein
&\omega = \, \sqrt{2}\,\omega_p  + {{\omega_B}^4 \over 2 \, \omega_p^4} \, 
(k-\, k_0)c \, , \hskip .3 truein   k_0^2 \,\approx 2 \, \omega_p^2/c^2
& \mbox { if $\theta \, \ll 2  \omega_p^2/{\omega_B}^2$}
\mbox{} \nonumber \\ \mbox{}
\hskip -.5 truein
& \omega = \, k c -  { {{4\,\left( k - {\it k_0} \right)c\,{{{\it \omega_p}}^4}\,
\over 
      {{\sin  ^2 \theta}}\,{{\cos ^2 \theta}}} {{{{\it {\omega_B}}}^4}}} } ,
 \hskip .1 truein
 k_0^2 c^2 \,= 2 \, \omega_p^2 +  \, {\omega_B}^2\, \sin ^2 \theta
& \mbox { if $ \theta \, \gg  \,  2 \omega_p^2/{\omega_B}^2  $}
\label{fjhi14}
\end{eqnarray}
 For angles smaller than 
 $\theta _c$ we can generally use the approximation of 
parallel propagation when considering the dispersion relations of the
waves, while for larger angles we must take into account the effects
of oblique propagation.

The other limits when the dispersion relations for the O and Alfv\'{e}n
waves may be obtained in closed form are the asymptotic limits 
far from the cross-over point. The
 large and small wave vector asymptotic solutions are 
\begin{eqnarray}
&& 
\hskip -.4 truein
\omega^2 =  \left\{  \begin{array}{ll}
 \phantom{ {{{ {a\over b} \over {a\over b} } } \over {{ {a\over b} \over 
{a\over b} } } } }
{ k^2 c^2 }\,\left( 1 - {{2\,{{{\it  \omega_p}}^2}\,{{\cos  ^2 \theta}}}\over
         {{{{\it  {\omega_B}}}^2}}} \right)  +
   2\,{{{\it  \omega_p}}^2}\,{{\sin  ^2 \theta}}  &\mbox{ O wave}
\\
\phantom{ {{{ {a\over b} \over {a\over b} } } \over {{ {a\over b} \over 
{a\over b} } } } }
 2\,{{{\it  \omega_p}}^2}\,{{\cos  ^2 \theta}}\,
\left(1 - {{2\,{{{\it  \omega_p}}^2}\,{{\sin  ^2 \theta}}}\over {{ k^2 c^2 }}} -
      {{2\,{{{\it  \omega_p}}^2}\,{{\sin  ^2 \theta}}}\over
        {{{{\it  {\omega_B}}}^2}}} \right)  & \mbox{ Alfv\'{e}n wave}
\end{array}\right.
 \hskip -.3 truein
 \mbox{ if $ kc \, \gg \omega_p$}
\label{fhdu12}
\mbox{} \\ \mbox{}
&& \hskip -.4 truein
\omega^2 =  \left\{  \begin{array}{ll}
 \phantom{ {{{ {a\over b} \over {a\over b} } } \over {{ {a\over b} \over 
{a\over b} } } } }
 2\,{{{\it  \omega_p}}^2} + { k^2 c^2 }\,
   \left( 1 - {{{ k^2 c^2 }\,{{\cos  ^2 \theta}}}\over {{{{\it  \omega_p}}^2}}}
        \right) \,{{\sin  ^2 \theta}}  & \mbox{ O wave }
\\
 \phantom{ {{{ {a\over b} \over {a\over b} } } \over {{ {a\over b} \over 
{a\over b} } } } }
 { k^2 c^2 }\,{{\cos  ^2 \theta}}\,
    \left( 1 - {{2\,{{{\it  \omega_p}}^2}}\over {{{{\it  {\omega_B}}}^2}}} -
    {{{ k^2 c^2 }\,{{\sin  ^2 \theta}}}\over {2\,{{{\it  \omega_p}}^2}}} \right)
& \mbox{ Alfv\'{e}n wave}
\end{array}\right.
\hskip -.2 truein
\mbox{ if $ kc \, \ll \omega_p$}
\label{fhdu122}
\end{eqnarray}
\subsection{Infinite Magnetic Field}

In the limit of infinitely strong magnetic field 
 the dispersion relations for the O 
(plus sign)  and Alfv\'{e}n modes (minus sign) are
(\cite{arons1}) 

\begin{equation}
\omega^2 = {{k^2 c^2 }\over 2} + {{{\it  \omega_p}}^2} \pm
 {{{\sqrt{{k^4} c^4 + 4\,{{{\it  \omega_p}}^4} -
          4\,k^2\, c^2{{{\it  \omega_p}}^2}\,\cos (2\,\theta)}}}\over 2}
\label{fhdu139}
\end{equation}
The short and long wave length asymptotics are then given by 
 (\ref{fhdu12}) and (\ref{fhdu122}) with the magnetic field set to infinity.

An important point in considering the wave excitation in the superstrong
magnetic field is that we {\it cannot} neglect the very large but
finite magnetic field. In the approximation of the  infinitely strong magnetic field 
the O mode is always superluminous and thus cannot be excited by the
Cherenkov-type 
resonant wave-particle  interaction.  In this limit any instability
would occur on the Alfv\'{e}n waves which are strongly damped as they
propagate out in the pulsar magnetosphere. When the finite
magnetic field is taken into account, O wave becomes subluminous
for the small angles of propagation and can be resonantly  excited by the
Cherenkov, cyclotron or Cherenkov-drift 
interaction with the fast particles.

\subsection{CMA Diagram for Cold Pair Plasma}
\label{CMADiagram}

CMA diagrams (e.g. \cite{Budden}) are
 useful tools in considering wave propagation.
It is a plot of the refractive index versus some functions of wave,
plasma and cyclotron  frequencies.
We chose the following coordinates for  CMA diagram:
\begin{equation}
W = {1\over Y^2} = \left({\omega \over \omega_B}\right)^2,
\hskip .2 truein Z= {X\over  Y^2} =  \left({\omega_p\over \omega_B}\right)^2
\label{ab1}
\end{equation}
where $Y$ and $X$ are the standard quantities in the magnetosonic theory.
With this choice of coordinates the lines of constant  $Z$ are
the lines of constant density and are independent
of wave frequency. The lines of constant $W $  are
the lines of constant wave frequency and  are independent
of the  density.
The regions on the CMA  diagram are separated by the resonance,
where $n \rightarrow \infty$, and cutoffs, where $n \rightarrow 0$.

Using Eqs (\ref{fjhi1}) and (\ref{fjhi}) we find  resonance
\begin{eqnarray}
& W =1 & \mbox{X modes}
\label{ab2}
\mbox{} \\ \mbox{}
& W = {1\over 2} + Z \pm \sqrt{ \left(  {1\over 2} + Z \right)^2 - 2 Z \cos^2
\theta }
 & \mbox{ O \& Alfv\'{e}n modes}
\label{ab3}
\end{eqnarray}
and reflection points
\begin{eqnarray}
& W= 1+ 2 Z  & \mbox{X modes}
\label{ab4}
\mbox{} \\ \mbox{}
& \left\{
\begin{array} {cc}
  W= 1+ 2 Z&\\
W= 2 Z
\end{array} \right.  & \mbox{ O \& Alfv\'{e}n modes}
\label{ab5}
\end{eqnarray}

\begin{figure}
\psfig{file=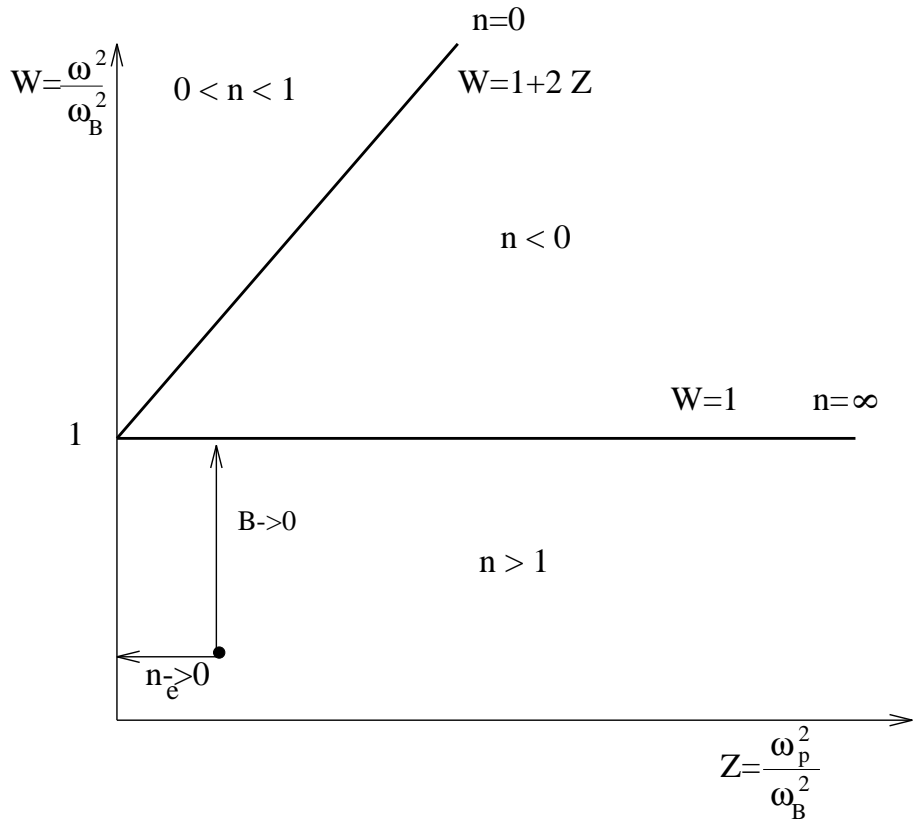,width=12.0cm}
\caption[CMA diagram for the X mode]{
CMA diagram for the X mode.The  vacuum case corresponds to $Z=0$.
On the axis $W=0$ refractive index is $n=1$.
Resonance occurs at $W=1$ ( $n = \infty$)
and reflection occurs at $W=1+ 2 Z$.
Typical X  waves  in the pulsar magnetosphere have $ Z \ll 1$,
$ W \ll 1$  and $ n > 1$ deep in the magnetosphere.
Arrows indicate the adiabatic tracks for the constant density and
decreasing magnetic field ($ B \rightarrow 0$) and
constant magnetic field and decreasing density ($ n_e \rightarrow 0$).
\label{CMAX}}
\end{figure}

\begin{figure}
\psfig{file=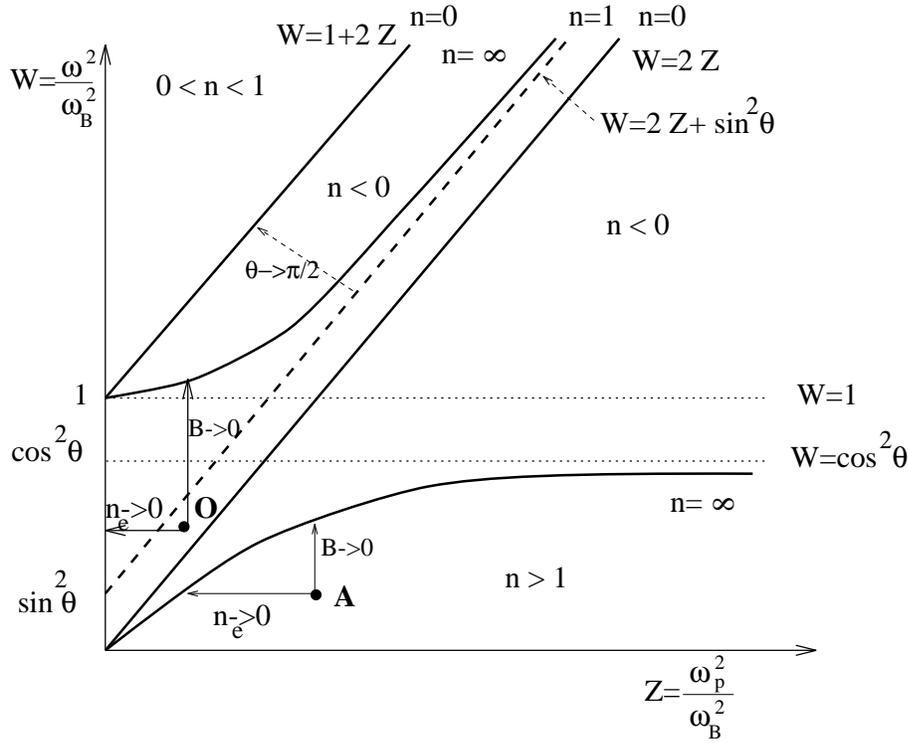,width=12.0cm}
\caption[CMA diagram for the O and Alfv\'{e}n modes]{
CMA diagram for the O and Alfv\'{e}n modes.  Vacuum case corresponds to $Z=0$.
On the axis $W=0$ refractive index is $n=1$.
Resonances  occur at  $n = \infty$ and reflections
 occur  at $W=1+ 2 Z$ and $W=2 Z$ ( $n = 0$). The curve
 $W=1+ 2 Z$ corresponds to the upper hybrid wave $ \omega^2 = \omega_B^2+
2  \omega_p^2$ and the  curve $W=2 Z$ corresponds to the plasma wave
 $ \omega^2 = 2  \omega_p^2$.
Typical O
 waves (denotes by $ {\bf O}$)  in the pulsar magnetosphere have $ Z \ll 1$,
$ 2 Z <  W \ll 1$.
Typical Alfv\'{e}n modes (denotes by $ {\bf A}$)
in the pulsar magnetosphere have $ W \ll 1 $ and $n >1 $.
The arrows $B \rightarrow 0$,  $ n_e \rightarrow 0$
and $\theta \rightarrow \pi/2$ indicate correspondingly adiabatic
tracks for constant density and decreasing magnetic field,
constant magnetic field and decreasing density and
increasing angle of propagation.
\label{CMAO}}
\end{figure}

For the $X$ wave the curve $n=1$ corresponds to $Z=0$ (vacuum case).
For the
 O
mode $n=1$ at $ W= 2 Z + \sin ^2 \theta$ (cross-over point) and
$Z=0$ (vacuum case).
Other useful relations for the resonances of the  coupled
 O
and Alfv\'{e}n modes are
\begin{eqnarray}
& {1\over 2} + Z +  \sqrt{ \left(  {1\over 2} + Z \right)^2 - 2 Z \cos^2
\theta }
= & \left\{
\begin{array} {cc}
1  & \mbox{ $\theta=0$}\\
1+ 2 Z  &  \mbox{ $\theta=\pi/2$} \\
1+ 2 Z \sin^2 \theta  &  \mbox{ $Z \ll 1$}\\
2 Z +  \sin^2 \theta &  \mbox{ $Z \gg 1$}
\end{array} \right.
\mbox{} \nonumber \\ \mbox{}
&
{1\over 2} + Z  -  \sqrt{ \left(  {1\over 2} + Z \right)^2 -2 Z \cos^2
\theta }
= & \left\{
\begin{array} {cc}
2 Z & \mbox{ $\theta=0$}\\
0 &  \mbox{ $\theta=\pi/2$} \\
2 Z \cos^2 \theta  &  \mbox{ $Z \ll 1$}\\
\cos^2 \theta  &  \mbox{ $Z \gg 1$}
\end{array} \right.
\label{ab6}
\end{eqnarray}
The CMA diagrams are plotted in Figs. \ref{CMAX} and  \ref{CMAO}.

\subsection{Polarization of waves in cold plasma}

To find the polarizations of the waves we construct 
a matrix of cofactors of $\Lambda$ (\cite{Melrosebook1}) :
\begin{eqnarray}
&&
\lambda _{\alpha \beta} = n^4 k_{\alpha} k_{\beta} -
n^2 \left(  k_{\alpha} k_{\beta} \epsilon_{\gamma \gamma}  +
\delta_{\alpha \beta} k_{\gamma} k_{\eta} \epsilon_{\gamma \eta} -
k_{\alpha}  k_{\gamma} \epsilon_{\gamma \beta} -
k_{\beta}  k_{\gamma} \epsilon_{ \alpha \gamma} \right)
\mbox{} \nonumber \\ \mbox{}
&&
 \hskip .3 truein+
{1\over 2} \delta_{\alpha \beta} \left(
\epsilon_{\gamma \gamma}^2 - \epsilon_{\gamma \eta} \epsilon_{ \eta \gamma}
\right) + \epsilon_{\alpha \gamma} \epsilon_{\gamma \beta} -
\epsilon_{\gamma \gamma}  \epsilon_{\alpha \beta}
\label{fhdu1391}
\end{eqnarray}
Then 
the polarization vectors may be chosen as columns of $\lambda_{\alpha \beta}$.

For  cold plasma,  the elements of $\lambda _{\alpha \beta}$ are
\begin{eqnarray}
&& 
\lambda _{xx} = 
 \left( -1 + n^2 - {\frac{2\,\omega_p^2}
       {-{{\omega}^2} + {{\omega_B}^2}}} \right) \,
   \left( -1 + {\frac{2\,\omega_p^2}{{{\omega}^2}}} +
     n^2\,{{\sin^2\theta}} \right)
\mbox{} \nonumber \\ \mbox{}
&&
\lambda _{x z} =
 n^2\,\left( -1 + n^2 + {\frac{2\,\omega_p^2}
       {{{\omega}^2} - {{\omega_B}^2}}} \right) \,\cos \theta\,
   \sin \theta = \lambda _{zx}
\mbox{} \nonumber \\ \mbox{}
&&
\lambda _{yy} =
 -\left( \left( -1 + n^2 + {\frac{2\,\omega_p^2}
          {{{\omega}^2}}} \right) \,
      \left( 1 + {\frac{2\,\omega_p^2}
          {-{{\omega}^2} + {{\omega_B}^2}}} \right)  \right)  +
 {\frac{2\,n^2\,{{\omega_B}^2}\,\omega_p^2\,
       \cos^2\theta}{{{\omega}^2}\,
       \left( -{{\omega}^2} + {{\omega_B}^2} \right) }}
\mbox{} \nonumber \\ \mbox{}
&&
\lambda _{z z} =
 \left( 1 - n^2 + {\frac{2\,\omega_p^2}
       {-{{\omega}^2} + {{\omega_B}^2}}} \right) \,
   \left( 1 + {\frac{2\,\omega_p^2}
       {-{{\omega}^2} + {{\omega_B}^2}}} -
     n^2\,\cos^2\theta \right)
\label{fhdu1393}
\end{eqnarray}
We note that these relations are exact in frequency.

For the X mode using (\ref{fjhi1}) for the refractive
index in (\ref{fhdu1393})  we find the polarization vector for the
X mode
$e_X = (0,1,0)$.
For the O and Alfv\'{e}n modes using (\ref{fjhi})   for the refractive
index in (\ref{fhdu1393})
we obtain the ratio of the electric field components in the wave:
\begin{equation}
{E_x \over E_z} =
 -{\frac{\left( {{\omega}^2} - {{\omega_B}^2} \right) \,
       \left( {{\omega}^2} - 2\,\omega_p^2 \right) \,
       \cot \theta}{{{\omega}^2}\,
       \left( {{\omega}^2} - {{\omega_B}^2} -
         2\,\omega_p^2 \right) }}
\label{fhdu1394}
\end{equation}

For the point far from the cross-over point we can use the approximation of
a very strong magnetic field to find
\begin{equation}
{E_x \over E_z} =
 \left( -1 + {\frac{2\,\omega_p^2}{{{\omega}^2}}} \right) \,
 \left( 1 - 2 {\omega_p^2 \over \omega_B^2} \right)   \cot \theta
\label{fhdu1395}
\end{equation}

Using the relation (\ref{fhdu1394}) we can estimate the polarization of the
O wave at the cross-over point. We find that 
 \begin{equation}
{E_x \over E_z} \approx 
 {\frac{2\,{{\omega_p}^2}\,\theta}{{{\omega_B}^2}}}
\label{fhdu13961}
\end{equation}
For the angles smaller than ${ \omega_p^2 / \omega_B^2}$ the 
O wave is quasi-longitudinal at the  cross-over point and for
larger angles it is quasi-transverse.

Relations (\ref{fhdu1395}) and (\ref{fhdu13961}) 
 allow us to find the normalized  polarization vectors:
\begin{eqnarray}
e_O=&&  \left\{  \begin{array}{ll}
\hskip -.2 truein 
 \phantom{ {{{ {a\over b} \over {a\over b} } } \over {{ {a\over b} \over 
{a\over b} } } } }
 \left\{\cos \theta\,\left( 1 - {\frac{2\,{{\omega_p}^2}\,
          {{\sin^2\theta}}}{{{\omega}^2}}} \right) ,0,
   \left( 1 + {\frac{2\,{{\omega_p}^2}\,{{\cos^2\theta}}}
        {{{\omega}^2}}} \right) \,\sin \theta \right\}
+  O\left( { \omega_p^2  \over  k^2 c^2 } \right) 
 & \mbox{ $ kc \, \gg \omega_p$}   
\\
\hskip -.2 truein
\phantom{ {{{ {a\over b} \over {a\over b} } } \over {{ {a\over b} \over 
{a\over b} } } } }
 \left\{  {\frac{-\left( k^2\,\sin (2\,\theta) \right) }
     {4\,\omega_p^2}},0,1\right\} +
 O\left( {  k^2 c^2  \over  \omega_p^2  } \right)
 & \mbox{ $kc \, \ll \omega_p$}
\end{array}\right.
\label{fhdu13951}
\mbox{} \\ \mbox{}
e_O=&&  \left\{  \begin{array}{ll}
\phantom{ {{{ {a\over b} \over {a\over b} } } \over {{ {a\over b} \over 
{a\over b} } } } }
\left\{   {\frac{{{\omega_B}^2}\,\theta}{{{\omega_0}^2}}},0,-1\right\}
        &
\mbox{ $ \theta \ll { 2 \omega_p^2 \over \omega_B^2} $ }\\
\phantom{ {{{ {a\over b} \over {a\over b} } } \over {{ {a\over b} \over 
{a\over b} } } } }
\left\{   1,0,-{\frac{{{\omega_0}^2}\,\csc \theta\,\sec \theta}
       {{{\omega_B}^2}}}\right\} &
\mbox{ $ \theta \gg  { 2 \omega_p^2 \over \omega_B^2} $}
\end{array}\right. \hskip .2 truein
\mbox{ $ \omega \approx \omega_0$}
\label{fhdu139511}
\mbox{} \\ \mbox{}
e_A =&&  \left\{  \begin{array}{ll}
\hskip -.2 truein
\phantom{ {{{ {a\over b} \over {a\over b} } } \over {{ {a\over b} \over 
{a\over b} } } } }
\left\{  \left( 1 + {\frac{2\,{{\omega_p}^2}\,{{\cos^2\theta}}}
        {k^2}} \right) \,\sin \theta,0,
   \cos \theta\,\left( 1 - {\frac{2\,{{\omega_p}^2}\,
          {{\sin^2\theta}}}{k^2}} \right)\right\} +
 O\left( { \omega_p  ^2 \over
k^2} \right)
& \mbox{ $ kc \, \gg \omega_p$} 
 \\
\hskip -.2 truein
\phantom{ {{{ {a\over b} \over {a\over b} } } \over {{ {a\over b} \over 
{a\over b} } } } }
\left\{ 1,0,{\frac{{{\omega}^2}\,\tan \theta}{2\,\omega_p^2}}\right\}
 + O\left( {  k^2 c^2  \over  \omega_p^2  } \right)
 & \mbox{ $kc \, \ll \omega_p$}
\end{array}\right.
\label{fhdu1396}
\end{eqnarray}
which are  accurate  to $ O\left( { \omega_p^2 \over
 {\omega_b}^2} \right)$.

\section{Waves  in a Cold Pair Plasma in Pulsar Frame}
\label{ColdPulsar}

In the pulsar magnetosphere the plasma is moving along the field lines
with a bulk Lorenz factor $\gamma_p \approx 10-1000$. 
In this frame the waves propagating in different directions have different
dispersion relations. So that the dispersion 
equation for the X mode becomes a fourth order equation for
$\omega({\bf k} )$. 
To simplify the consideration we will use the low frequency approximation
from the very beginning, i.e. we expand all the relevant relations
in $ 1/\omega_B$. In what follows,
the quantities measured in the pulsar frame will be denoted with primes.

For the  forward propagating waves, which in the plasma 
frame has $\theta \ll 1$,  
we obtain 
\begin{eqnarray}
&\hskip -1 truein
\omega_X ^{\prime} = k ^{\prime}
 c( 1 - \mbox{ {\large $
 {\omega_p^{\prime \, 2} \over 4 \gamma_p^3 \omega_B^2}$}})
&  \mbox{} \nonumber \\ \mbox{}
&
\omega_O ^{\prime}=
\left\{  \begin{array}{ll}
\phantom{ {{{ {a\over b} \over {a\over b} } } \over {{ {a\over b} \over 
{a\over b} } } } }
 k ^{\prime}
 c \left( 1 -
\mbox{ {\large $
 {\omega_p^{\prime \, 2} \over 4 \gamma_p^2 \omega_B^2} 
+ {\frac{  \, \gamma_p{{\, \omega_p}^{\prime \, 2} 
}\,{{\sin^2\theta ^{\prime} }}}
        {{c^2}\,{k^{\prime \, 2} }\,
           }} $}} \right)
& \mbox{ if $ k ^{\prime} c \,\gg  \,\gamma_p  \omega_p ^{\prime} $ }\\
\phantom{ {{{ {a\over b} \over {a\over b} } } \over {{ {a\over b} \over 
{a\over b} } } } }
v_p k^{\prime}   +
{ \sqrt{2} \omega_p^{\prime} \over \gamma_p^{3/2} } 
& \mbox{ if $ k ^{\prime} c \,\ll \,\gamma_p  \omega_p ^{\prime} $ }
\end{array} \right.
\mbox{} \nonumber \\ \mbox{} 
& \omega_A ^{\prime} = 
\left\{  \begin{array}{ll}
\phantom{ {{{ {a\over b} \over {a\over b} } } \over {{ {a\over b} \over 
{a\over b} } } } }
k ^{\prime} c \cos\theta ^{\prime} 
 \left(1 -  \mbox{ {\large $
 { \, \omega_p^{\prime \, 2}\,
\over 4  \,  \gamma_p^3  \, \omega_B^2 }   \,
   - {\frac{{c^2}\,{k^{\prime \, 2} }\,{{\sin^2\theta^{\prime}  }}}
        {4\,\gamma_p {{\, \omega_p}^{\prime \, 2}}}}  $}} \right)
& \mbox{ if $ k ^{\prime} c \,\ll  \,\gamma_p  \omega_p ^{\prime} $ }
\\
\phantom{ {{{ {a\over b} \over {a\over b} } } \over {{ {a\over b} \over 
{a\over b} } } } }
v_p k^{\prime}   +
\mbox{ {\large $
{ \sqrt{2} \omega_p^{\prime} \cos \theta ^{\prime} \over \gamma_p^{3/2} }
$}}
& \mbox{ if $ k ^{\prime} c \,\gg \,\gamma_p  \omega_p ^{\prime} $ }
\end{array} \right.
\label{qw}
\end{eqnarray}
The plasma frequencies in the two
frames are related by $\omega_p^{\prime} = \omega_p/\sqrt{\gamma_p}$.

The cross-over point, where the O mode becomes luminal, is 
\begin{equation}
\omega_0^{\prime \, 2} = { 2 \omega_p ^{\prime  \, 2}
\over \gamma_p} + { 4 \gamma_p^2 
\omega_B^2 \theta ^{\prime \, 2}  }
\label{qw1}
\end{equation}

\section{Waves in a Relativistically Hot  Pair  Plasma
 }
\label{WavesHot}

\subsection{Effects of Thermal Motion on Wave Dispersion}
\label{Effec} 

In this section we consider  wave  propagation
 in the relativistically hot,
strongly magnetized electron-positron plasma. 
The thermal motion of plasma particles affects
considerably the 
dispersion of the Alfv\'{e}n mode at frequencies $\omega \geq \omega_p$  and
 the dispersion of the O mode frequencies $\omega \approx \omega_p$.
Another important {\it quantitative } modification is in the
 dispersion relation of the X mode.
An important factor for the excitation of the 
X mode is the
 difference of its phase speed and the speed of light.
 This difference  is 
roughly proportional to $< 1/\gamma^3 >$ (Eq. \ref{liafsq1}).
It is
 decreased considerably by the bulk streaming of the
plasma.
In the relativistically hot streaming plasma there are 
more particles with low Lorentz factors, that contribute to 
 $< 1/\gamma^3 >$, than in the cold  plasma streaming with  with the same 
average velocity. 
So for a given streaming velocity relativistically hot plasma
has larger   $< 1/\gamma^3 >$ and larger growth rate.

\subsection{Distribution functions}
\label{dis}

To estimate the thermal effects on the dispersion of the plasma mode we use the   two
kinds of distribution functions: 
(i)
waterbag distribution
\begin{equation}
f(p_z) = \, \left\{  \begin{array}{ll}
{n_p\over 2 p_T}, &  \mbox{ if $- p_T \, < p_z \, < p_T$} \\
0, &  \mbox{ otherwise}
\end{array} \right.
\label{flioj}
\end{equation}
(here $p_T \approx m c \gamma_T$ is the scatter in moments)
and (ii)
relativistic Maxwellian distribution (see also Appendices
\ref{RelativisticMax} and \ref{Cutoff} for the calculations of the relevant 
moments of the distribution)
\begin{equation}
f(p_z) = 
{n_p \over 2  K_1 (\beta_T) } \exp \left\{ - \beta_T p_{\mu} U^{\mu} \right\}
\label{flio1j}
\end{equation}
here  $\beta_T=1 /T_p$,
$T_p$ is the invariant temperature,
$p_{\mu} $ is a four-momentum of
 the particle,
 $U ^{\mu} $ is four velocity of the
reference frame, $ K_1$ is a modified Bessel function.
In most of the calculations to follow 
we will assume that the plasma is very hot: $p_T/mc \gg 1$ and
$ \, T_p  =\sqrt{p_T^2/(mc)^2+1} \gg 1$.

Both these distributions are "fast falling" at large moments.
This is an important factor for the dispersion relation
of plasma waves (see below). The advantage of the  water bag
distribution is  that the various moments of the
distribution can be easily calculated. The relativistic 
 Maxwellian distribution is explicitly Lorentz-invariant (see  Appendix
\ref{RelativisticMax} for details of Lorentz transformation).

The relevant moments of the distributions
are summarized in Table \ref{WaterTbl} for the water bag
distribution (in the plasma frame only) and in Table \ref{Maxwe} 
for the relativistic
 Maxwellian distribution (in both plasma and pulsar frame).

 \begin{table}
\begin{tabular}{|c|c|c|c|} \hline
$\phantom{ {{{ {a\over b} \over {a\over b} } } \over {{ {a\over b} \over
{a\over b} } } } }$ 
$<\gamma>$ & 
 $<p v> $ &
 $ \mbox{ {\large $ < {1\over \gamma} > $ }}  $ &
 $ \mbox{ {\large $ < {1\over \gamma^3} > $ }}  $    \\ \hline
 $\phantom{ {{{ {a\over b} \over {a\over b} } } \over {{ {a\over b} \over
{a\over b} } } } }$ $ \gamma_T/2 $ &
$ \gamma_T/2 $ &
{\large $ {\ln \gamma_T \over \gamma_T } $ }&
{\large $ {1 \over \gamma_T } $ } \\
\hline
\end{tabular}
\caption[Moments of the  waterbag distribution]{ Relevant moments of  the water bag
distribution in its rest frame (dimensionless units). It is assumed that $p_T /m_e c
\approx \gamma_T \gg 1$.}
\label{WaterTbl}
\end{table}

\begin{table} \begin{center}
\caption[Moments of the one-dimensional relativistic Maxwellian distribution]
{ Moments of the  one-dimensional relativistic Maxwellian distribution}
\psfig{file=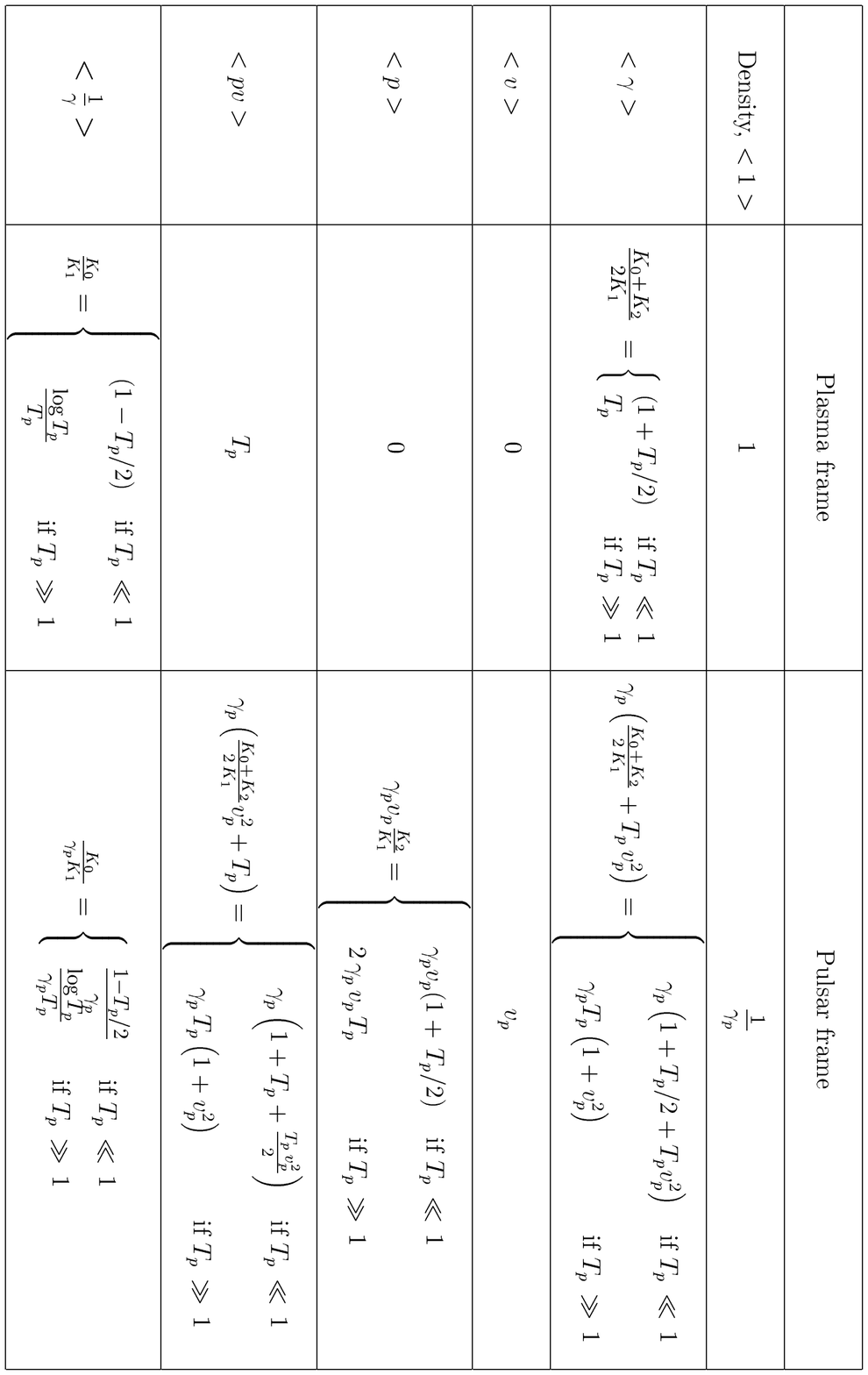,width=18.0cm}
\label{Maxwe}
\end{center} \end{table}

The water bag distribution
 is generally a good approximation for the account of the
thermal motion of the particles. Its major drawback  is the absence of 
a tail of
high energy particles, that can resonate with the waves in the plasma.
The Cherenkov resonance on the tail particles will result in a
strong  damping
of the  waves. The cyclotron resonance on the tail particles may
result in a wave excitation if the distribution function is asymmetric
with   a
long high energy tail.   
The condition, that the Cherenkov resonance is  unimportant,  is that the
phase speed of the waves in plasma is much larger, than the 
thermal velocity of the particles. In the case of the idealized 
water  bag distribution this condition has to be put in by hand. 
Whenever the phase speed of the wave becomes comparable to the 
thermal velocity the waves should be considered strongly damped and
nonexistent. Therefore, we expect that the high frequency branch of the 
Alfv\'{e}n wave, which in the limit of cold plasma had a very low phase
velocity, will be strongly damped.

Here, we should also mention a long standing controversy 
\label{controversy} about the 
dispersion of the longitudinal waves and the 
possibility of the two stream 
instabilities in the relativistic plasma. 
In the initial work (\cite{Silin})  and later works   
(\cite{SuvorovChugunov}) it was stated that the relativistic
plasma does not support subluminous longitudinal waves. This 
problem has been considered anew (\cite{TsytovichKaplan})
who found subluminous waves. The controversy has been resolved
by \cite{LominadzeMikhailovskii} who demonstrated the existence
of the subluminous waves in the range $ 0 < n-1< 1/< \gamma>^2  $ 
($ n  \approx 1$), provided that the third moment of the 
distribution ($< \gamma^3>$) is finite (here 
$n$ is the refractive index and $< \gamma> \gg 1$ is the average
Lorentz factor of the plasma particles). 
Thus, when the distribution function falls off at large momenta slower than
$1\over \gamma^4$ subluminous plasma waves do not exist.

For the water bag distribution, the dispersion of the plasma waves
for the parallel propagation is given by  Eq. (\ref{liafs3}). 
We find that $n-1$ becomes larger than 
$1/\, T_p ^2 $ for $ \omega >  \mbox{ several  times } \omega_0$.
For larger frequencies the Longitudinal plasma waves are
either strongly
damped or do not exist at all (\cite{Silin}).

\subsection{Dispersion Relations in Relativistic Pair Plasma}
\label{dess}

To simplify the analysis we will use the low frequency
approximation $\omega \, \ll {\omega_B}$ and the assumption
of a very strong magnetic field $ {\, T_p  \omega_p^2 \over {\omega_B}^2}
\, \ll 1$ from the very beginning. The dielectric tensor is then given by
\begin{eqnarray}
\epsilon_{xx}&&=   1 +   d\,\, T_p \,
      \left( 1 + n^2\,\beta_T^2\,\cos^2\theta \right)\, 
\, =\epsilon_{yy}
\mbox{} \nonumber \\ \mbox{}
\epsilon_{z z }&&=  1 - {\frac{2\,n^2\,\omega_p^2}
       {\, T_p \,\left( 1 - n^2\,\beta_T^2\,\cos^2\theta
            \right) }} +
  d\,\, T_p \,n^2\,{{\sin^2\theta}}
\mbox{} \nonumber \\ \mbox{}
\epsilon_{xy}&&= 
 \epsilon_{yx}=
\epsilon_{x z}=
\epsilon_{z x}=
\epsilon_{y z}=
\epsilon_{z y}
\label{det1}
\end{eqnarray}
where
\begin{equation}
d = {\omega_p^2 \over \omega_B^2}, \hskip .3 truein \beta_T =
\sqrt{ 1-{ 1\over T_p^2}}
\label{0}
\end{equation}

The normal modes of a hot
 plasma  are given by the solution
of (\ref{dettt}) with the dielectric tensor (\ref{det1}).
 Similarly to the cold case, equation (\ref{det1}) 
factorizes into a dispersion relation for the X mode and a coupled
equation for the Alfv\'{e} and O modes.

\subsection{Parallel Propagation}

In the case of parallel propagation the dispersion equation
gives  two transverse wave with the
dispersion
\begin{equation}
\omega^2= k^2 \, c^2 \left( 1 - d \, \, T_p 
 ( 1 + \beta_T^2) \right)
\label{liafs2}
\end{equation}
and a plasma wave 
\begin{equation}
\omega^2= { 2\, \omega_p^2 \over \, T_p }+ k^2 c^2 \beta_T^2 
\label{liafs3}
\end{equation}

It is also useful to represent the dispersion relations for the plasma waves
near the cross-over point in the form (\ref{fjhi3}).
 For the relativistic  plasma
components we find
\begin{eqnarray}
&&
 \left. \left( { \partial K( \omega,k)
 \over \partial k}\right)
 \right/
 \left( { \partial K ( \omega,k)  \over \partial \omega } \right)=
\left( 1- { < \gamma (1+v)^2> \over < \gamma^3 (1+v)^3>} \right) 
\mbox{} \nonumber \\ \mbox{}
&&
\omega \approx 
k c - {1\over T^2} (k -k_0)
\label{liafss1}
\end{eqnarray}

The  phase speed for the high frequency asymptotic of the plasma wave
(\ref{liafs3})  approaches the phase speed of the thermal particles 
$ c \beta_T$. For the more realistic distribution
function, these parts of the dispersion relation will be strongly 
damped on the Cherenkov resonance with the thermal tail particles.
 The  high frequency asymptotic of the plasma wave
belongs to the 
Alfv\'{e}n wave. From this we make a conclusion that the high frequency
($ kc \geq  \, T_p   \omega_p $)  part of the Alfv\'{e}n wave
is strongly damped and  does not propagate.

\subsection{Oblique Propagation}

\begin{figure}
\psfig{file=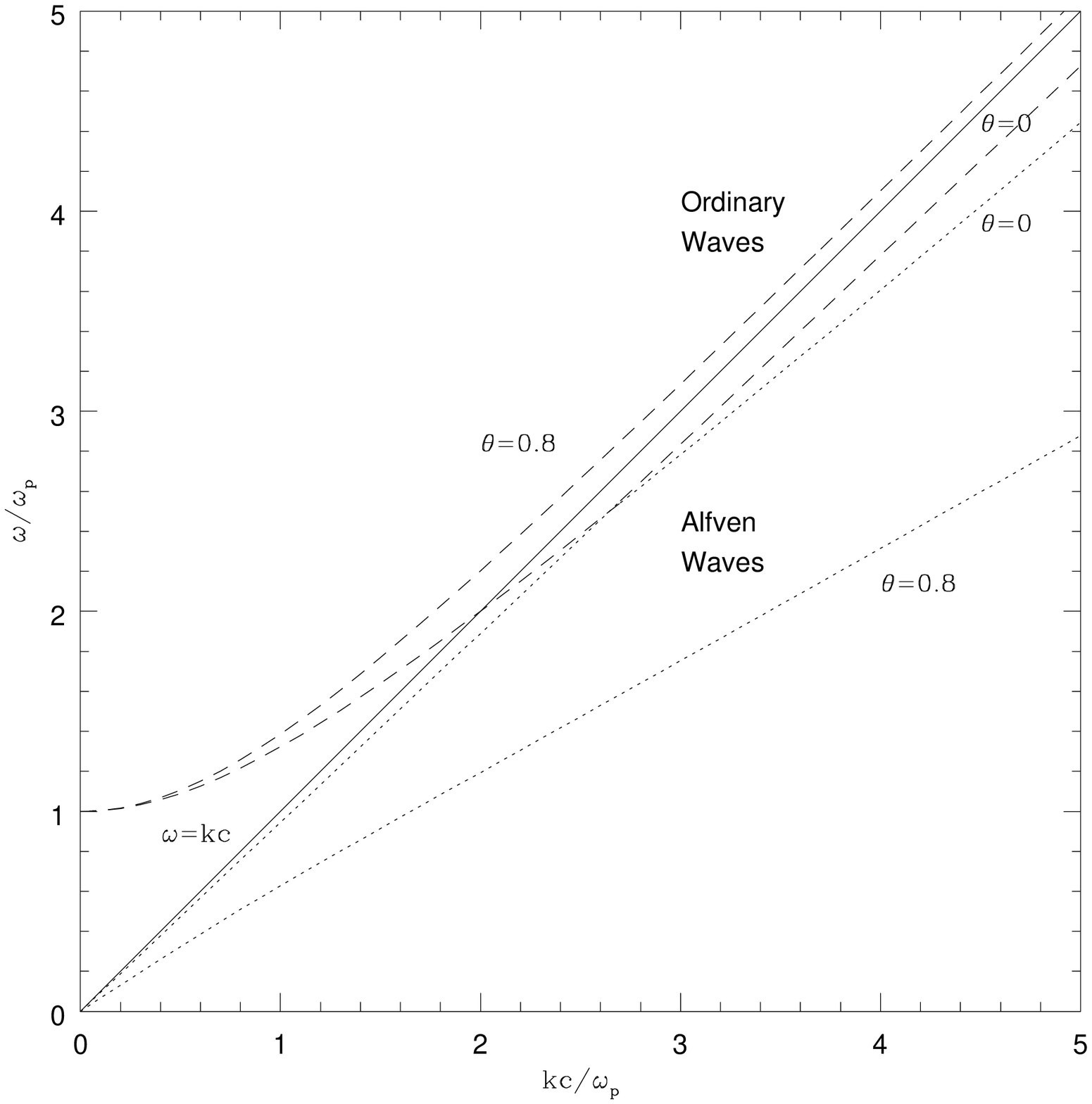,width=12.0cm}
\caption[ Dispersion curves in a hot electron-positron
plasma]{
 Dispersion curves for the waves in a hot electron-positron
plasma in the plasma frame in the limit $\omega \ll \omega_B$.
Only Alfv\'{e}n (dotted) and O
(dashed)  modes are shown.
The dispersion curve for the X mode is very similar to the
cold case. For the illustrative purposes we have chosen $\, T_p  =2$.
The dispersion curves of the  O and Alfv\'{e}n modes intersect
only for parallel propagation.
\label{fig5}
}
\end{figure}

 The dispersion relation for the X mode is 
\begin{equation}
\omega^2= \, k^2 c^2 \left(1  -  \,d\, T_p \,
     \left( 1 + \beta_T^2\,\cos^2\theta  \right)
\right)
\label{det2}
\end{equation}
The dispersion relations for the O and  Alfv\'{e}n
 waves in a hot pair plasma
are plotted in Fig. \ref{fig5}.

It is  possible to obtain the asymptotic expansion
of the dispersion relation of  Alfv\'{e}n and O modes
in the limits of very small and very large wave vectors.
In the limit $ kc \,\gg \sqrt{\, T_p } \omega_p $ we have
\begin{equation}
\omega^2\,=\, 
 \left\{ \begin{array}{ll}
\hskip -.2 truein
\phantom{ {{{ {a\over b} \over {a\over b} } } \over {{ {a\over b} \over 
{a\over b} } } } }
 {c^2}\,k^2\,\beta_T^2\,\cos^2\theta\,
    \left( 1 - 
\mbox{{\large $
{\frac{2\,\omega_p^2}
        {{c^2}\,{\, T_p ^3}\,k^2\,\beta_T^2\,
          \left( -1 + \beta_T^2\,\cos^2\theta \right) }} 
$}}
\right) &
\hskip -.2 truein
\mbox{ Alfv\'{e}n wave} \\
\hskip -.2 truein
\phantom{ {{{ {a\over b} \over {a\over b} } } \over {{ {a\over b} \over 
{a\over b} } } } }
 {c^2}\,k^2\,\left( 1 - d\,\, T_p \,
       \left( 1 + \beta_T^2\,\cos^2\theta \right)  \right) \,
    \left( 1 -
\mbox{{\large $
 {\frac{2\,\omega_p^2\,{{\sin^2\theta}}}
        {{c^2}\,\, T_p \,k^2\,
          \left( -1 + \beta_T^2\,\cos^2\theta \right) }}
$}} \right) &
\mbox{ O-wave} 
\end{array} \right. 
\label{liafss}
\end{equation}
while in the opposite limit $ kc \,\ll\,\sqrt{\, T_p } \omega_p $
\begin{equation}
\omega^2\,=\,
 \left\{ \begin{array}{ll}
\phantom{ {{{ {a\over b} \over {a\over b} } } \over {{ {a\over b} \over 
{a\over b} } } } }
 {c^2}\,k^2\,\cos^2\theta\,
    \left( 1 - d\,\, T_p \,
       \left( 1 + \beta_T^2\,\cos^2\theta \right)  \right) \,
    \left( 1 - {\frac{{c^2}\,k^2\,{{\sin^2\theta}}}
        {2\,\, T_p \,\omega_p^2}} \right)
& \mbox{ Alfv\'{e}n wave} \\
\phantom{ {{{ {a\over b} \over {a\over b} } } \over {{ {a\over b} \over 
{a\over b} } } } }
  \frac{2\,\omega_p^2}{ \, T_p } +
    {c^2}\,k^2\,\left( \beta_T^2\,\cos^2\theta + {{\sin^2\theta}}
        \right) &
\mbox{ O-wave}
\end{array} \right. 
\label{liafs}
\end{equation}

The
X mode is   always superluminous  and the
 Alfv\'{e}n mode is always subluminous.
The O  mode is superluminous for small small vectors
 $ kc \,\ll \sqrt{T_p }\omega_p$
and may become subluminous for very small angles of propagation
$\theta \ll \sqrt{T_p }\omega_p / \omega_B$.

The cross-over point (where  the phase speed of the O  mode
become equal to the speed of light)  is now   $\omega_0^2\,=k_0^2 c^2\,
\approx \, 2 \, \, T_p  \omega_p^2+{ 
{\omega_B}^2 \sin ^2 \theta } $.
Using relation (\ref{fjhi02}) we can approximate
the dispersion relation near the cross-over point as
\begin{equation}
\omega = k c -   \kappa  (k- k_0) , \hskip .2 truein
\kappa = { 1\over T^2} -  {  (\omega_B^4 +
4 T_p^3 \omega_B^2 \omega_p^2 ) \theta^2 \over
16 T_p^6 \omega_p^4}
\label{liafs01}
\end{equation}

\subsection{Polarization of waves in a hot plasma}
\label{Polarizationhot}

In the case of a hot plasma the matrix of cofactors
$\lambda^{(h)} _{\alpha \beta}$ is quite complicated and is not given here.
Simple relations may be obtained in the limit $\omega_B = \infty$
and near the cross-over point for the O wave.
In the limit $\omega_B = \infty$ we find the elements of the 
matrix $\lambda^{(h)} _{\alpha \beta}$
 \begin{eqnarray}
&& \lambda^{(h)} _{xx} = 
 \left( -1 + n^2 \right) \,\left( -1 -
     {\frac{2\,\omega_p^2}
       {\, T_p \,{{\omega}^2}\,
         \left( -1 + n^2\,v_0^2\,\cos^2\theta \right) }} +
     n^2\,{{\sin^2\theta}} \right)
\mbox{} \nonumber \\ \mbox{}
&& \lambda^{(h)} _{xz}  =
 n^2\,\left( -1 + n^2 \right) \,\cos \theta\,\sin \theta
 = \lambda^{(h)} _{zx} 
\mbox{} \nonumber \\ \mbox{}
&& \lambda^{(h)} _{yy}  =
 1 - n^2 - {\frac{2\,\omega_p^2\,
       \left( -1 + n^2\,\cos^2\theta \right) }{\, T_p \,
       {{\omega}^2}\,\left( -1 + n^2\,v_0^2\,\cos^2\theta \right) }}
\mbox{} \nonumber \\ \mbox{}
&& \lambda^{(h)} _{zz}  =
 \left( -1 + n^2 \right) \,\left( -1 + n^2\,\cos^2\theta \right)
\label{q1}
\end{eqnarray}

For the X mode we find that the
polarization vector is 
$e_X=(0,1,0)$, while for the O mode
\begin{equation}
{E_x \over E_z} =
 {\frac{n^2\,\cos \theta\,\sin \theta}
    {-1 + n^2\,\cos^2\theta}}
\label{q2}
\end{equation}
These relations are valid for the points not close to the cross-over point
of the O wave (near the cross-over point
 the  approximation  $\omega_B = \infty$
is not applicable). Near the cross-over point, $ n=1$, we find
\begin{equation}
{E_x \over E_z} =
  -{\frac{{{\omega_B}^2} \theta}
     {{ 4 T_p^3 {\omega_p}^2}}}
\label{q21}
\end{equation}

For oblique propagation the behavior of the O mode 
at the cross-over point changes
at 
\begin{equation}
\theta \approx {  4 T_p^3 d}
\label{liafs02}
\end{equation}
For smaller angles the O mode 
is quasiparallel at the cross-over point while,
for large angles, it is quasitransverse.

The polarization vectors for the O and Alfv\'{e}n modes are then given
by
\begin{eqnarray}
 &&
e_O^{(h)} =  \left\{  \begin{array}{ll}
\left\{  \cos \theta\,\left( 1 - {\frac{2\, T_p \,
          \omega_p^2\,{{\sin^2\theta}}}{c^2 k^2}} \right) ,0,
   -\left( \left( 1 + {\frac{2\, T_p \,\omega_p^2\,
            \cos^2\theta}{c^2 k^2}} \right) \,\sin \theta \right)\right\},
& \mbox{  $ kc \, \gg \omega_p$}\\
\left\{  {\frac{{{\omega_B}^2}\,\theta}{{{\omega_0}^2}}},0,-1\right\}
        & \hskip -2  truein
\mbox{ $ \theta \ll { 2 \, T_p  \omega_p^2 \over \omega_B^2},   \, \,  $  
$\omega= \omega_0^{(h)}$ }\\
\left\{  1,0,-{\frac{{{\omega_0}^2}}
       {{{ \sin \theta \cos \theta \omega_B}^2}}}\right\} &
\hskip -2 truein
\mbox{ $ \theta \gg  { 2  \, T_p  \omega_p^2 \over \omega_B^2}, \, \,  $
$\omega= \omega_0^{(h)}$ }
\end{array}\right.
\label{q3}
\mbox{} \\ \mbox{} &&
e_A^{(h)} =  \left\{  1,0,{\frac{{{\omega}^2}\,\tan \theta}
     {2\,\, T_p \,\omega_p^2}}\right\} 
\hskip 1 truein  \mbox{  $ kc \, \ll  \omega_p$}
\label{q4}
\end{eqnarray}

\subsection{ Dispersion Relation for Hot Pair Plasma in Pulsar Frame}
\label{PulsarframeHot}

The  dispersion relations for the forward propagating modes
in the pulsar frame in the limit $\, \omega ^{\prime} \ll \, \omega_B$ 
are
\begin{eqnarray}
&
\omega  ^{\prime}_X \,=\,   k ^{\prime} c  \left(
1- {\, \omega_p^{\prime \, 2} T_p \over 4  \, \omega_B^2 \,  \gamma_p^3}
\right)  &
\mbox{} \nonumber \\ \mbox{}
\hat{ \omega}_O \,=\, &  k ^{\prime} c  \left(
1 - { \, \omega_p^{\prime \, 2} \,T_p\,
\over 4 \,  \gamma_p^3  \, \omega_B^2 }    \,
+ {\frac{ \gamma_p{{\, \omega_p}^{\prime \, 2}} T_p\,{{\sin^2\theta ^{\prime}}}}
        {{c^2}\,{k^2}\,
           }} \right)
&
\mbox{ if $ k^{\prime} c \,\gg \sqrt{ T_p} \,\gamma_p \omega_p^{\prime}
$ and $\theta  ^{\prime} < 1/T_p$}
\mbox{} \nonumber \\ \mbox{}
\hat{ \omega}_A \,=\, &  k ^{\prime} c  \cos \theta ^{\prime}
\left(1 - { \, \omega_p^{\prime \, 2} \,T_p\,
\over 2  \,  \gamma_p^3  \, \omega_B^2 }  \,
   - {\frac{{c^2}\,{k^{\prime \, 2} } \,{{\sin^2\theta^{\prime}}}}
        {2\,T_p\,  \, \gamma_p {{\, \omega_p}^{\prime \, 2}}}} \right)
 &
\mbox{ if $ kc \,\ll\,\sqrt{T_p }\, \gamma_p  \omega_p $}
\label{liafsq1}
\end{eqnarray}

These relationships are valid for the frequencies satisfying
the inequality
\begin{equation}
\, \omega \ll \,  \gamma_p \, \omega_B /T_p
\label{x}
\end{equation}
This  is a condition 
 that in the reference frame of the plasma the
frequency of the waves is much smaller that the typical
cyclotron frequency of the particles $\, \omega_B /T_p$.

The cross-over point in this frame  is
\begin{equation}
\omega_0^{\prime \, 2} = { 2 \omega_p ^{\prime  \, 2}
T_p \over \gamma_p} + { 4 \gamma_p^2
\omega_B^2 \theta ^{\prime \, 2}  }
\label{qw3}
\end{equation}

\section{Hydrodynamic and Kinetic Instabilities}
\label{HydrodynamicOrkinetic}

The description of the beam-plasma instabilities is based on the scheme
used to solve the general problem of linear oscillations in plasma.
The initial equations  are the linearized kinetic equations for the
particles in a self-consistent electromagnetic field and Maxwell's 
equations. When the unperturbed state of the beam and plasma
is stationary and spatially uniform, we can use Eq. (\ref{dettt})
to find the normal modes of a medium.

For the beam-plasma system the dielectric tensor 
$\epsilon_{\alpha \beta}(\omega,{\bf k})$  may be represented as a sum
of contributions from  plasma and beam. 
\begin{equation}
\epsilon_{\alpha \beta}(\omega,{\bf k}) =
 \delta_{\alpha \beta} + {4 \pi c \over \omega}
 \sigma _{\alpha \beta}^{plasma} + {4 \pi c \over \omega}
 \sigma _{\alpha \beta}^{beam}
\label{a1a12}
\end{equation}
where $ \sigma _{\alpha \beta}^{plasma}$ and $ \sigma _{\alpha \beta}^{beam}$
are the conductivity tensors of plasma and beam.

Sometimes it is possible
to consider beam as a weak perturbation to the system. Then, in the zeroth
approximation,  the  normal modes of the medium will be determined
from (\ref{dettt}) with $\sigma _{\alpha \beta}^{beam}$ set to zero.
This will produce a set of  normal modes of the medium 
$\left\{  \omega({\bf k})^l \right\}$.

If  the plasma alone  is stable, then the frequency of the normal
modes will have a zero imaginary part. In the first approximation,
dispersion relation (\ref{dettt}) may be expanded taking into account a
small contribution to the dielectric tensor from the beam. 
The frequency shift $\Delta({\bf k})^l$ of the normal
mode $ \omega({\bf k})^l$ is then determined from
\begin{equation}
\Delta({\bf k})^l \left. \left[ {\partial \over \partial \omega} 
K_p (\omega,{\bf k}) \right] \right|_{\omega({\bf k})^l}
 + K_b  (\omega,{\bf k})=0
\label{a2}
\end{equation}
where $K_p (\omega,{\bf k})$ and $K_b  (\omega,{\bf k})$ are the plasma
and beam parts of Eq.(\ref{dettt}). For stable plasma without a
beam, $K_p (\omega,{\bf k})$ and 
$\left\{ \omega({\bf k})^l \right\}$ 
are real.

Two separate cases may be distinguished here depending on  whether
the complex part of  
the beam contribution to the dispersion relations (\ref{a2})
 $K_b  (\omega,{\bf k})$
is zero or nonzero. If Im($K_b  (\omega,{\bf k})=0$ then equation
(\ref{a2}) has real coefficients. The complex solutions of Eq.(\ref{a2})
(if any) are complex conjugates. Solutions with the positive 
complex part correspond to the growing waves. These are
hydrodynamic instabilities. In  hydrodynamic  instabilities,  all the
particles of the beam resonate with the normal mode of the plasma.
This requires that the growth rate of the instability
be greater than the intrinsic bandwidth of the growing waves:

\begin{equation}
|{\bf k \cdot \delta v}| \ll {\rm Im} (\Delta ({\bf k})).
\label{a3}
\end{equation}

Here ${\bf k}$ is the resonant wave vector, ${\bf \delta v}$ is the 
scatter in the velocity of the beam particles. This is satisfied
for a very small scatter in the velocity of the beam particles,
so that all the particles from the beam resonate with the beam.

Alternatively, if the complex part of the
the beam contribution to the dispersion relations (\ref{a2})
 $K_b  (\omega,{\bf k})$
is  nonzero, the frequency shift $\Delta ({\bf k})$ will always
have a complex part. If the complex part of $\Delta ({\bf k})^l $ is 
larger than zero, then the corresponding normal mode $\omega ({\bf k})^l $
will be growing at the expense of the beam energy, while
for negative $\Delta ({\bf k})^l $ the  mode will
be damped on the resonant particles of the beam.
This case corresponds to the kinetic instability. The requirement that the
frequency shift $\Delta ({\bf k})^l $ due to the complex
part of  $K_b  (\omega,{\bf k})$ dominates over the shift due to the
large real part of $K_b  (\omega,{\bf k})$ requires that the growth
rate be much less than the the intrinsic bandwidth of the growing waves
(reversed inequality (\ref{a3})). This is satisfied for a very large
 scatter in the velocity of the beam particles, so that at any given moment
only a small fraction of the beam particles is in resonance with the wave.

Though the physical interpretations of the 
kinetic and hydrodynamic instabilities are quite different, they
may be considered
as two limiting cases of a general beam instability. 
 For a relativistic beam traveling along a magnetic field
with average Lorentz factor $\gamma_b$,  scatter in parallel Lorentz factors
$\Delta \gamma$,  and average pitch angle $\psi$ the condition
of the hydrodynamic approximation (\ref{a3}) takes the form
\begin{equation}
k_{\parallel} c\,
\left({\psi^2\over 2}+ {\Delta \gamma \over \gamma^3} \right)\,
+ k_{\perp} c \,\psi + 
{s \omega_B \Delta \gamma \over \gamma^2 } \, \ll \Gamma
\label{Gammm}
\end{equation}
where $s$ is the harmonic number ($s=0$ for Cherenkov resonance,
$s \ne 0$ for cyclotron resonance) and $\Gamma$ is a growth rate of 
an instability. For the kinetic instability, this
inequality is reversed.

From (\ref{Gammm}) it follows that there exist a critical pitch
angle
\begin{equation}
\psi_{crit} = {1\over \gamma } \, \sqrt{{ \Delta \gamma \over \gamma}} 
\label{a5}
\end{equation}

For $\psi > \psi_{crit}$ the scatter in pitch angles dominates over
longitudinal velocity spread. For $\psi > \psi_{crit}$ the 
average "longitudinal" mass of the beam particles decreases
by the factor of $(\psi \gamma_b)^2$ so that the 
instabilities whose growth rate is inversely proportional to the
"longitudinal" mass of the particles (like Cherenkov instability 
of plasma waves) may be enhanced considerably.

Relativistic particles propagating along the   curved magnetic field of  a pulsar
magnetosphere 
initially are in the ground quantum state (zero pitch angle).
They can   develop a finite  pitch angle
by (i) particle-particle collisions, (ii) interaction with the electromagnetic
field (Compton scattering on the diffuse thermal photons or recoil due the
emission of electromagnetic waves at anomalous cyclotron resonance),
 and (iii) when the adiabatic approximation
for the propagation breaks down (when the Larmor radius becomes comparable
with the size of the inhomogeneity). The pitch angle of the particles is then
determined by the balance of these forces on one hand and radiation damping at 
the
normal synchrotron resonance and the force due to the conservation of adiabatic 
invariant on the other hand.

In  magnetosphere,
 the particle-particle collision time is  very long compared with the
dynamical time because of 
the relatively low density of particles, 
high speed and the  one-dimensional character
of the motion. We also assume that the Compton scattering 
on the diffuse 
thermal photons is unimportant and that the adiabatic approximation
for the propagation of particles is satisfied. The  transverse component
of the  force due to the radiation 
damping at normal synchrotron resonance dominates the 
transverse motion of the particles near the neutron star, making the 
pitch angles equal to zero. 
Then the pitch angles  will remain zero throughout the
region where the above conditions are satisfied.

In what follows we assume that 
 plasma is one-dimensional before the development of instabilities.
The condition of hydrodynamic approximation (\ref{Gammm}) is then
\begin{equation}
k_{\parallel} c\, {\Delta \gamma \over \gamma^3} +
{s \omega_B \Delta \gamma \over \gamma^2 } \, \ll \Gamma
\label{Gam112}
\end{equation}
In the kinetic regime this inequality is reversed.

For an instability to be important as a possible source of 
coherent emission generation, its growth rate, evaluated in the
pulsar frame, should be much larger than the pulsar rotation frequency
$\Omega$. The growth rates in the pulsar and  plasma frames
are related by

\begin{equation}
\Gamma^{\prime}= { \Gamma \over \gamma_p}.
\label{vg0}
\end{equation}

So the requirement of a fast growth in the plasma frame is
\begin{equation}
 { \Gamma \over \gamma_p \Omega } \gg 1.
\label{vg9}
\end{equation}

Another, more stringent  requirement on the  growth rate comes from the angular
dependence of a growth rate. The emitting plasma propagates in a
curved magnetic field. If an instability has a considerable growth
inside a characteristic angle $\delta \theta ^{\prime}$,
 then the growth length
should be larger than $ \delta \theta R_c$, where $R_c$ is the curvature
of the magnetic fields. In the plasma frame this requirement is

\begin{equation}
 \Gamma \gg { c \gamma_p^2  \over R_c \, \delta \theta},
\label{vgg}
\end{equation}
where we used $ \delta \theta ^{\prime} \approx \delta \theta /\gamma_p$.

\section{Cold Pair Plasma: Resonances} 
\label{Resonances}

\begin{figure}
\psfig{file=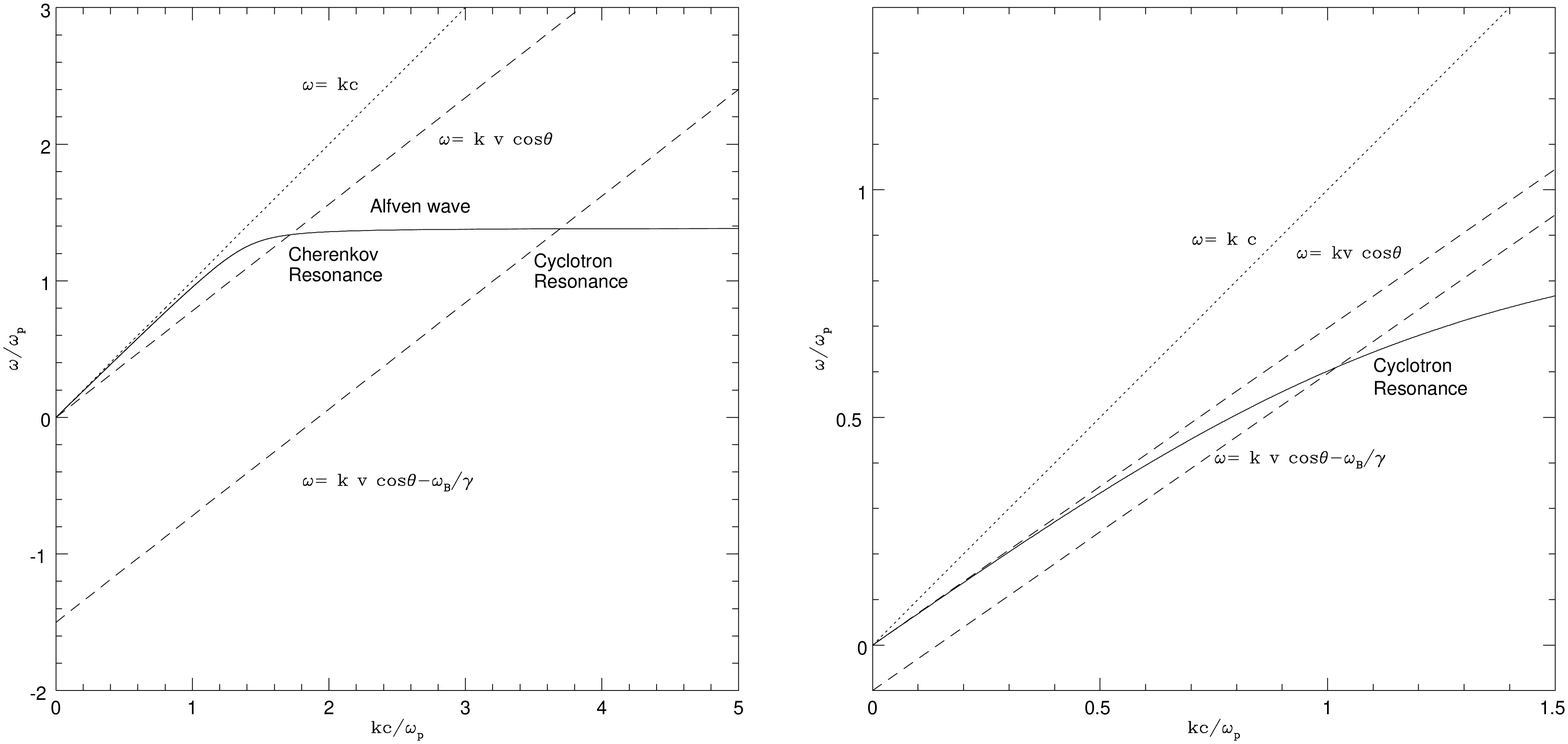,width=12.0cm}
\hskip 1 truein 
(a)
\hskip 2 truein
(b)
\vskip .2 truein 
\caption{ (a)
resonances of the Alfv\'{e}n mode in the cold plasma for $\mu < 1$,
(b) resonances of the Alfv\'{e}n mode in the cold plasma for $\mu> 1$.
\label{fig2}}
\end{figure}

\begin{figure}
\begin{center}
\psfig{file=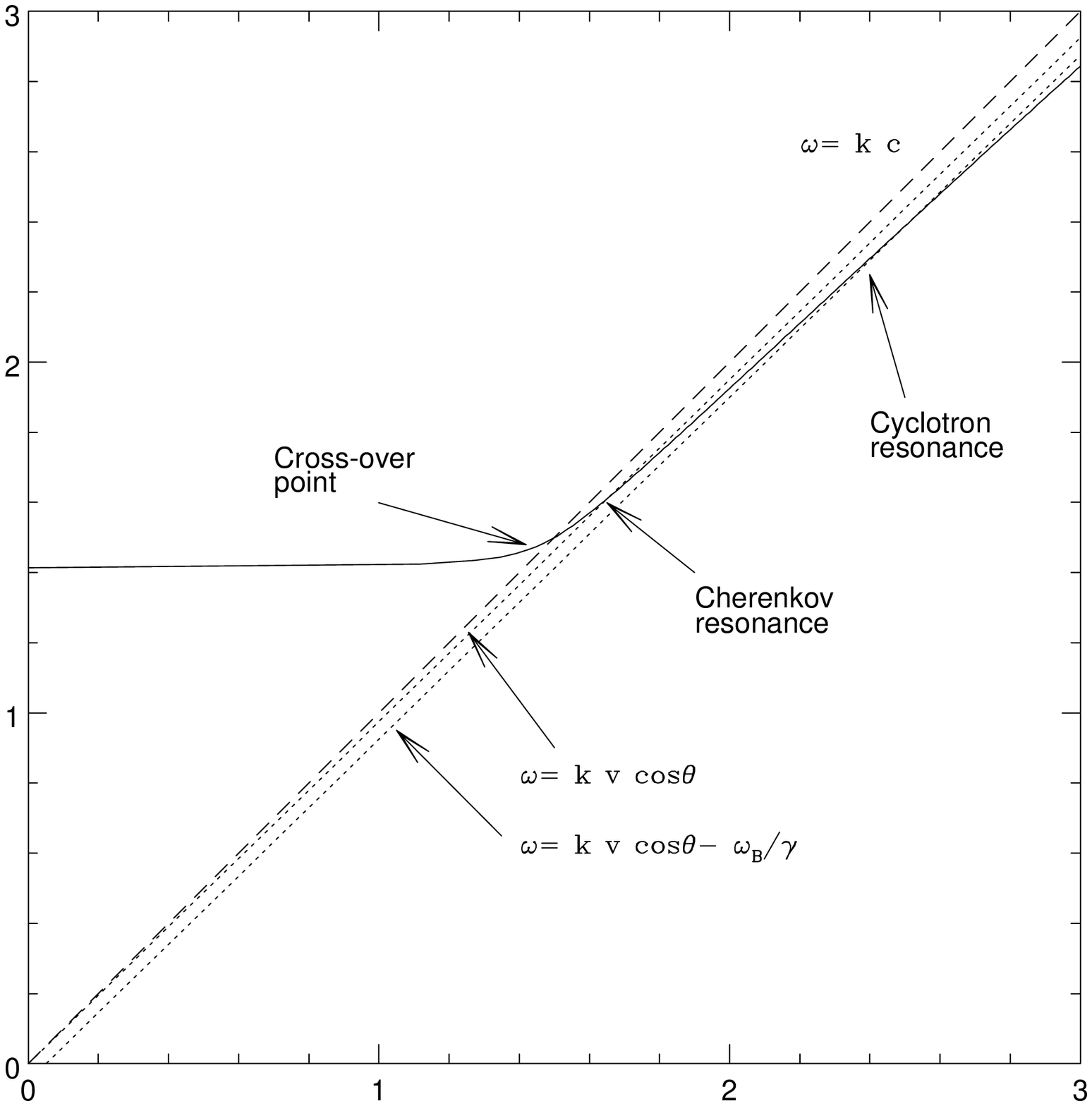,width=10.0cm}
\caption[Resonances on the O  mode in the cold plasma for $\mu> 1$]{
Resonances on the O  mode in the cold plasma for $\mu> 1$.
\label{fig4}
}
\end{center}
\end{figure}

In the cold plasma approach the resonant interaction between the fast
particles and the plasma may be considered as the  interaction of the
waves in the plasma with the waves in the beam. The interaction is the
strongest when the dispersion relations of the waves intersect.
Consequently,
we are looking for the possible resonances between the waves in the
plasma (\ref{fjhi}) and the waves in the beam (see Fig. \ref{fig2}
 and \ref{fig4}):
\begin{eqnarray}
&& \omega= v_b\, k_{z} 
\label{uiaj}
\mbox{}  \\ \mbox{}
&& \omega= v_b\, k  \cos \theta \pm {\omega_B \over \gamma_b}
\label{uiaj1}
\end{eqnarray}

As we will see in Section \ref{Coldplasma},
 the resonant interaction of the plasma
waves with the Cherenkov waves in the beam (\ref{uiaj}) is described by the
cubic equation for the frequency shift, which always has complex conjugate
solutions. This implies that the Cherenkov resonant interaction
of the waves in the beam and in the plasma is always unstable.

In contrast, the frequency shift due to the cyclotron interaction of the waves 
in the beam and in the plasma (\ref{uiaj1}) is 
described by a quadratic equation,  which has two real
solutions for the plus sign in (\ref{uiaj1}) and two 
 complex solutions for the minus sign in (\ref{uiaj1}). Thus, only the 
minus sign in (\ref{uiaj1}) will contribute to the instability
growth rate. The resonance (\ref{uiaj1}) with the minus sign
is called anomalous Doppler resonance. This corresponding instability may
be considered as the interaction of the negative energy wave in the beam with
the positive energy wave in plasma. Due to the resonant coupling, the 
amplitudes of  both waves grow exponentially.

Now let us consider the condition for the resonances (\ref{uiaj}) and
 (\ref{uiaj1}) to occur.
From the low frequency asymptotics of the
 Alfv\'{e}n waves (\ref{fhdu122}) we infer
that the possibility of the
 Cherenkov excitation of the Alfv\'{e}n waves depends on the parameter
\begin{equation}
\mu\,=\,
{ 2 \,\gamma_b \, \omega_p \over {\omega_B}}
\label{uiaj21}
\end{equation}

If $\mu \, < \, 1$, then Alfv\'{e}n waves can be excited by Cherenkov
resonance.\footnote{ In the case of cold plasma this may be considered
as a sufficient condition for the Cherenkov excitation of Alfv\'{e}n waves.
In the case of hot plasma this is only a necessary condition (see below).}
However, if $\mu \, > \, 1$ then Alfv\'{e}n waves cannot be excited by Cherenkov
resonance. Instead, resonance  can occur for an  O mode subject to the
requirement of sufficiently small angles of propagation Fig. \ref{fig4}.

For the  cold plasma in the region of open field lines we have
\begin{equation}
\mu\,=\,  \gamma_b \sqrt{ {2 \, \lambda \, \Omega \over  \gamma_p  {\omega_B}} }
=\,
\sqrt{ {2 \, \lambda \, \Omega \over  {\omega_B}} } { \gamma_b^{\prime} \over 
\gamma_p ^{3/2} } \,=\, 
5 \times 10^{-3} \left({r\over R_{NS}} \right)^{3/2}  \,=\,
\left\{ \begin{array}{ll}
< 1, & \mbox{ if $  \left({r\over R_{NS}} \right) < 43 $} \\
> 1,  & \mbox{ if $  \left({r\over R_{NS}} \right) > 43 $}
\end{array}\right.
\label{uiaj22}
\end{equation}

So, at small radii ($\mu \ll 1$) it is the Alfv\'{e}n wave that is excited by the
Cherenkov resonance, while for larger radii ($\mu \geq 1$) it is the O-mode
that can be excited by the Cherenkov resonance. 
In the outer parts of magnetosphere ($ r \geq 100 R_{NS} $) the
parameter $\mu $ becomes much larger than unity:
$\mu \gg 1$.

For the parallel propagation (and only in this case)
  the parts of the O and Alfv\'{e}n modes that
have longitudinal polarization may be considered as forming a
single plasma wave with a dispersion $ \omega \,=\, \sqrt{2} \omega_p$.
In this particular case, the excitation of either
O or Alfv\'{e}n part of the longitudinal plasma mode
is very similar. But as the waves propagate in the curved magnetic
field lines, the parts of the plasma mode corresponding to the 
O or Alfv\'{e}n wave will evolve differently resulting in a
different observational characteristics of the emergent radiation.

In what follows we consider separately the two possible
 cases of Cherenkov resonances: $\mu >1 $ and
$\mu< 1$.

We also note that the  X
wave cannot be excited by the Cherenkov resonance. Though the formal
intersection of the  Cherenkov wave in the beam  (\ref{uiaj})
with the dispersion relation of the X mode is possible 
for all frequencies if $\mu =1$, the transverse polarization of the
X mode excludes a resonant interaction with 
particles streaming along the  magnetic field.

The cyclotron resonance on the X modes occurs at 
$\omega _{\rm  res} \,\, \ll {\omega_B}$,
provided that
\begin{equation}
{ \omega_p^2 \gamma_b \over {\omega_B}^2} \gg 1
\label{fj6}
\end{equation}
Using the fiducial plasma parameters 
of the cold plasma,  we find
\begin{equation}
{ \omega_p^2 \gamma_b \over {\omega_B}^2} \,=\, \lambda \, \gamma_b
{2\Omega \over \gamma_p  {\omega_B}}=
{ \lambda \gamma_b^{\prime} \over 
\gamma_p^2 } { \Omega \over {\omega_B}} 
  \,=\, 1.3 \times
10^{-10} \left({r\over R_{NS}}\right)^3
\label{fhd35}
\end{equation}
which implies that the X mode can be excited by the
cyclotron resonance only in the outer parts of magnetosphere for radii
satisfying 
\begin{equation}
\left({r_{\rm  res} \,\over R_{NS}}\right) \, >
\left({ \omega_B^{\ast}  \gamma_p^2 \over \lambda \gamma_b^{\prime} 
\Omega}
\right)^{1/3} \approx   2 \times 10^3  \,
\label{fhd315}
\end{equation}

The location of the cyclotron resonance on the X mode
 is quite sensitive  to the
choice of the bulk streaming energy.
Comparing the resonant frequency (Table \ref{rescold1}) with the plasma
frequency, we find
\begin{equation}
{ \omega_{\rm res,X, cycl} \over \omega_p} \,=\, { {\omega_B}^3  \over \gamma_b \,
\omega_p^3} \,=\, {\gamma_p^{3/2} 
 \over \gamma_b \lambda^{3/2}  } \left({\omega_B
 \over  2 \Omega } \right) ^{3/2}\, \gg 1
\label{fhd36}
\end{equation}
which implies that the X mode is always excited
with the frequencies much larger than the plasma frequency.

Locations of the resonances in the cold plasma are given in Table
\ref{rescold1}.

\begin{table} \begin{center}
\caption
{ Resonances in cold pair plasma}
\psfig{file=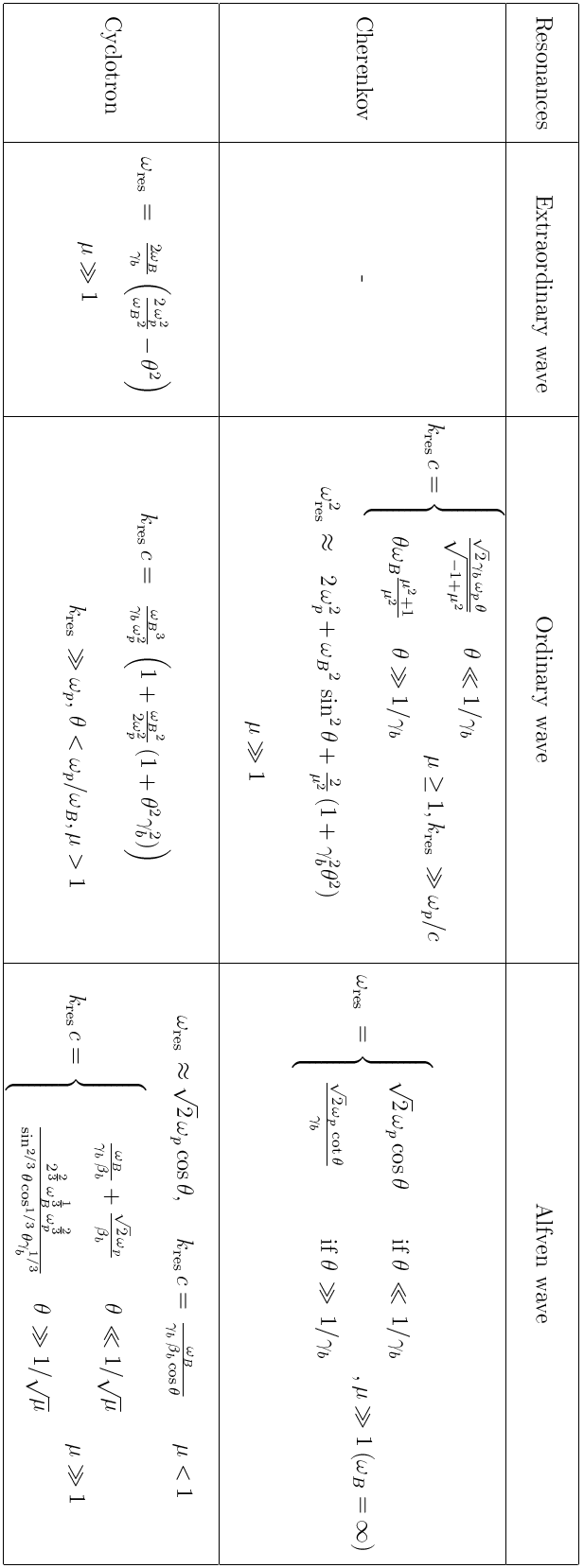,width=16.0cm}
\label{rescold1}
\end{center} \end{table}

\section{Hydrodynamic Instabilities In Cold Plasma}
\label{Parallelpropagation1}

\subsection{Dielectric Tensor for Cold Beam-Plasma System}
\label{PDieleTensBeam-Plasma}

The dielectric tensor for beam of the density $n_b$ propagating with the
velocity $v_b$ along the magnetic field $B$ through a plasma of the  density $n$
can be found from a general expression (Eq \ref{epsilon}) with zero
drift velocity $u_{\alpha} =0$
and  distribution function
$f_{\alpha}(p_{z}) =n_p \, \delta(p_{z}) + n_b  \delta(p_{z} - p_b)$
($n_b$ is a density of a beam and $p_{z}$ is a momentum of beam particles): 
\begin{eqnarray}
\epsilon_{xx}&&=   1 + {{2\,{{{\it \omega_p}}^2}}\over
       {-{{\omega }^2} + {{{\it {\omega_B}}}^2}}} -
     {{{{{\it  {\omega_b}}}^2}\,{{{\it  \hat{\omega}}}^2}}\over
       {{\it\gamma_b}\,{{\omega }^2}\,{{{\it  \tilde{{\omega}}}}^2}}}
\, =\epsilon_{yy}
\mbox{} \nonumber \\ \mbox{}
\epsilon_{xy}&&= {{-i\,{{{\it  {\omega_b}}}^2}\,{\it {\omega_B}}\,{\it
  \hat{\omega}}}\over
  {{{{\it\gamma_b}}^2}\,{{\omega }^2}\,{{{\it  \tilde{{\omega}}}}^2}}}
=- \epsilon_{yx}
\mbox{} \nonumber \\ \mbox{}
\epsilon_{x z}&&=
 -{{k\,{{{\it  {\omega_b}}}^2}\,{\it  \hat{\omega}}\,{\it v_b}\,\sin  \theta}
\over
         {{\it\gamma_b}\,{{\omega }^2}\,{{{\it  \tilde{{\omega}}}}^2}}}=
\epsilon_{z x}
\mbox{} \nonumber \\ \mbox{}
\epsilon_{y z}&&=
{{i\,k\,{{{\it  {\omega_b}}}^2}\,{\it {\omega_B}}\,{\it v_b}\,\sin  \theta}\over
      {{{{\it\gamma_b}}^2}\,{{\omega }^2}\,{{{\it  \tilde{{\omega}}}}^2}}}
 =
-\epsilon_{z y}
\mbox{} \nonumber \\ \mbox{}
\epsilon_{z z}&&=  1 - {{2\,{{{\it \omega_p}}^2}}\over {{{\omega }^2}}} -
     {{{{{\it  {\omega_b}}}^2}}\over
       {{{{\it\gamma_b}}^3}\,{{{\it  \hat{\omega}}}^2}}} -
   {{k^2\,{{{\it  {\omega_b}}}^2}\,{{{\it v_b}}^2}\,{{\sin  ^2 \theta}}}\over
   {{\it\gamma_b}\,{{\omega }^2}\,{{{\it  \tilde{{\omega}}}}^2}}}
\label{epsilonnn}
\end{eqnarray}
where $ \hat{\omega} \,=\,
\omega - \, k\, v_b \, \cos \theta$,  $ \tilde{{\omega}}^2 \,=\,
(\omega - \, k\, v_b \, \cos \theta)^2 - {\omega_B}^2/\gamma_b^2$
and $\gamma_b=1/\sqrt{1-{v_b^2\over c^2} }$.

We will always assume that beam can be considered as a weak perturbation,
so that we can employ the expansion procedure described in Section
\ref{HydrodynamicOrkinetic}.

\subsection{ Parallel Propagation}
\label{Parallelpropagation}

In this section we calculate the growth rates for the beam instabilities
for the waves propagating along the magnetic field, that we will use later
as guide lines for the general case of oblique propagation.

For the propagation along the magnetic field the dispersion relation
(\ref{dettt})  with a dielectric tensor  \ref{epsilonnn}  factorizes:
\begin{eqnarray}
&&
-1 + {2\,\omega_p^2\over \omega ^2} +
   {\omega_b^2\over
 \gamma_b^3\, \hat{\omega} ^2}
=0
\label{poja}
\mbox{} \\ \mbox{}
&&
-1 + n^2 + {{2\,{{{\it \omega_p}}^2}}\over {{\omega^2} -
 {{{\it {\omega_B}}}^2}}} +
   {{{{{\it {\omega_b}}}^2}\,{\it \hat{\omega} }}\over
    {{\it \gamma_b}\,{\omega^2}\,
\left(\pm \, {{{\it {\omega_B}}} / {{\it \gamma_b}}} +
         {\it \hat{\omega}} \right) }}=0
\label{oh}
\end{eqnarray}

Equation (\ref{poja}) describes hydrodynamic excitation 
of longitudinal plasma waves. As discussed above, this may be 
a longitudinal part of Alfv\'{e}n or O mode depending on the 
parameters of plasma. 

Equation (\ref{oh}) describes the cyclotron excitation
of the O and X modes. For the parallel
propagation the cyclotron excitation of the Alfv\'{e}n wave
does not occur.

\subsection{Cherenkov Excitation of Plasma Waves  for $\theta=0$}

We now look for the correction to the relations (\ref{fhdu1334}) and
 (\ref{uiaj})  when the two intersect.
\begin{eqnarray}
\omega \,=&&\,  \sqrt{2} \omega_p + \Delta
\mbox{} \nonumber \\ \mbox{}
\omega \,= && v_b\, k \cos \theta+ \Delta
\label{ksajhf}
\end{eqnarray}

Expanding in small  $\Delta$, we find that the frequency shift
satisfies a third order equation:
\begin{equation}
 -{\frac{{\sqrt{2}}\,{{\Delta}^3}}{\omega_p}} +
    {\frac{{{\omega_b}^2}}{{{\gamma_b}^3}}} = 0
\label{ddda1}
\end{equation}

Equation (\ref{ddda1}) always has one real and two complex
conjugated roots.
The  complex root with the positive complex  part corresponds to the
instability.

Solving Eq. (\ref{ddda1}), we find
 the complex part of the  frequency shift:
\begin{equation}
{\rm Im}(\ \Delta) = {{{\sqrt{3}}\,{{{\it  \omega_p}}^{{1\over 3}}}\,
  {{{\it  {\omega_b}}}^{{2\over 3}}}}\over {{2^{{7\over 6}}}\,{\it\gamma_b}}}=
{\sqrt{3} \lambda^{1/6} \sqrt{\Omega {\omega_B}}\over
2^{2/3} \gamma_b \sqrt{\gamma_p} }
\label{lkfjs}
\end{equation}

This is a growth rate for the Cherenkov excitation of plasma waves
(c.f. \cite{Godfrey},  \cite{Egorenkov}).

We can compare the importance of the  Cherenkov excitation 
of plasma waves
by 
evaluating growth rate (\ref{lkfjs}) for the set of fiducial parameters
of a cold plasma  and comparing it with the dynamical time (\ref{vg0}):

\begin{equation}
 { {\rm Im} (\ \Delta) \over \gamma_p \Omega } \approx
{ \lambda^{1/6}  \over  \gamma_b \gamma_p^{3/2}}\sqrt{ {\omega_B}\over \Omega}=
86 \left({y \over R_{NS} }\right)^{-3/2} 
\label{lkfjsp}
\end{equation}
 
From which it follows that this instability may  be important
for $r \, \leq 20$.
We will see in 
 Section \ref{Coldplasma} that the second criterion (\ref{vgg}) is not satisfied
for the Cherenkov excitation of Alfv\'{e}n or O waves, so that the Cherenkov
instability  does not develop.

\subsection{Cyclotron Excitation of  Transverse Waves  for $\theta = 0$ }

We expect that the 
hydrodynamic instability will be strongest at small wave vectors.
We can then use the low frequency approximation (\ref{X})
to the dispersion of the transverse waves.
We seek  the correction to the relations (\ref{X}) and
(\ref{uiaj1}) when the two intersect.
\begin{eqnarray}
\omega \,=&&\, k c\, \left(1 -{ \omega_p^2\over {\omega_B}^2}\right) +
\ \Delta
\mbox{} \nonumber \\ \mbox{}
\omega \,=&&\, k \, v_b \, \cos \theta - {{\omega_B}\over \gamma_b}  +
\ \Delta
\label{daf}
\end{eqnarray}

Expanding in small  $\Delta$, we find that the frequency shift 
satisfies the  quadratic  equation:

\begin{equation}
 -{\frac{\Delta\,k^2 c^2 }{{{\omega}^3}}} \pm 
    {\frac{\omega_B\,{{\omega_b}^2}}
      {\Delta\,\gamma_b\,{{\omega}^2}}} = 0
\label{daf21}
\end{equation}

The $ \pm $ sign in (\ref{daf21}) corresponds to the two signs in 
(\ref{oh}). For the normal Doppler resonance (plus sign in (\ref{oh}) and
(\ref{daf21})) the resulting frequency shift is real. For the
anomalous  Doppler resonance, the  frequency shift is complex:
\begin{equation}
 \Delta  \,= \, \pm
 {{i\,{\sqrt{\omega }}\,{\sqrt{{\it  {\omega_B}}}}\,{\it  {\omega_b}}}\over
 2   {{\it \gamma_b}\,k c}}
\label{dafd}
\end{equation}
which gives near the resonant frequency (Table \ref{rescold1})
\begin{equation}
 {\rm Im}  (\Delta)  = \,
 {{i\,{\it  \omega_p}\,{\it  {\omega_b}}}\over
 2\,  {{\sqrt{{\it \gamma_b}}}\,{\it  {\omega_B}}}} =
{ \sqrt{ \lambda} \Omega \over \gamma_p \sqrt\gamma_b }
\label{dafslj}
\end{equation}

For the parameters of a cold plasma the growth rate (\ref{dafslj}) is 
much longer than the dynamical time everywhere inside the
light cylinder.

\begin{equation}
 { {\rm Im} (\ \Delta) \over \gamma_p \Omega } \approx
{\sqrt{ \lambda} \over \gamma_p^2 \sqrt{ \gamma_b }}
= 10^ {-4} \ll  1 
\label{dafslj1}
\end{equation}
Which implies that hydrodynamic regime of the cyclotron instability is
 unimportant.

\subsection{Perpendicular Propagation}

Next,  we consider the hydrodynamic instabilities for the
waves propagating perpendicular to the magnetic field
(magnetized Wiebel  instability (\cite{Weibel1959})).
The normal modes of plasma without a beam for $\theta=\,\pi/2$
follow from Eq. (\ref{dettt}):
\begin{eqnarray}
&&
{{\omega }^2} = k^2 + 2\,{{{\it  \omega_p}}^2}
\mbox{}  \\ \mbox{}
&&  {{\omega }^2} = {{{\it  {\omega_B}}}^2} + 2\,{{{\it  \omega_p}}^2}
\label{opjad1}
\mbox{} \\ \mbox{}
&&n^2 = 1 + {{2\,{{{\it  \omega_p}}^2}}\over
       {-{{\omega }^2} + {{{\it  {\omega_B}}}^2}}}
\label{opjad}
\end{eqnarray}

In the limit $   {\omega_B} \, \rightarrow \, \infty$ the two solution 
of the biquadratic  Eq.
(\ref{opjad})  may be expanded in large  $   {\omega_B}$: 
\begin{eqnarray}
&&
{{\omega }^2} =  {{{\it  {\omega_B}}}^2} + 2\,{{{\it  \omega_p}}^2} +
{{2\,{c^2}\,k^2\,{{{\it  \omega_p}}^2}}\over {{{{\it  {\omega_B}}}^2}}}
\label{opjad13}
\mbox{} \\ \mbox{}
&&  {{\omega }^2} = {c^2}\,k^2\,
     \left(1 - {{2\,{{{\it  \omega_p}}^2}}\over {{{{\it  {\omega_B}}}^2}}} \right)
\label{opjad14}
\end{eqnarray}

In the  limit $kc \ll {\omega_B}$ 
the dispersion curves (\ref{opjad13}) and  (\ref{opjad1}) 
approach each other near the upper hybrid frequency:
 $ {{\omega }^2} =  {{{\it  {\omega_B}}}^2} +
 2\,{{{\it  \omega_p}}^2}$. 
We   then expand determinant (\ref{epsilonnn}) for $\theta = \pi/2$ 
near the upper hybrid frequency $ \omega  = \sqrt{  {\omega_B}^2 + 2\, 
   \omega_p^2} + \ \Delta$ keeping the terms up to the second order in 
$ \ \Delta$.
\begin{equation}
  -{\frac{{{\Delta}^2}\,{{\omega_B}^2}}
       {{\omega_p^4}}} +
    {\frac{{c^2}\,\Delta\,k^2}
   {{\omega_B}\,\omega_p^2}} -
    {\frac{{c^2}\,k^2\,{\omega_b}^2}
      {\gamma_b\,{{\omega_B}^4}}} = 0
\label{opjad21}
\end{equation}

Solving Eq. (\ref{opjad21}), 
 we find the
frequency shift 
\begin{equation}
\ \Delta =   {{k\,{{{\it  \omega_p}}^2}\,\left({\sqrt{{\it \gamma_b}}}\,k -
        {\sqrt{{\it \gamma_b}\,k^2 - 4\,{{{\it  {\omega_b}}}^2}}} \right) }\over
    {2\,{\sqrt{{\it \gamma_b}}}\,{{{\it  {\omega_B}}}^3}}}
\label{fsdhuo}
\end{equation}
which shows an instability for $kc \, < \, {{2\,{\it  {\omega_b}}}/{{\it \gamma_b}}}$
with a maximum growth rate 
\begin{equation}
 {\rm Im} (\ \Delta)_{max} \approx \, 
 {{{{{\it  \omega_p}}^2}\,{{{\it  {\omega_b}}}^2}}\over
    {{{{\it \gamma_b}}^{{3\over 2}}}\,{{{\it  {\omega_B}}}^3}}}
\label{fsdhuo1}
\end{equation}
which is negligible  for all reasonable pulsar plasma parameters.

\section{Oblique Wave Excitation  in  Cold Plasma in the Hydrodynamic
Regime}
\label{Coldplasma}

In this section we develop a general theory of the hydrodynamic 
weak beam 
instabilities in the cold magnetized electron-positron plasma.
We 
expand (\ref{epsilonnn}) in small  ${\omega_b}$ keeping only first terms.
After considerable algebra we obtain
\begin{eqnarray}
\hskip -.2 truein 
&& \left(1 - n^2 + {{2\,{{{\it \omega_p}}^2}}\over
        {-{{\omega }^2} + {{{\it {\omega_B}}}^2}}} \right) \,
    \left(\left(-1 + n^2 +
         {{2\,{{{\it \omega_p}}^2}}\over {{{\omega }^2}}} \right) \,
       \left(1 - {{2\,{{{\it \omega_p}}^2}}\over
          {{{\omega }^2} - {{{\it {\omega_B}}}^2}}} \right) 
\right.
\mbox{} \nonumber \\ \mbox{}
&&
\left. \hskip 2 truein  +
      {{2\,n^2\,{{{\it {\omega_B}}}^2}\,{{{\it \omega_p}}^2}\,
          {{\cos ^2 \theta}}}\over
        {{{\omega }^2}\,\left({{\omega }^2} - {{{\it {\omega_B}}}^2} \right) }}
       \right) +
\mbox{} \nonumber \\ \mbox{}
&&
 {{\omega_b}^2\over
      \gamma_b^3\,  \hat{\omega^{  2} }}
  \left(-1 + n^2 \cos ^2 \theta + {2\, \omega_p^2\over
         \omega ^2 -  {\omega_B}^2}  \right) \,
 \left(-1 + n^2 +{ 2\, \omega_p^2\over
         \omega ^2 -  {\omega_B}^2 }\right) +
\mbox{} \nonumber \\ \mbox{}
&&
 {{{{{\it {\omega_b}}}^2}}\over \gamma_b
 {{{{\it  \tilde{{ \omega}}^{2} }}}}}  \left(
 {{2\,k c \,n^2\,\left(-1 + n^2 +
   {{2\,{{{\it \omega_p}}^2}}\over {{{\omega }^2} - {{{\it {\omega_B}}}^2}}}
\right) \,{\it \hat{\omega}}\,{\it \beta_b}\,\cos \theta\,{{\sin ^2 \theta}}
        }\over {\,{{\omega }^2}}} +
\right.
\mbox{} \nonumber \\ \mbox{}
&&
    {{k^2 c^2\,\left(-1 + n^2 \right) \,{{{\it \beta_b}}^2}\,
   \left(-1 + n^2\,{{\cos ^2 \theta}} \right) \,{{\sin ^2 \theta}}
        }\over {\,{{\omega }^2}}} +
\mbox{} \nonumber \\ \mbox{}
&&
 {2\,k^2 c^2\,{{{\it \omega_p}}^2}\,{{{\it \beta_b}}^2}\,
        \left({{2\,\left(-{{\omega }^2} + {{{\it {\omega_B}}}^2} +
                {{{\it \omega_p}}^2} \right) }\over
            {{{\omega }^2}\,{{\left(-{{\omega }^2} + {{{\it {\omega_B}}}^2}
                    \right) }^2}}} +
       {{n^2\,\left(1 + {{\cos ^2 \theta}} \right) }\over
        {{{\omega }^2}\,\left({{\omega }^2} - {{{\it {\omega_B}}}^2} \right)}
            } \right) \,{{\sin ^2 \theta}}} +
\mbox{} \nonumber \\ \mbox{}
&&
\left.
{ \hat{ \omega}^{2} \over \omega ^2 } 
    \left(  \phantom{{{{{a\over b}\over{a\over b}}}\over{{{a\over b}\over{a\over b}}}}}
\left(-1 +
            { 2\, \omega_p^2\over \omega ^2} \right) \,
           \left(-2 + n^2 + {{4\,{{{\it \omega_p}}^2}}\over
               {{{\omega }^2} - {{{\it {\omega_B}}}^2}}} +
             n^2\,{{\cos ^2 \theta}} \right)  
\right. \right. 
\mbox{} \nonumber \\ \mbox{}
&&
\left. \left. 
\phantom{{{{{a\over b}\over{a\over b}}}\over{{{a\over b}\over{a\over b}}}}} +
      n^2\,\left(-2 + n^2 +
             {4\, \omega_p^2\over
               \omega ^2 - \omega_B^2} \right) \,
           \sin ^2 \theta \right)  
\right)
\label{fhshafk}
\end{eqnarray}

The term containing $1/ \hat{  \omega^{  2}}$ contribute
to Cherenkov excitation and the  term containing $1/  \tilde{{ \omega }}^ 2$ 
contribute to the cyclotron excitation.

To find the growth rates we expand the plasma part of (\ref{fhshafk}) near the
plasma modes (Eqs.  (\ref{fjhi1}) and (\ref{fjhi}))
 and the beam part near the resonances 
$\hat{\omega}\,=\,0$ (for Cherenkov excitation)
or
$ \tilde{\omega} \,=\,0$ (for cyclotron instability).
The expansion of the   plasma part of (\ref{fhshafk}) near the
plasma modes is done according to the relation
\begin{equation}
\omega\,=\, \omega^{(0)} + \Delta \left({ \partial K_p \over \partial \omega}
\right)
\left|_{\omega^{(0)}} \phantom{{a\over b}}  \right.
\label{fhshafk2}
\end{equation}
where $K_p$ is the plasma
part of the determinant (\ref{epsilonnn})
\begin{eqnarray}
&&
K_p\,=\,  
 \left(1 - n^2 + {{2\,{{{\it \omega_p}}^2}}\over
        {-{{\omega }^2} + {{{\it {\omega_B}}}^2}}} \right) 
 \mbox{} \nonumber \\ \mbox{}
&& \times
\,
    \left(\left(-1 + n^2 +
         {{2\,{{{\it \omega_p}}^2}}\over {{{\omega }^2}}} \right) \,
       \left(1 - {{2\,{{{\it \omega_p}}^2}}\over
          {{{\omega }^2} - {{{\it {\omega_B}}}^2}}} \right)  +
      {{2\,n^2\,{{{\it {\omega_B}}}^2}\,{{{\it \omega_p}}^2}\,
          {{\cos ^2 \theta}}}\over
        {{{\omega }^2}\,\left({{\omega }^2} - {{{\it {\omega_B}}}^2} \right) }}
       \right) 
\label{fhshafk3}
\end{eqnarray}
and $\omega^{(0)}$ are the solutions  of the equation $K_p=0$.

Simultaneously, in the beam part of the Eq. (\ref{fhshafk})
we should use the  normal modes of the medium for the estimates
of $\omega$ and the refractive index $n$.

\subsection{ Excitation of Alfv\'{e}n waves}
\label{AlfvenCold}

We recall that the 
Cherenkov excitation of Alfv\'{e}n waves is possible only for $\mu < 1$
(Section \ref{Resonances}).
Expanding in small $\Delta$ near the Alfv\'{e}n wave dispersion relation 
in the limit of infinite magnetic field and
using the resonant frequency (Table \ref{rescold1})
 we can find the growth rate
\begin{equation}
\Delta =  
\left\{  \begin{array}{ll}
\phantom{{{{{a\over b}\over{a\over b}}}\over{{{a\over b}\over{a\over b}}}}}
 {\frac{{\sqrt{3}}\,{\omega_p^{{\frac{1}{3}}}}\,
       {{\omega_b}^{{\frac{2}{3}}}}\,
       {{\cos \theta}^{{\frac{1}{3}}}}}{{2^{{\frac{7}{6}}}}\,
       \gamma_b}} & 
\mbox{ if $\theta \,\ll \, 1/\gamma_b$} \\
\phantom{{{{{a\over b}\over{a\over b}}}\over{{{a\over b}\over{a\over b}}}}}
 {\frac{{\sqrt{3}}\,{\omega_p^{{\frac{1}{3}}}}\,
       {{\omega_b}^{{\frac{2}{3}}}}\,\cot \theta}{
       {2^{{\frac{7}{6}}}}\,{\gamma_b^{{\frac{8}{3}}}}}} &
\mbox{ if $\theta \, \gg  \, 1/\gamma_b$}
\end{array} \right.
\label{fhdur2}
\end{equation}

The hydrodynamic growth rate of the Alfv\'{e}n waves has a maximum
for parallel propagation.
The numerical estimate of the  maximum  
rate for Cherenkov excitation of Alfv\'{e}n wave is given by (\ref{lkfjsp})
subject to the condition $\mu < 1$ ( Section \ref{Resonances}).

We can now compare the instability growth length with the 
coherence length (\ref{vgg}). Using the estimate $\delta\theta \sim
1/\gamma_b$ we have
\begin{equation}
{R_c \delta\theta {\rm Im} \Delta\over c \gamma_p^2} \approx 
 {R_c \Omega \over c} { \lambda^{1/6} \over \gamma_p^3 \gamma_b^2} 
\sqrt{ { \omega_B \over \Omega }}
\ll 1
\label{kf}
\end{equation}
which implies that the Cherenkov instability on the
Alfv\'{e}n waves is unimportant.

\subsubsection{Cyclotron excitation of Alfv\'{e}n waves}

Near the cyclotron resonance $ \tilde{\omega } =0$ we keep in 
(\ref{fhshafk}) only the terms proportional to $1/  \tilde{\omega }^ 2$.
We consider two separate case depending on whether
$\mu$ is larger or smaller than unity.

For $\mu < 1$  (i.e. near the neutron star surface)
using the short wave length
asymptotics of the Alfv\'{e}n branch (\ref{fhdu12}) and the resonance condition
(\ref{uiaj1}) we find 
a pair of complex solutions
\begin{equation}
   \Delta \,=\, \pm i
  {\frac{{\sqrt\omega_p}\,{\omega_b}\,
       {\sqrt{\cos \theta}}\,\sin \theta}{{2^{{\frac{5}{4}}}}\,
       {\sqrt{\omega_B}}}} =\,
{\lambda ^{1/4} \Omega  {\sqrt{\cos \theta}}\,\sin \theta
 \over \sqrt{2} \gamma_p^{3/4}} 
\left( {{\omega_B} \over  \Omega} \right)^{1/4}
\label{fhd118}
\end{equation}
This frequency shift agrees with the $\delta_-^{C3}$ of the Table 3,
region 3 of
Godfrey \cite{Godfrey}.

The ratio of the growth rate in equation (\ref{fhd118}) to the 
dynamical time is
\begin{equation}
{  {\rm Im}  ( \Delta) \over \Omega \gamma_p } = 
{\gamma_b^{1/4}  {\sqrt{\cos \theta}}\,\sin \theta 
\over \sqrt{2} \gamma_p^2} \left( {{\omega_B} \over  \Omega} \right)^{1/4}
\leq 150 \left( { r \over R_{NS} } \right) ^{-3/4} 
\label{f1}
\end{equation}

The criteria of a short growth length (\ref{vg9}) gives
\begin{equation}
{ R_c \delta \theta {\rm Im}  ( \Delta) \over c \gamma_p^2}
\approx { R_c \Omega \over c} {\lambda^{1/4} \over \gamma_p^{11/4} }
 \left( {{\omega_B} \over  \Omega} \right)^{1/4}
=  1.6 \left( { r \over R_{NS} } \right) ^{-3/4}
\label{f01}
\end{equation}
where we have approximated the trigonometric functions by unity.

The equations (\ref{f1}) and (\ref{f01})
imply that the cyclotron excitation
of Alfv\'{e}n wave in the hydrodynamic regime 
may be marginally important near the neutron star
(see also Sections \ref{HydorHotAlfvenCyclo} and Eq. \ref{gioj1}).

For $\mu \gg 1$  and $\theta \gg 1/\sqrt{\mu}$ (this case corresponds
to the resonant frequency much smaller than $\omega_p$)
 using the short wave length
asymptotics of the Alfv\'{e}n branch (\ref{fhdu12}) and the resonance condition
(\ref{uiaj1}) we find  
a pair of complex solutions
\begin{equation}
\Delta=\pm  i
 {\frac{{{\omega_B}^{{\frac{1}{3}}}}\,
      {\omega_b}\,{{\tan \theta}^{{\frac{1}{3}}}}}{{2^
        {{\frac{5}{6}}}}\,{\gamma_b^{{\frac{5}{6}}}}\,
      {\omega_p^{{\frac{1}{3}}}}}} = \pm  i
{ \Omega  \tan ^ {1/3} \theta \over 2^{3/2} \lambda^{1/6} \gamma_b ^{5/6}
 \gamma_p^{1/3}}
\left( {{\omega_B} \over  \Omega} \right)^{2/3} 
\label{fhd115}
\end{equation}

Comparing this growth rate with the dynamical time we obtain
\begin{equation}
{  {\rm Im}  ( \Delta) \over \Omega \gamma_p } =
{ \tan \theta ^{1/3}  \over 2^{3/2}  \lambda^{1/6} \gamma_b ^{5/6}
 \gamma_p^{4/3} } 
\left( {{\omega_B} \over  \Omega} \right)^{2/3}
\approx 3 \times 10^4 \tan \theta ^{1/3} \left( { r \over R_{NS} } \right)^{-2}
\label{fhd115a}
\end{equation}
where we estimated the typical angle of emission by unity.
The ratio (\ref{fhd115a})  is larger than unity for 
$ r/  R_{NS} \geq 43$.

 The
second criteria on the growth rate (\ref{vgg}) becomes
\begin{equation}
{ R_c \delta \theta {\rm Im}  ( \Delta) \over c \gamma_p^2}
\approx { R_c \Omega \over c} {1\over \lambda^{1/6} \gamma_b^{5/6}
\gamma_p^{7/3} } \left( {\omega_B \over \Omega} \right)^{2/3}=
120 \left( {r\over R_{NS} }\right)^{-2}
\label{fa}
\end{equation}
which implies that the cyclotron excitation
of Alfv\'{e}n waves is efficient in the region where $\mu > 1$ 
($ { r \over R_{NS} } \geq 43 $).

\subsection{Excitation of the X wave}

Using (\ref{fhshafk}) we see that 
X mode {\it is not emitted} by the Cherenkov resonance
since the corresponding term is zero if evaluated at the
X mode (\ref{fjhi1}).

Near the cyclotron resonance we  obtain for the X mode
 a pair of complex solutions
\begin{equation}
{\it  \Delta} = \pm {i\over 2}\,{\sqrt{  {\omega_B}}\,{\it  {\omega_b}}\over
      \gamma_b\,\sqrt{\omega }}
\label{fhd22}
\end{equation}
where we used $kc \, \approx \omega$. 
Using   Table \ref{rescold1}
we obtain a growth rate for the cyclotron excitation
of the X mode
\begin{equation}
{\it  \Delta} = \pm {{{i}\,{\it  \omega_p}\,{\it  {\omega_b}}}\over
 2\,   {{\sqrt{{\it \gamma_b}}}\,{\it  {\omega_B}}}}
\label{fhd5}
\end{equation}

The growth rate is almost constant inside the cone $\theta < {2\omega_p
\over {\omega_B}}$ and is zero for larger $\theta$.  The growth rate
(\ref{fhd5}) was also obtained for the case of parallel propagation
(\ref{dafslj}). As we saw in 
Section \ref{Parallelpropagation} 
the hydrodynamic instability on the X
mode is unimportant.

\subsection{Excitation of the O mode  }
\label{ColdplasmaOrdinary}

\subsubsection{ Cherenkov resonance (  $\mu >1$)}
\label{Coldplasmaordinary}

Starting with (\ref{fhshafk}) 
we expand the plasma part near the dispersion relation
for O mode  
(\ref{fjhi}) and the beam part near Cherenkov resonance (\ref{uiaj})
 $ n = 1/(\beta_b \cos\theta) $.
In the low frequency approximation $\omega \ll \omega_B$ O 
wave can be excited by Cherenkov resonance only if $ \mu > 1$.
We distinguish two cases: $\mu \geq 1$ and $ \mu \gg 1$. In the former
case the resonance occurs at approximately cross-over point for the
O wave while in the latter case the resonance occurs at 
$\omega \gg \omega_0$.
 
In the case $\mu \geq 1$  with $\omega_{\rm  res} \,$ given in Table \ref{rescold1},
the growth rate is 
\begin{equation}
\Delta=  {\frac{{\sqrt{6}}\,{{\omega_B}^{{\frac{2}{3}}}}\,
      \omega_p\,{{\omega_b}^{{\frac{2}{3}}}}\,
      \sin \theta}{2\,
      {{\gamma_b}^{{\frac{2}{3}}}}\,
 {{\left( -1 + {{\mu}^2} \right) }^{{\frac{1}{6}}}}\,
      {{\left( 8\,{{\omega_p}^4} +
           2\,{{\omega_B}^2}\,{{\sin^2\theta}} \right) }^
        {{\frac{1}{3}}}}}}
\label{liafs112}
\end{equation}

The growth rate is proportional to the angle of propagation with
respect to the magnetic field. This is  due to the increase of the  potential 
part of the O mode with the angle. The case of $\mu = 1$ corresponds
to the unlikely case when the beam velocity is exactly equal to the
Alfv\'{e}n velocity. The growth rate (\ref{liafs112}) is valid for 
$\theta < 1/\gamma_b$. For  all practical purposes this growth rate is 
very small.

In the case $\mu \gg 1$ the complex shifts of the frequencies near the 
 Cherenkov resonance 
$ \hat{\omega} = \Delta$ 
are given by
\begin{eqnarray}
{\rm Im}( \Delta) \,=\, &&
 {\frac{{\sqrt{3}}\,{\omega_p^{{\frac{2}{3}}}}\,
      {{\omega_b}^{{\frac{2}{3}}}}\,
      {\sqrt{2\,\omega_p^2 +
          {{\omega_B}^2}\,{{\sin^2\theta}}}}}{2\,
    \gamma_b\,{{\left( 4\,{\omega_p^4} +
           {{\omega_B}^4}\,{{\sin^2\theta}} \right) }^
        {{\frac{1}{3}}}}}} \,=
\mbox{} \nonumber \\ \mbox{}
&&
 \left\{ \begin{array}{ll}
\phantom{{{{{a\over b}\over{a\over b}}}\over{{{a\over b}\over{a\over b}}}}}
 { \sqrt{3} {\omega_b}^{2/3} \, \omega_p^{1/3} \over
2^{7/6} \gamma_b } & 
\mbox{ if $ \theta \, \ll \, { 2 \omega_p^2 \over {\omega_B}^2} $} \\
\phantom{{{{{a\over b}\over{a\over b}}}\over{{{a\over b}\over{a\over b}}}}}
 \sqrt{ {3\over 2}} \, { \omega_p^{5/3}\, {\omega_b}^{2/3} \over
\gamma_b {\omega_B}^{4/3} \sin^{2/3} \theta}  &
 \mbox{ if $ \theta \, \gg \, { 2 \omega_p^2 \over {\omega_B}^2}$}
\end{array} \right. 
\mu \gg 1
\label{fhd1191}
\end{eqnarray}

The maximum growth rate is reached for the parallel propagation, while for oblique
propagation the growth rate falls off as $\theta ^{-2/3}$.
We have already estimated the growth rate for the parallel propagation
(\ref{lkfjsp}) and found that it can be marginally  efficient.
We can apply the second requirement on the growth rate which
relates the characteristic angle of emission and the radius of curvature
(\ref{vgg}). Estimating $\delta \theta \approx {\omega_p^2 \over {\omega_B}^2}$
we find
\begin{equation}
{  R_c \delta \theta   {\rm Im}  \Delta \over  c \gamma_p^2} =
\left( { \lambda ^{4/3} \over \gamma_p^3} \right)
\left( { \Omega \over {\omega_B}}  \right)
\left( { R_c \Omega \over c }  \right)
\ll 1 
\label{l1}
\end{equation}
which implies that the Cherenkov instability on the O mode
does not develop.

\subsubsection{ Cyclotron excitation of the O mode}

The growth 
rate for the cyclotron excitation of the O mode in the hydrodynamic
regime may be estimated from the growth 
rate for the cyclotron excitation of the X mode 
(\ref{fhd22})  with the resonant frequency given in 
Table \ref{rescold1}.
The maximum growth rate is reached for the parallel propagation and is equal
to the cyclotron growth rate of the X mode (\ref{fhd22}).
The growth rate (\ref{fhd22}) decreases with the increase of the 
resonant frequency. Since the  cyclotron resonance on the O mode 
happens at the frequencies larger than the the  cyclotron resonance on the
X mode,  the corresponding growth
rate of the O mode are  smaller for oblique propagation.

We conclude this section by the table of the hydrodynamic growth rate
in a cold plasma (Table \ref{COLDGROWTH}).

 \begin{table} \begin{center}
\caption
{ Hydrodynamic growth rates in cold plasma}
\psfig{file=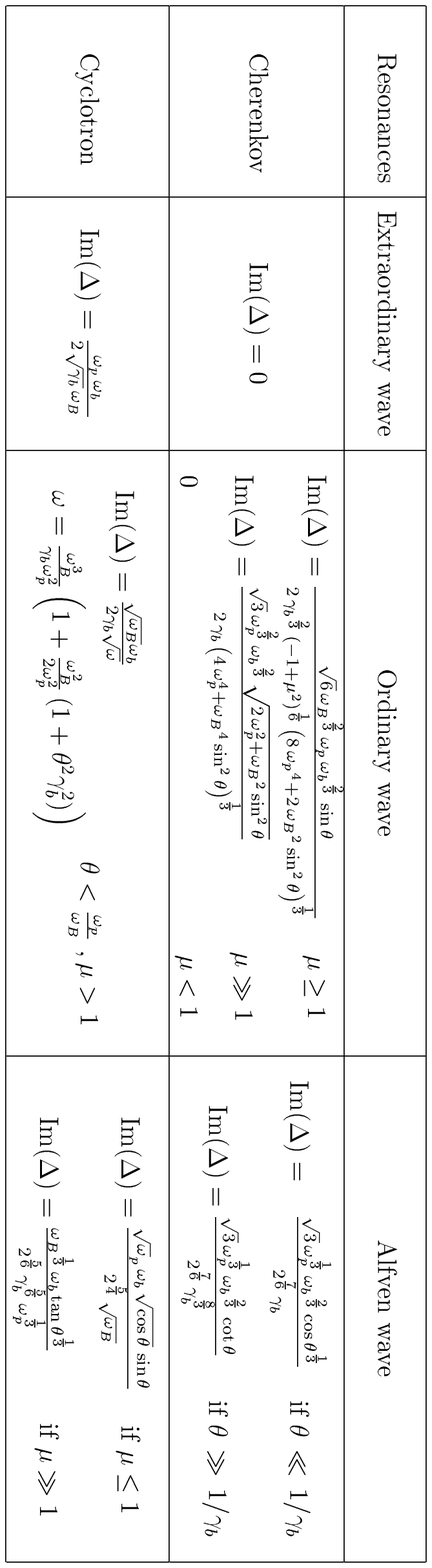,width=18.0cm}
\label{COLDGROWTH}
\end{center} \end{table}

\section{ Relativistic  Pair
Plasma: Resonances}
\label{Resonances1}

From the low frequency approximation to the  Alfv\'{e}n waves
 dispersion we find that
the  possibility of Cherenkov excitation of Alfv\'{e}n wave 
in a relativistic  hot plasma 
depends on the parameter
\begin{equation}
\mu_h = \sqrt{
{ 2 \gamma_b^2 \, T_p   \omega_p^2 (1+ \beta_T^2) \over {\omega_B}^2} }
\approx { 2 \gamma_b \sqrt{ \, T_p } \omega_p \over {\omega_B}}
\label{liafs1}
\end{equation}
(compare with (\ref{uiaj21})).

Using our fiducial numbers parameter $\mu$ may be estimated
\begin{equation}
\mu_h= 
 2 \gamma_b \sqrt{ { \, T_p  \lambda \Omega  \over \omega_B \gamma_p} } =
  = 5 \times
10^{-3} \left({r\over R_{NS}}\right)^{3/2}
 \,=\,
\left\{ \begin{array}{ll}
< 1, & \mbox{ if $  \left({r\over R_{NS}} \right) < 43 $} \\
> 1,  & \mbox{ if $  \left({r\over R_{NS}} \right) > 43 $}
\end{array}\right.
\label{liafss2}
\end{equation}
Numerically $\mu_h$ and $\mu$ are equal for 
the chosen  set  of the fiducial numbers for the cold and  hot cases.

Similarly to the cold case,
 the parameter $\mu_h$ determines the possibility
of the excitation of the Alfv\'{e}n and O waves.
If $ \mu_h \,<  \, 1$ then the O wave cannot be excited by Cherenkov
resonance. In this case the Alfv\'{e}n wave may be excited by the Cherenkov
interaction subject to the condition that the resonance occurs on the 
parts of the dispersion curve that are not strongly damped (see below).
If  $ \mu_h \, >  \, 1$ then the O wave may excited by Cherenkov
resonance
 for the angles of propagation $ \theta < { 
\sqrt{ \, T_p } \omega_p \over {\omega_B}}$. 

Another limitation on the possible resonance comes from the requirement that
the waves in the plasma are not strongly damped at the location of the 
resonance. This is an important constraint on the resonance
of the Alfv\'{e}n wave, which is strongly damped at large wave vectors.

Using the dispersion relation for the  Alfv\'{e}n waves in the limit
$kc \ll \omega_p$, we find that the cyclotron resonance on the
 Alfv\'{e}n wave occurs at $kc \ll \omega_p$ for the
angles of propagation larger than 
\begin{equation}
\theta^2 = { \omega_B \sqrt{T_p} \over \gamma_b \omega_p}
\label{ves}
\end{equation}

\begin{figure}
\psfig{file=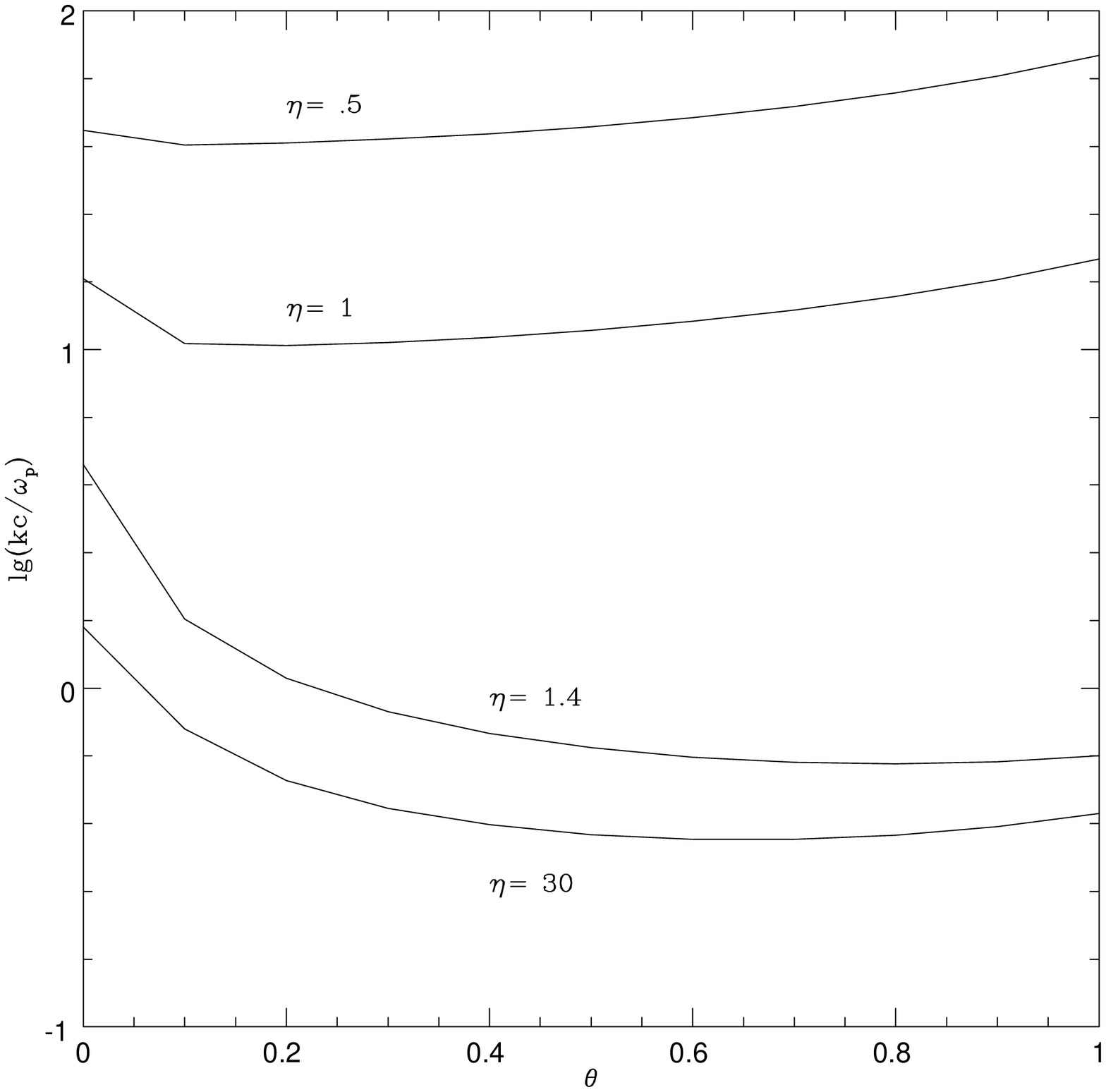,width=12.0cm}
\caption[Cyclotron resonance of the Alfv\'{e}n wave] {
 Location of a cyclotron resonance of the Alfv\'{e}n wave. 
For $\eta \leq 1$ (very hot plasma) cyclotron resonance on Alfv\'{e}n waves
 occurs at $ kc \gg \, T_p ^{1/2}  \omega_p $, 
 where the waves  are strongly damped.
\label{fig7}
}
\end{figure}

For smaller angles the  location of the cyclotron
resonance  on the Alfv\'{e}n wave 
depends on the parameter 
\begin{equation}
\eta\,=\,
{\frac{\gamma_b\,\omega_p}
    {{\, T_p ^{{\frac{3}{2}}}}\,{\omega_B}}}
\label{fhdu26}
\end{equation}
If $\eta \ll 1$ (very hot plasma), then
the  cyclotron
resonance  on the Alfv\'{e}n wave
occurs in the region
$\omega \gg \omega^{(0)}$, where Alfv\'{e}n waves are strongly damped.
If, on the other hand,  $\eta \gg 1$ (warm plasma),
the  cyclotron
resonance  on the Alfv\'{e}n wave
occurs at approximately $\omega^{(h)}_0$, wher 
 Alfv\'{e}n waves are not damped (Fig. \ref{fig7}).

Since Alfv\'{e}n wave cannot escape to infinity, they
should be converted to electromagnetic modes before they are
damped on the thermal particles. The Alfv\'{e}n waves with
large angles, which are generated with the frequency
$\omega \ll \omega^{(h)}_0$ would have more time for the
nonlinear processes to convert them into escaping modes, than
the  Alfv\'{e}n waves generated in a warm
plasma with $\omega \approx \omega^{(h)}_0$ and small angles of propagation.
 Thus, the cyclotron
resonance  on the Alfv\'{e}n wave
is likely to produce waves propagating in a cone around magnetic field.

The resonances in the relativistic pair plasma are given in Table
\ref{reshot1}.

\begin{table} \begin{center}
\caption
{ Resonances in hot pair plasma}
\psfig{file=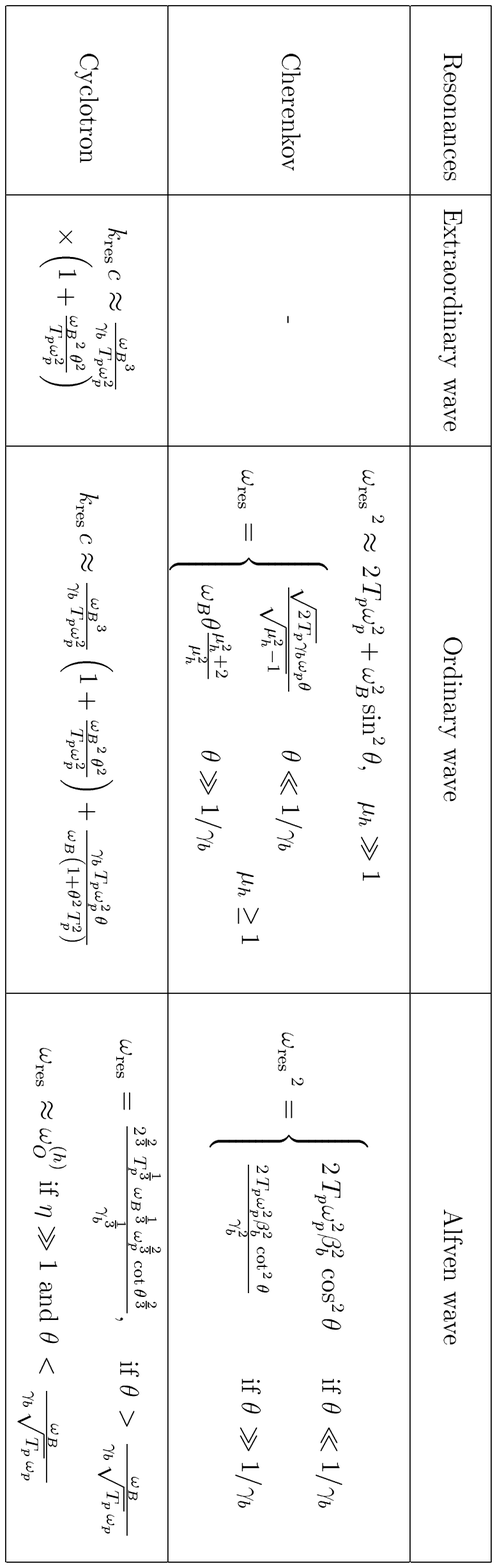,width=18.0cm}
\label{reshot1}
\end{center} \end{table}

\section{ Hydrodynamic Wave Excitation in Relativistic  Pair Plasma}
\label{Hydroexcitrelativistic}

\subsection{Dielectric Tensor for the Beam-Hot Plasma System  }
\label{DielTe}
To simplify the analysis we will use the low frequency
approximation $\omega \, \ll {\omega_B}$ and the assumption
of a very strong magnetic field $ {\, T_p  \omega_p^2 \over {\omega_B}^2}
\, \ll 1$ from the very beginning. The dielectric tensor is then given by
\begin{eqnarray}
\hskip -.2 truein 
\epsilon_{xx}&&=   1 + d\,\, T_p \,
      \left(1 + n^2\,\beta_T^2\,\cos^2\theta \right)\,
-
 {{{{{\it  {\omega_b}}}^2}\,{{{\it  \hat{\omega}}}^2}}\over
       {{\it\gamma_b}\,{{\omega }^2}\,{{{\it  \tilde{{\omega}}}}^2}}}
\, =\epsilon_{yy}
\mbox{} \nonumber \\ \mbox{}
\epsilon_{xy}&&= {{-i\,{{{\it  {\omega_b}}}^2}\,{\it {\omega_B}}\,{\it
 \hat{\omega}}}\over
  {{{{\it\gamma_b}}^2}\,{{\omega }^2}\,{{{\it  \tilde{{\omega}}}}^2}}}
=- \epsilon_{yx}
\mbox{} \nonumber \\ \mbox{}
\epsilon_{x z}&&=
 d\,\, T_p \,n^2\,\beta_T^2\,\cos \theta\,\sin \theta
-{{k\,{{{\it  {\omega_b}}}^2}\,{\it  \hat{\omega}}\,{\it \beta_b}\,\sin  \theta}
\over
         {{\it\gamma_b}\,{{\omega }^2}\,{{{\it  \tilde{{\omega}}}}^2}}}=
\epsilon_{ z x}
\mbox{} \nonumber \\ \mbox{}
\epsilon_{ y  z}&&=
{{i\,k\,{{{\it  {\omega_b}}}^2}\,{\it {\omega_B}}\,{\it \beta_b}\,\sin  \theta}\over
      {{{{\it\gamma_b}}^2}\,{{\omega }^2}\,{{{\it  \tilde{{\omega}}}}^2}}}
 =
-\epsilon_{ z y}
\mbox{} \nonumber \\ \mbox{}
\epsilon_{ z  z }&&=  1 - {\frac{2\,n^2\,\omega_p^2}
       {\, T_p \,\left(1 - n^2\,\beta_T^2\,\cos^2\theta
            \right) }} +
 d\,\, T_p \,n^2\,{{\sin^2\theta}} -{\frac{\omega_b^2}
      {{\gamma_b^3}\, \hat{\omega}^{2}}} -
  {\frac{k^2\,{\omega_b}^2\,{{\beta_b}^2}\,
 {{\sin^2\theta}}}{\gamma_b\,{{\omega}^2}\,
       \tilde{\omega}^{2}}}
\label{dettt1}
\end{eqnarray}

\subsection{Parallel Propagation}
\label{ParalPropa}

For parallel propagation Eq. (\ref{dettt}) with the dielectric
tensor (\ref{dettt1}) factorizes:
\begin{eqnarray}
&& 
 1  -
   {\frac{2\,n^2\,\omega_p^2}
     {\, T_p \,\left(1 - n^2\,\beta_T^2 \right) }} 
- {\frac{\omega_b^2}
     {{\gamma_b^3}\, \hat{\omega^2}}} =0
\label{liafs5}
\mbox{}  \\ \mbox{}
&& 
  1 - n^2 + {\frac{{\omega_b}^2\, \hat{\omega}}
      {\gamma_b\,{{\omega}^2}\,
        \left({\omega_B}/\gamma_b - \hat{\omega} \right) }} +
    d\,\, T_p \,\left(1 + n^2\,\beta_T^2 \right) =0
\mbox{}  \\ \mbox{}
\label{liafs6}
&&
 1 - n^2 - {\frac{{\omega_b}^2\,\hat{\omega} }
      {\gamma_b\,{{\omega}^2}\,
        \left({\omega_B} /\gamma_b  + \hat{\omega} \right) }} +
    d\,\, T_p \,\left(1 + n^2\,\beta_T^2 \right) =0
\label{liafs4}
\end{eqnarray}

Following the same procedure of expanding the dispersion 
relations in small frequency shifts  $\Delta$ near the
intersection of the two resonant curves, 
we find  from
(\ref{liafs5}) the growth
rate for the Cherenkov excitation of plasma waves:

\begin{equation}
{\rm Im} (\Delta) =  {\frac{{\sqrt{3}}\,{\omega_p^{{\frac{1}{3}}}}\,
      {{\omega_b}^{{\frac{2}{3}}}}}{{2^{{\frac{7}{6}}}}\,
      \gamma_b\,{\sqrt{\, T_p }}}}=
\, { \sqrt{3} \sqrt{ \Omega \omega_B} \lambda ^{1/6}  \over
2^{2/3} \gamma_b \sqrt{\gamma_p \, T_p } }
\label{liafs8}
\end{equation}
(cf. with Egorenkov et al. \cite{}).

Using the relations between parameters of the hot plasma (Eq. \ref{vg} with
$<\gamma> =2 T_p \gamma_p$),
the condition of a fast growth (\ref{vg9}) for the growth rate (\ref{liafs8})
takes the form
\begin{equation}
{ {\rm Im} (\Delta) \over \gamma_p \Omega }\approx \,
 { \lambda ^{1/6} \over \gamma_b \gamma_p ^{3/2} \, T_p ^{1/2} } 
\sqrt{ { \omega_B \over \Omega}} = 20 
\left({r\over R_{NS} }\right)^{-3/2}
\label{liafs001}
\end{equation}

For the fixed values of $\gamma_b$ and $\gamma_p$ 
the growth rate for the 
Cherenkov excitation of plasma waves in a hot plasma 
is smaller by the factor $ \, T_p ^{2/3}$ as compared with the cold plasma.

Solving Eq. (\ref{liafs6}),
 we find the growth rate for the 
cyclotron excitation of transverse waves

\begin{equation}
\Delta= i
 {\frac{{\sqrt{\, T_p }}\,\omega_p\,
      {\omega_b}}{{ 2\,\sqrt\gamma_b}\,{\omega_B}}}=
\sqrt{{  \lambda \, T_p  \over  \gamma_b} } {\Omega \over \gamma_p}
\label{liafs132}
\end{equation}

Comparison of this growth rate with the dynamical time gives

\begin{equation}
 { {\rm Im} ( \Delta) \over \Omega \gamma_p} =
{1\over \gamma_p^2} \sqrt{{  \lambda \, T_p  \over  \gamma_b} } 
 \approx
{1\over \gamma_p^{2} }  \left({y \over R_{NS} }\right)^{3/2}
= 10^ {-4}  \left({y \over R_{NS} }\right)^{3/2} < 1
\label{dafj1}
\end{equation}

From (\ref{dafj1}) and (\ref{dafslj1}) if follows that
the cyclotron excitation of the transverse waves
in the hydrodynamic regime
is not affected by the relativistic temperature of the plasma
particles and is not important in the pulsar magnetosphere.

Similarly to the cold case we omit the details of the calculations
of the growth rates and 
 conclude this section by the table of the hydrodynamic
growth rates in the relativistic hot pair plasma
Table \ref{HOTGROWTH}.

\begin{table} \begin{center}
\caption[ Hydrodynamic growth rates  in hot  plasma]
{ Hydrodynamic growth rates in hot  plasma}
\psfig{file=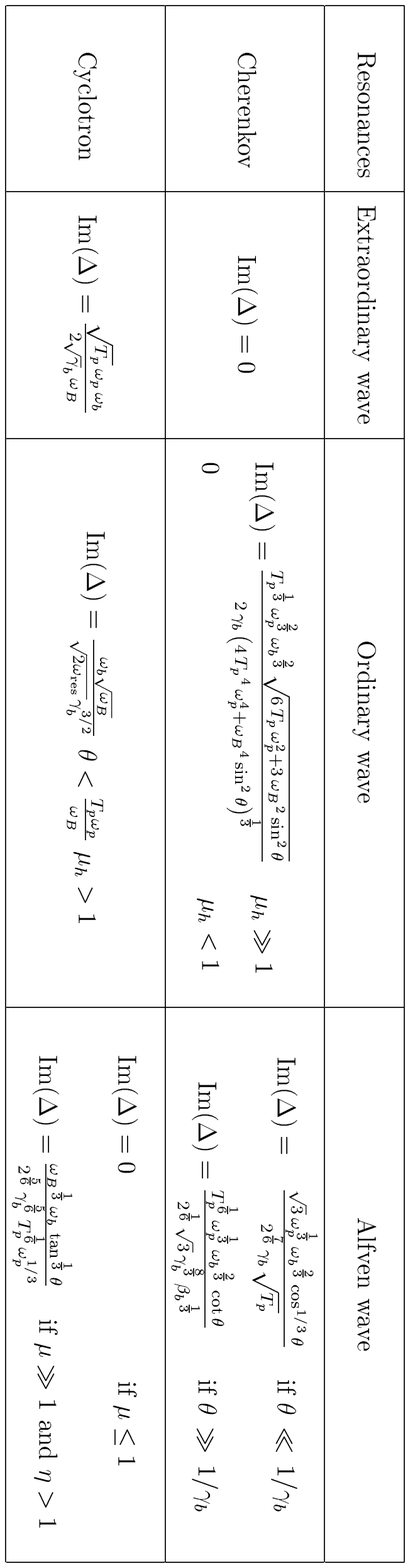,width=18.0cm}
\label{HOTGROWTH}
\end{center} \end{table}

\section{ Excitation of Oblique Waves in Relativistic  Pair Plasma}
\label{Hydroexcitrelativistic1}

\subsection{Excitation of Alfv\'{e}n Waves}

\subsubsection{ Cherenkov excitation of the Alfv\'{e}n  mode ( $ \mu _h
< 1$)}
\label{Lil}

We give here the growth rates for the 
Cherenkov excitation of the Alfv\'{e}n  mode in the limit
$ \mu _h \, \ll 1$. Then we can use the infinite
magnetic field approximation to the dispersion of  Alfv\'{e}n waves.
We then expand the plasma part of the  the dielectric tensor
(\ref{det1}) in large gyrofrequency keeping the zeroth order ( $d=0$).

The complex part of the frequency shift is
\begin{equation}
 {\rm Im} (\Delta) = 
\left\{  \begin{array}{ll}
 {\frac{{\sqrt{3}}\,{\omega_p^{{\frac{1}{3}}}}\,
      {{\omega_b}^{{\frac{2}{3}}}} \cos^{1/3} \theta }{{2^{{\frac{7}{6}}}}\,
      \gamma_b\,{\sqrt{\, T_p }}}} 
\phantom{{{{{a\over b}\over{a\over b}}}\over{{{a\over b}\over{a\over b}}}}}
 & \mbox{ if $\theta \, \ll  1/ \gamma_b $} \\
 {\frac{{\, T_p ^{{\frac{1}{6}}}}\,
      {\omega_p^{{\frac{1}{3}}}}\,
      {{\omega_b}^{{\frac{2}{3}}}}\,\cot \theta}{{2^
        {{\frac{1}{6}}}}\,{\sqrt{3}}\,{\gamma_b^{{\frac{8}{3}}}}\,
      {{\beta_b}^{{\frac{1}{3}}}}}}
\phantom{{{{{a\over b}\over{a\over b}}}\over{{{a\over b}\over{a\over b}}}}}
 & \mbox{ if $\theta \, \gg  1/ \gamma_b $}
\end{array}\right.
\label{liafsp11}
\end{equation}

\subsubsection{ Cyclotron excitation of Alfv\'{e}n wave}
\label{HydorHotAlfvenCyclo}

Following the same procedure as in Section \ref{Lil}
we find equation governing the cyclotron excitation of Alfv\'{e}n wave
in the relativistic pair plasma 
in the case when the plasma part of the dispersion relation for
the Alfv\'{e}n wave is calculated in the infinite magnetic field limit.
Assuming that $ 1 - n\,\cos \theta \gg  {\gamma_b^{-2}}$
we can set $\beta_b$ to unity.

Following the discussion in Section \ref{Resonances1} we are interested in the
Alfv\'{e}n wave 
cyclotron instabilities occurring  at frequencies $ \omega \leq 
\sqrt{ T_p }\omega_p$ since for larger frequencies the Alfv\'{e}n wave
is strongly damped. This is satisfied only for warm plasma ( $\eta > 1$).
We can analytically find the growth rate in this  case for the 
angles of propagation larger than $\mu_h$ ($\mu_h \ll 1$).
 In this case, using $\omega \approx kc \cos \theta$,
we find 
two complex solutions for $\Delta$
\begin{equation}
\Delta= \pm i
  \frac{{\omega_B}^{\frac{1}{3}}\,{\omega_b}\,
      \tan  ^{\frac{1}{3}}  \theta}{2^{\frac{5}{6}}\,
      \gamma_b^{\frac{5}{6}}\,
      \, T_p ^{\frac{1}{6}} \omega_p^{1/3} }=
{ \Omega  \tan ^ {1/3} \theta \over 2^{3/2} \lambda^{1/6} 
\, T_p ^ {1/6} \gamma_b ^{5/6}
 \gamma_p^{1/3}}
\left( {{\omega_B} \over  \Omega} \right)^{2/3}
\label{gioj1}
\end{equation}

Using the relations between parameters of the hot plasma
we conclude that the
cyclotron excitation of Alfv\'{e}n wave is unaffected
by the relativistic thermal spread of the plasma particles.

\subsection{Cyclotron excitation  of the X mode }

Following the same procedure and using the resonant condition 
(Table \ref{reshot1}),
we find the following  frequency
shifts describing the cyclotron excitation  of the X mode
in a relativistic pair plasma:
\begin{equation}
\Delta= 
 {\frac{{\frac{i}{2}}\,{\sqrt{\, T_p }}\,\omega_p\,
      {\omega_b}}{{\sqrt\gamma_b}\,{\omega_B}}}
\label{liafs1321}
\end{equation}

\subsection{Excitation of the  O mode}

\subsubsection{Cherenkov excitation of O mode ($ \mu _h > 1$)}

Like in the case of the plasma, the 
Cherenkov excitation of the O mode case occur in two different
regimes: $\mu _h \geq 1$ and $\mu _h \gg 1$. In the former case the 
resonant frequency is much larger that the cross-over frequency. In this
limit the  O mode is quasi transverse and 
 Cherenkov excitation is unimportant.

In the case  $ \mu _h \gg 1$ the Cherenkov resonance occurs approximately
at the cross-over point.
 The resonant frequency may be
approximated by the cross-over point:
$ \omega^2 = 2 \, T_p  \omega_p^2 + \omega_B^2 \sin ^2 \theta$.
The complex part of the frequency shift is given by
\begin{eqnarray}
&&
{\rm Im} \Delta =
 {\frac{{{\, T_p }^{{\frac{1}{3}}}}\,
       {\omega_p^{{\frac{2}{3}}}}\,
       {{\omega_b}^{{\frac{2}{3}}}}\,
       {\sqrt{6\,\, T_p \,\omega_p^2 +
         3\,{{\omega_B}^2}\,{{\sin^2\theta}}}}}{2\,
       \gamma_b\,{{\left( 4\,{{\, T_p }^4}\,
             {\omega_p^4} +
            {{\omega_B}^4}\,{{\sin^2\theta}} \right) }^
         {{\frac{1}{3}}}}}} =\,
 \mbox{} \nonumber \\ \mbox{}
&&
\left\{  \begin{array}{ll}
 {\frac{{\sqrt{3}}\,{\omega_p^{{\frac{1}{3}}}}\,
       {{\omega_b}^{{\frac{2}{3}}}}}{2\,{2^{{\frac{1}{6}}}}\,
       \gamma_b\,{\sqrt{\, T_p }}}}
& \mbox{ if $\theta \, \ll 
 {\frac{{{\, T_p }^2}\,\omega_p^2}
     {{{\omega_B}^2}}} $} \\
 {\frac{{\sqrt{{\frac{3}{2}}}}\,{{\, T_p }^{{\frac{5}{6}}}}\,
       {\omega_p^{{\frac{5}{3}}}}\,
       {{\omega_b}^{{\frac{2}{3}}}}}{\gamma_b\,
       {{\omega_B}^{{\frac{4}{3}}}}\,
       {{\sin \theta}^{{\frac{2}{3}}}}}}
& \mbox{ if $ {\frac{{{\, T_p }^2}\,\omega_p^2}
     {{{\omega_B}^2}}} \, \ll \theta \, \ll
 {\frac{{\sqrt{2}}\,{\sqrt{\, T_p }}\,\omega_p}
     {\omega_B}} $}
\end{array}\right.
\label{pds}
\end{eqnarray}

The maximum growth rate is reached for the parallel propagation
(\ref{liafs01}). In Section \ref{Coldplasmaordinary} we found that in cold
plasma the Cherenkov growth of the O mode is
unimportant due to the very short coherence length, which is,
in turn, limited by the small range of angles of the growing
waves. Since the hydrodynamic growth rate of the O mode 
in hot plasma is smaller than in cold plasma, we can make
a conclusion that this instability is unimportant.

\subsubsection{ Cyclotron excitation of the O mode}

Similarly to the case of a cold plasma, the growth rate for the
cyclotron excitation of the O mode may be
estimated from the growth rate of the X mode
with the resonant frequency given in Table \ref{reshot1}.
The growth rate has a maximum for the parallel propagation and
decreases with the angle due to the sharp increase of the 
resonant frequency.

We conclude this section by the table of the hydrodynamic
growth rates in the relativistically hot pair plasma
(Table \ref{HOTGROWTH}).

\section{ Kinetic Instabilities}
\label{Kineticinstabilities}

As we have discussed in Section \ref{HydrodynamicOrkinetic}, 
 a general beam instability may be
 treated analytically in the hydrodynamic and kinetic
limiting cases. We have considered hydrodynamic beam instabilities 
in pair plasma in Sections \ref{Coldplasma} and \ref{Hydroexcitrelativistic}. 
Now we turn to the kinetic 
regime of instabilities. The condition for the kinetic consideration
to apply is the opposite of the condition (\ref{a3}). It requires a substantial
scatter in the velocities of the resonant particles.
In what follows 
we assume that distribution of the  beam particles 
is described by the 
 relativistic, one-dimensional Maxwellian distribution:
\begin{equation}
f(p_{z})=n_b \,
{1\over 2\, K_1({1\over T_b }) \gamma_b }
\exp\left(-{p_{\mu} U^{\mu}  \over T_b } \right)
\label{fralativ21}
\end{equation}
where $n_b$ is the density of the beam measured in the laboratory frame
(the Lorentz invariant proper density is $ n_b \gamma_b$), 
$U^{\mu} = (\gamma_b , \beta_b \gamma_b)$ is the four velocity
 of the rest frame of the beam, $T_b$ is the beam temperature
  in units $mc^2$, $K_1$ is a modified Bessel function.

This function may be simplified in the limit of
 cold beam (in its frame)  $T_b  \ll  1$ and large streaming 
velocity
 $\gamma_b \gg 1$.
We find then 
\begin{equation}
f(p_{z})= { n_b \over \sqrt{2 \pi} p_t } 
 \exp\left(-{ (p_{z} -p_b)^2 \over 2  p_t^2}
 \right)
\label{qq1}
\end{equation}
where $p_t^2 =  \gamma_b^2 T_b m c $ is the scatter in parallel moments.

In case of kinetic instabilities the growth rate is  given by 
(e.g., \cite{Melrosebook1})
\begin{equation}
\Gamma = \left.- { (e_{\alpha}^{\ast} \epsilon^{\prime \prime}_{\alpha \beta}
 e_{\beta}) \over {1\over \omega^2 } { \partial \over \partial \omega }
\omega^2 (e_{\alpha}^{\ast} \epsilon^{\prime}_{\alpha \beta}
 e_{\beta}) } \right|_ { \omega= \omega({\bf k})}
\label{piajd44}
\end{equation}
where $ \epsilon^{\prime}_{\alpha \beta}$ and $
\epsilon^{\prime \prime}_{\alpha \beta}$ are hermitian and antihermitian
parts of the dielectric tensor, $ \omega({\bf k})$  is the frequency 
of the excited normal modes of the medium,
and $ e_{\alpha}$ is its polarization vector.
The antihermitian parts of the  dielectric tensor are due to the 
resonant interaction of the particles from the beam at Cherenkov (\ref{uiaj}) 
and cyclotron resonances (\ref{uiaj1}). Using the Plemelj formula we find
\begin{eqnarray}
\epsilon^{\prime \prime}_{xx}&&= - i { 2 \pi^2 e^2 \over  \omega^2 m}
\int {dp_{z} \over \gamma} \hat{\omega} f(p_{z})  
\delta\left( \hat{\omega} - {{\omega_B}\over \gamma} \right)
=\epsilon^{\prime \prime}_{yy}
\mbox{} \nonumber \\ \mbox{}
\epsilon^{\prime \prime}_{z z}&& = i
{ 4 \pi^2 e^2 \over  \omega } \int dp_{z}
v_{z}  { \partial f(p_{z}) \over \partial
p_{z}} \delta \left(\hat{\omega}\right) - 
 i { 2 \pi^2 e^2 \sin^2 \theta^2 k^2 c^2 \over  \omega^2 \omega_B}
\int dp_{z} \gamma v_{z} ^2 f(p_{z}) 
\delta\left(\hat{\omega} - {{\omega_B}\over \gamma} \right)
\mbox{} \nonumber \\ \mbox{}
\epsilon^{\prime \prime}_{x z}&&=- i { 2 \pi^2 e^2 k \sin \theta \over
m \omega^2 \omega_B} \int dp_{z} \hat{\omega} v_{z} f(p_{z})
\delta\left(\hat{\omega} - {{\omega_B}\over \gamma} \right)
=\epsilon^{\prime \prime}_{z x}
\mbox{} \nonumber \\ \mbox{}
\epsilon^{\prime \prime}_{x y}&& \approx 0= \epsilon^{\prime \prime}_{yx} =
 \epsilon^{\prime \prime}_{y z} = \epsilon^{\prime \prime}_{z y}
\label{piajd1}
\end{eqnarray}

Using the   polarization vectors (\ref{q3}),(\ref{q4}) we find
that for the quasitransverse waves (O mode $\omega \gg
\omega_0^{(h)}$, Alfv\'{e}n mode $\omega \ll
\omega_0^{(h)}$ and O mode $\omega \approx 
\omega_0^{(h)}$, $\theta \gg \omega_B^2 / (T_p \omega_p^2) $),
while for the O mode at the cross-over point and
$\theta \ll \omega_B^2 / (T_p \omega_p^2)$ 
\begin{equation}
{1\over \omega^2}
 { \partial \over \partial \omega }
\omega^2 ({\bf e \cdot \epsilon^{\prime} \cdot
 e  }) = 
\left\{ \begin{array}{ll}
{ 2 \over \omega } & \mbox{cold plasma}\\
{ T_p \omega \over \omega_p^2}  & \mbox{hot  plasma}
\end{array} \right.
 \label{piajd12}
\end{equation}

With  polarization vectors  (\ref{fhdu13951}) and
 (\ref{fhdu1396})  we find from  (\ref{piajd1}), that 
for  quasitransverse parts of the waves
\begin{eqnarray}
&&  ({\bf e_X \cdot \epsilon^{\prime \prime } \cdot  e_X} )=
- i { 2 \pi^2 e^2 \over  \omega^2 m}
\int {dp_{z} \over \gamma}  \hat{\omega} f(p_{z})
\delta\left(\hat{\omega} - {{\omega_B}\over \gamma} \right)
\mbox{} \label{hh1} \\ \mbox{}
&&
 ({\bf e_O \cdot \epsilon^{\prime \prime } \cdot  e_O}) =
{ 4 \pi^2 e^2 \over m \omega } \int dp_{z}
v_{z}  { \partial f(p_{z}) \over \partial
p_{z}} \delta \left(\hat{\omega}\right)  \sin ^2 \theta
\mbox{} \nonumber \\ \mbox{}
&& \hskip .3 truein
 +
{ 2 \pi^2  e^2  \over \omega^2 \omega_B m } 
\int dp_{z} \left(k v_{z} - \omega \cos \theta \right)^2 \, f(p_{z}) 
\delta\left(\hat{\omega} - { \omega_B \over \gamma} \right) =
\epsilon^{\prime \prime \, Ch}_O +
\epsilon^{\prime \prime \, C}_O
\mbox{} \label{hh2} \\ \mbox{}
&&
({\bf e_A  \cdot \epsilon^{\prime \prime } \cdot  e_A}) =
{  \pi^2 e^2 \over m \omega } \, {\omega^4 \over \omega_p^4} 
 \int dp_{z}
v_{z}  { \partial f(p_{z}) \over \partial
p_{z}} \delta \left(\hat{\omega}\right)  \tan  ^2 \theta
\mbox{} \nonumber \\ \mbox{}
&& \hskip .3 truein
  +
{ 2 \pi^2  e^2  \over \omega^2 \omega_B m }
\int dp_{z} \left(\omega - k v_{z} \cos \theta \right)^2 \, f(p_{z})
\delta\left(\hat{\omega} - { \omega_B \over \gamma} \right) =
\epsilon^{\prime \prime \, Ch}_A+
\epsilon^{\prime \prime \, C}_A
\label{piajd21}
\end{eqnarray}
where we split the antihermitian part for the O and Alfv\'{e}n  modes
 in two parts:
$\epsilon^{\prime \prime \, Ch}$ is  due to the  Cherenkov resonance
and $\epsilon^{\prime \prime \, C}$ is  due to the cyclotron resonance.

Most of the relations (\ref{piajd21}), excepting   $\epsilon^{\prime \prime \, Ch}_A$,
are valid for both cold and hot plasma.
For hot plasma we have
\begin{equation}
\epsilon^{\prime \prime \, Ch\, (h)}_A\equiv 
({\bf e_A  \cdot \epsilon^{\prime \prime } \cdot  e_A})^{(h)} =
{  \pi^2 e^2 \over m \omega } \, {\omega^4 \over \, T_p ^2 \omega_p^4}
 \int dp_{z}
v_{z}  { \partial f(p_{z}) \over \partial
p_{z}} \delta \left(\hat{\omega}\right)  \tan  ^2 \theta
\label{piajd22}
\end{equation}

For the Cherenkov excitation of the O mode in the limit
$\mu _ h \gg 1$ (when the resonance occurs at the cross-over point) we
find
\begin{equation}
 ({\bf e_O \cdot \epsilon^{\prime \prime } \cdot  e_O}) _{Ch}=
 \left\{  \begin{array}{ll}
 { 4 \pi^2 e^2 \over m \omega } \int dp_{z}
v_{z}  { \partial f(p_{z}) \over \partial
p_{z}} \delta \left(\hat{\omega}\right)
\phantom{{{{{a\over b}\over{a\over b}}}\over{{{a\over b}\over{a\over b}}}}}  &
\mbox{ $ \theta  \ll { 2 \omega_p^2 \over \omega_B^2} $ } \\
 { 4 \pi^2 e^2 \over m \omega } \,
{ \omega _0^4 \over  \omega _B^4 \cos^2 \theta \sin ^2 \theta }
 \int dp_{z}
v_{z}  { \partial f(p_{z}) \over \partial
p_{z}} \delta \left(\hat{\omega}\right) 
\phantom{{{{{a\over b}\over{a\over b}}}\over{{{a\over b}\over{a\over b}}}}} &
\mbox{ $ \theta  \gg  { 2 \omega_p^2 \over \omega_B^2} $ }
\end{array} \right.
\label{piajd221}
\end{equation}
The calculations of the integrals 
in (\ref{hh1} - \ref{piajd221}) are given in Appendix \ref{Integrals}

\subsection{Parallel Propagation}

We first consider an important, separate case  of 
parallel propagation. 

Using the polarization vectors ${\bf e_l}=(0,0,1)$ for longitudinal waves and
 ${\bf e_t}=(1,0,0)$ for 
 transverse waves
we find 
\begin{eqnarray}
&&
({\bf e_t \cdot \epsilon^{\prime \prime } \cdot  e_t}) =
- i { 2 \pi^2 e^2 \over  \omega^2 m}
\int {dp_{z} \over \gamma}  \hat{\omega} f(p_{z})
\delta\left(\hat{\omega} - {{\omega_B}\over \gamma_b} \right)
\mbox{} \label{hhh0} \\ \mbox{}
&&
 { \partial \over \partial \omega }
\omega^2 ({\bf e_t \cdot \epsilon^{\prime} \cdot
 e_t  })  \approx 2 \omega
\mbox{} \label{hhh} \\ \mbox{}
&&
({\bf e_l \cdot \epsilon^{\prime \prime } \cdot  e_l}) =
 i
{ 4 \pi^2 e^2 \over  \omega } \int dp_{z}
v_{z}  { \partial f(p_{z}) \over \partial
p_{z}} \delta \left(\hat{\omega}\right)
\mbox{} \label{hhh1} \\ \mbox{}
&&
{1\over \omega^2}
 { \partial \over \partial \omega }
\omega^2 ({\bf e_l  \cdot \epsilon^{\prime} \cdot
 e_l }) = 
 \left\{  \begin{array}{ll}
{ 1\over \sqrt{2} \omega_p} , & \mbox{cold plasma}\\
{ \, T_p  \omega \over \omega_p^2 } , & \mbox{hot plasma}
\end{array} \right.
\label{piajd2211}
\end{eqnarray}

The corresponding growth rates are 
\begin{eqnarray}
&&
\Gamma_t= { \pi  \omega_{p,res}^2 \over 4 \omega }
\left( f  \right)_{\rm  res} \,
\mbox{} \label{hhh4} \\ \mbox{}
&&
\Gamma_l = { \pi\omega_p^2  \omega_{p,res}^2 \over \, T_p  k c  \omega^2} 
\left(\gamma^3 { \partial f \over \partial \gamma } \right)_{\rm  res} \,
\label{piajd22122}
\end{eqnarray}

With the distribution function of the form 
(\ref{qq1}) we find 
\begin{eqnarray}
&&
\Gamma_t \approx { \pi \omega_{p,res}^2 \over \omega \Delta \gamma},
\hskip .2 truein \mbox{ $\omega= { \omega_B^3 \over \gamma_b \, T_p  
 \omega_p^2 }$ } 
\mbox{} \label{hhh5} \\ \mbox{}
&&
\Gamma_l \approx  {n_b \over n_p} \,
{ \pi\omega_p \gamma_b^3 \over \, T_p ^{5/2} \Delta \gamma^2},
\hskip .2 truein 
\mbox{ $\omega = \omega_0 = \sqrt{ 2 \, T_p } \omega_p$}
\label{piajd22123}
\end{eqnarray}

The kinetic growth rates (\ref{hhh5}) and (\ref{piajd22123})
can be compared with growth rates in hydrodynamic regime (Eqns
(\ref{lkfjs}) and (\ref{dafd})).
In a hydrodynamic regime both cyclotron and Cherenkov growth rates
are proportional to the negative 
powers of the particle's Lorentz factor.
This is a significant factor for the primary beam and for 
the particles from the tail of plasma distributions.
In contrast, kinetic growth rates (\ref{hhh5}) and (\ref{piajd22123})
are not suppressed by the relativistic streaming of resonant particles.
On the other hand, kinetic growth rates (\ref{hhh5}) and (\ref{piajd22123})
scale linearly with a small ratio of the beam density to plasma density
while  hydrodynamic growth rates (\ref{lkfjs}) and (\ref{dafd}) 
are proportional to $1/3$ and $1/2$ power of this ratio.

\subsection{Excitation of Oblique Alfv\'{e}n Waves in a Kinetic Regime}
\label{ExObli}

\subsubsection{Cherenkov Resonance}

Using (\ref{piajd44}),  (\ref{piajd21}), (\ref{q1}), (\ref{piajd22}) 
 and (\ref{piajd44}),
we find a growth rate for the Cherenkov excitation of Alfv\'{e}n wave 
in a cold plasma:
\begin{equation}
\Gamma ={\pi \over 8} { \omega_b^2 \over  kc \cos \theta} 
{ \omega^4 \over \omega_p^4} \tan^2 \theta {\gamma^3 \over \Delta \gamma^2}
\label{piajd23}
\end{equation}
with  the resonant $\omega$ and $k$ given in Table \ref{rescold1}
 for cold plasma
and Table \ref{reshot1} for the hot plasma.
In a hot plasma the growth rate is decreased by a factor  $\, T_p ^2$.

This growth rate is very small. Alfv\'{e}n waves in the limit $\omega \ll 
\omega_p$ are almost transverse and  are not excited effectively by the
Cherenkov resonance. A strong dependence on $\omega$ and $\theta$ 
corresponds to the increasing potential part of Alfv\'{e}n waves for larger
 $\omega$ and $\theta$.

\subsubsection{Cyclotron Resonance}

Using (\ref{piajd44}),  (\ref{piajd21}), (\ref{q1})  and (\ref{piajd44}),
the  growth rate for the cyclotron excitation
of Alfv\'{e}n waves is
\begin{equation}
\Gamma = {\pi \over 4} \, {\omega_b ^2 \over \omega_{\rm  res} \, \Delta \gamma} 
\label{piajd24}
\end{equation}
with the resonant frequency given in Table \ref{rescold1} in the cold case
or Table \ref{reshot1} in the warm case.

\subsection{Excitation of the Oblique Ordinary Waves in a Kinetic Regime}
\subsubsection{Cherenkov Excitation}

The  Cherenkov excitation of the O mode strongly depends on the 
parameter $\mu_h$ and the angle of propagation. Excitation is possible only
for $\mu _h > 1$. For  $\mu _h \geq 1 $ the resonance occurs at
$\omega \gg \omega_0^{(h)}$. Then, using the
 polarization vector  Eq. (\ref{fhdu13951}),
the
resonance frequency   (Tables \ref{rescold1} and \ref{reshot1})
we find from (\ref{piajd44}) 
\begin{equation}
\Gamma =  {\pi \over 2}  {\omega_b^2 \over k_{\rm  res} \, c } 
{\gamma_b^3 \sin ^2 \theta \over \Delta \gamma^2}
\label{piajd25}
\end{equation}

For $ \mu _h \gg 1  $ the Cherenkov  resonance occurs approximately
at the cross-over point $\omega_0^{(h)}$. Using the polarization vector
 (\ref{fhdu139511}), the
resonance frequency (\ref{fjhi2}) and Eq. (\ref{piajd1})
we find from Eq. (\ref{piajd44})
\begin{equation}
\Gamma =
 \left\{  \begin{array}{ll}
\phantom{{{{{a\over b}\over{a\over b}}}\over{{{a\over b}\over{a\over b}}}}}
 {\pi \over \sqrt{2}  2}  {\omega_b^2 \over \omega_p} 
{\gamma_b^3 \over \Delta \gamma^2}
 &
\mbox{ $ \theta  \ll { 2 \omega_p^2 \over \omega_B^2} $ } \\
\phantom{{{{{a\over b}\over{a\over b}}}\over{{{a\over b}\over{a\over b}}}}}
  {\pi \over   2}   {\omega_b^2 \omega_0^{(h) \,3}  \over
\omega_B^4  \sin ^2 \theta \cos  ^2 \theta }
{\gamma_b^3 \over \Delta \gamma^2}
 &
\mbox{ $ \theta  \gg  { 2 \omega_p^2 \over \omega_B^2} $ }
\end{array}
\right.
\label{piajd26}
\end{equation}

Equations (\ref{piajd25}) and (\ref{piajd26}) imply that the 
Cherenkov excitation of the O mode is effective
only if $ \mu _h \gg 1 $ and in the narrow angle $\theta \ll 
{ 2 \omega_p^2 \over \omega_B^2} $. This condition may be satisfied 
only in the outer regions of the pulsar magnetosphere.  The 
growth rate of the Cherenkov excitation of the O mode in the
kinetic regime is proportional to the density of the resonant particles.
In the outer parts of pulsar magnetosphere, the density has decreased
considerably which prevents the development of the 
Cherenkov instability.
Numerically, it turns out that in the pulsar magnetosphere the
kinetic instabilities may be stronger than hydrodynamic.

\subsubsection{Cyclotron Excitation of the Ordinary Mode}

Using (\ref{hh2}),   (\ref{q1})  and (\ref{piajd44})
the  growth rate for the cyclotron excitation
of the O wave is 
\begin{equation}
\Gamma = {\pi \over 4} \, {\omega_{p,res}^2
 \over \omega_{\rm  res} \,  \cos ^2 \theta \Delta \gamma}
\label{piajd29}
\end{equation}
with the resonant frequency given in  
Table \ref{rescold1} in the cold case 
or Table \ref{reshot1} in the hot case. Here $\omega_{p,res}$ is the plasma
frequency of the resonant particles.
The angle of emission is limited by $\theta \leq \omega_ p /\omega_B$.
The maximum growth rate, which is  attained with   parallel
propagation, is estimated below.

\subsection{Excitation of the X Mode}

Using (\ref{hh1}),  (\ref{q1})  and (\ref{piajd44})
the  growth rate for the cyclotron excitation
of the X wave is
\begin{equation}
\Gamma = {\pi \over 4} \, {\omega_{p,res}^2 \over \omega_{\rm  res} \, \Delta \gamma}
\label{piajd291}
\end{equation}
Using the resonant frequency (Table \ref{reshot1}) we find
\begin{equation}
\Gamma = {\pi \over 4} \, {\omega_{p,res}^2 \omega_p^2 \gamma_b
\, T_p  \over \omega_B^3 \Delta \gamma} ={ \pi \lambda_{\rm  res} \, \lambda
\gamma_b \, T_p  \over  \Delta \gamma \gamma_p } \, {\Omega^2 \over \omega_B} 
\label{piajd292}
\end{equation}

The conditions of the fast growth are 
\begin{eqnarray}
 &&
{\Gamma \over \Omega} > 1, \hskip .2 truein
\mbox{ if} \, \left({ R \over R_{NS} } \right) > 300
\mbox{} \label{piajd2921} \\ \mbox{}
 &&
{R_c \delta \theta \Gamma \over c \gamma_p^2} =
{ \pi \lambda_{\rm  res} \,  \lambda^{3/2} \over \Delta \gamma} \,
{ R_c \Omega \over c} \, \left({ \Omega \over  \omega_B} \right)^{3/2}
\label{piajd293}
\end{eqnarray}

Since cyclotron instability develops in the outer regions of pulsar
magnetosphere, condition (\ref{piajd293}) can be satisfied for the 
regions close to the the magnetic axis with $ R_c \approx 10^{10} $ cm.
The lower streaming Lorentz factors increase the cyclotron
instability growth rate.

We conclude this section by the table of kinetic growth
rates (Table \ref{KineticInst}).

\begin{table} \begin{center}
\caption[ Kinetic growth rates  in a pair plasma]
{ Kinetic growth rates in a pair plasma}
\psfig{file=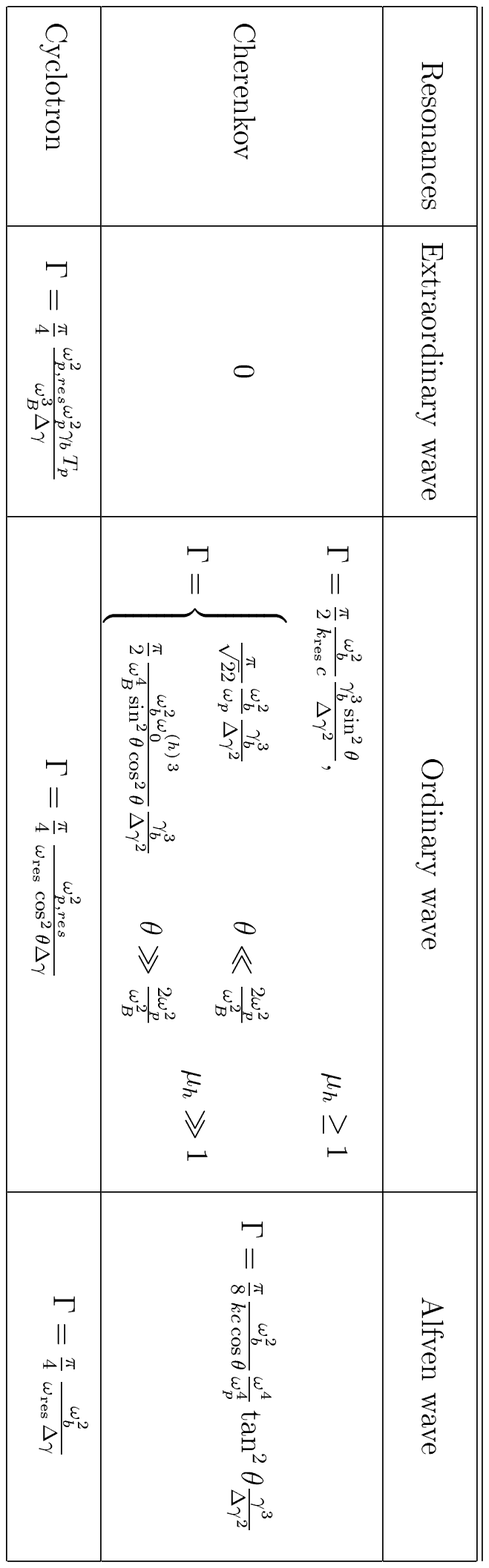,width=20.0cm}
\label{KineticInst}
\end{center} \end{table}

\section{Hydrodynamic Versus  Kinetic Instabilities}

Having calculated the growth rates for the hydrodynamic and kinetic regimes
of the Cherenkov and cyclotron instabilities, we can check whether the 
conditions of the corresponding regimes are satisfied.

\subsection{ Cherenkov Resonance} 

The condition of the  hydrodynamic regime for the Cherenkov
excitation is given by (\ref{Gammm}) with $\nu=0$ (the condition for the
kinetic regime is reversed). We can distinguish two separate cases:
when the scatter in velocity of the resonant particles is due to the 
the scatter in parallel  velocity or to the scatter in pitch angles.
In the former case condition (\ref{Gammm}) with the parallel growth rate
(\ref{liafs8}) gives the following requirement for the  hydrodynamic-type
Cherenkov instability:
\begin{equation}
{ \gamma_b^2 \over \sqrt{\, T_p }  \Delta \gamma \lambda^{1/3} } \gg 1
\label{b}
\end{equation}
which is well satisfied for the chosen plasma parameters.

In the case when the  scatter in pitch angles dominates over
the scatter in parallel  velocity the condition  for the  hydrodynamic type
Cherenkov instability reads
\begin{equation}
\psi^2 \ll {1\over \gamma_b \sqrt{ \, T_p  \lambda} }
\label{b1}
\end{equation}
This is not satisfied.
This implies that if the primary beam does not acquire any 
significant transverse gyrational energy 
as it propagates out in the pulsar magnetosphere, then the Cherenkov-type
instabilities occur in the hydrodynamic regime.

We can also verify that the condition for the kinetic growth of the 
beam without any scatter in pitch angles is not satisfied. 
The inverse of the  condition (\ref{Gammm})  with the parallel growth rate
in the kinetic regime (\ref{piajd26})  give the following condition for the
validity of the kinetic approximation
\begin{equation}
{\gamma_b^5 \over \lambda \, T_p ^3 \Delta \gamma^3 } \ll 1
\label{b2}
\end{equation}
which is not satisfied for the chosen plasma parameters.

We conclude that the Cherenkov instability for the 
parallel propagation is in the hydrodynamic regime.

\subsection{ Cyclotron  Resonance}

For the cyclotron  resonances the left-hand side of (\ref{Gammm}) is dominated
by the last term. For the cyclotron excitation of the X
mode condition (\ref{Gammm}) with the growth rate (\ref{dafslj}) give for the
  hydrodynamic type instability to apply
\begin{equation}
\Delta \gamma \ll \sqrt{  \, T_p   \lambda } 
 { \gamma_b^{3/2} \Omega \over \gamma_p \omega_B}  = 10^{-7}
\left({R \over R_{NS} }\right)^3 
\label{b3}
\end{equation}
which is most probably not satisfied even in the outer regions
of the pulsar magnetosphere. 

The condition for the kinetic approximation for the
  cyclotron excitation of the X
mode  follows from  (\ref{Gammm}) and  (\ref{piajd292}):
\begin{equation}
\Delta \gamma  \gg 
 \left({ \gamma_{\rm  res} \,^3   \lambda \lambda_{\rm  res} \, T_p \over
 \gamma_p } \right)^{1/2}
 { \Omega \over \omega_B } 
\label{b4}
\end{equation}
which is well satisfied inside the pulsar magnetosphere.

From these estimates we conclude that the cyclotron instability
in the pulsar magnetosphere occurs in the kinetic regime. This is
different from the electrostatic Cherenkov instabilities on the
primary beam, that
occur in a hydrodynamic regime.

This difference is very important for the theories of the pulsar radio
 emission. The kinetic instabilities,
in contrast to the hydrodynamic, are  not suppressed by the
large relativistic factor of the resonant particles. 
Thus, the  kinetic instabilities are more favorable as a possible 
source of the pulsar radio emission.

It is possible to illustrate graphically the
 difference between the  hydrodynamic regime
of the Cherenkov instability and the kinetic regime of the cyclotron
instability. On the frequency-wave vector diagram  for the O mode
(Fig. 
\ref{fig4}), the dispersion curves of the 
cyclotron wave in the beam $\omega= k v_b \cos \theta - \omega_B/\gamma_b$
is almost parallel to the dispersion curves of the excited waves in plasma
at the location of the resonance. Thus, a small change in the velocity of the
resonant particles results in a considerable change of the resonant 
frequency. This vindicates the kinetic approximation that requires a large
bandwidth of the growing waves. In contrast, for the very large 
streaming $\gamma$-factor of the primary beam (so that $\mu, \, \mu_h \,
\gg 1$), the  Cherenkov
resonances on the O and X modes occur approximately at the 
cross-over frequency in a narrow frequency band.

\section{Conclusion }
\label{Concl}

In this work we considered  normal modes and 
wave excitation in the strongly magnetized
electron-positron plasma of the pulsar magnetosphere.
 We found the location of resonances and 
calculated the growth rates for the Cherenkov and
cyclotron excitation of  the O, X and Alfv\'{e}n waves in
two limiting regimes of  hydrodynamic and kinetic instabilities
 taking into account angular
dependence of the growth rates.
 The main results of the paper
are \\
(i) Cherenkov instabilities develop in the hydrodynamic regime
while cyclotron instabilities develop in the kinetic regime.\\
(ii) Cherenkov instability on the primary beam develops on the
Alfv\'{e}n waves in the regions close to the stellar surface and on the
O mode in the outer regions of the pulsar magnetosphere
(\ref{uiaj22}), (\ref{liafss2}).\\
(iii) Cyclotron instability can develop on all three wave branches.
On the Alfv\'{e}n branch, the cyclotron instability does not develop in a very
hot plasma (\ref{fhdu26}).\\
(iv) The typical range of angles (in the plasma frame)
 with the highest growth rates are

$
\delta \theta \approx {\omega_p^2 / \omega_B^2} $
for Cherenkov excitation of the O mode

$
\delta \theta \approx {1/ \gamma_b}
 $
 for Cherenkov excitation of the Alfv\'{e}n mode

$
\delta \theta \approx \omega_p/ \omega_B $
for cyclotron excitation of the O and X modes

$
\delta \theta \approx 1 $
for cyclotron excitation of the Alfv\'{e}n mode\\
(v) We also note, that 
Cherenkov instability due to the relative drift of the plasma
particles can develop only on the Alfv\'{e}n mode.

These arguments suggest that electromagnetic cyclotron 
instabilities are more likely to develop in the pulsar 
magnetosphere than electrostatic.

\acknowledgments
I would like to thank Roger Blandford for his support 
and comments, George Machabeli and George Melikidze for
useful cooperation and
  Abastumani Astrophysics Observatory for
the hospitality during my stays in Tbilisi.
 This research was supported
by the NSF under grant No AST-9529170 and by the 
CITA fellowship.

\newpage

\appendix

\section{ Calclations of the Resonant Integrals}
\label{Integrals}

The calculations presented in this Appendix are used in all the
calculations of  the kinetic growth rate of the
instabilities.

\begin{equation}
 \int dp_{z}
v_{z}  { \partial f(p_{z}) \over \partial
p_{z}} \delta \left(  \hat{ \omega}\right) ={1\over k \cos \theta \,  m_e} 
\left(  v_{z} \gamma^3  { \partial f(p_{z}) \over \partial p_{z}} \right)_{\rm  res} \,
\label{ap1}
\end{equation}

\begin{eqnarray}
&& 
 \int dp_{z}
\left( k v_{z} - \omega \cos \theta \right)^2 \, f(p_{z})
\delta\left( \hat{\omega} - { \omega_B \over \gamma_b} \right) =
\mbox{} \nonumber \\ \mbox{}
&& \hskip .3 truein 
 \int dp_{z}
\left( k v_{z} - \omega \cos \theta \right)^2 \, f(p_{z})
{ 1\over \left| - { k c \cos \theta \over m \gamma^3} + { \omega_B p_{z} \over 
\gamma^3 m c} \right| } \delta( p_{z} -p_{\rm  res} \,)
\label{ap2}
\end{eqnarray}

For $ \omega_B \gg k c$ this reduces to
\begin{equation}
{ \gamma^2 \left(  \omega \sin \theta^2 - \omega_B/\gamma \right) \over
\omega_B v_{z} \cos^2 \theta} f(p_{z}) \approx
\left( {\omega_B  f(p_{z})  \over v_{z} \cos ^2 \theta} \right)_{\rm  res} \,
\label{ap3}
\end{equation}

Similarly we have
\begin{equation}
 \int dp_{z}
( \omega - k v_{z} \cos \theta)  f(p_{z}) \delta\left( \hat{\omega} 
- { \omega_B \over \gamma_b} \right) =
\left( {\gamma   f(p_{z})  \over v_{z}} \right)_{\rm  res} \,
\label{ap4}
\end{equation}
and
\begin{equation}
 \int dp_{z}
(\omega - k v_{z} \cos \theta)^2 f(p_{z}) \delta\left( \hat{\omega} - { \omega_B
\over \gamma_b} \right) =
\left( { \omega_B  f(p_{z})  \over v_{z}} \right)_{\rm  res} \,
\label{ap5}
\end{equation}

\section{Relativistic Maxwellian Distribution}
\label{RelativisticMax}

We seek an appropriate expression for the relativistic
one dimensional distribution.
The aim of this Appendix is to define the relevant physical
quantities measured in different systems.
The relation obtained in the Appendix are extensively
used in Section \ref{WavesHot} when considering the properties
of waves in a relativistically hot plasma.

Relativistic covariant
dispersion relations for plasma waves have been considered
by \cite{GodfreyDisp}, \cite{MelroseCovar} and others
(see reference in \cite{MelroseCovar}).
The general expression for the {\it frame-invariant}
distribution function is
\begin{equation}
f({\bf p}, {\bf r})^{ \rm inv}
 = { 1\over (2 \pi \hbar)^3 } \exp \left\{ \mu({\bf r})
 - \beta_T p^{\nu} U _{\nu}\right\}
\label{qa1}
\end{equation}
here $\mu$ is a chemical potential, $\beta_T =1 /T)$,
$T$ is invariant temperature,
${\bf p}$ is the momentum of
 the particle,
$p^{\nu}$ is a four-momentum of
 the particle
and $U _{\nu}$ is four velocity of the
reference frame (speed of light and
particle mass are set to unity in this Appendix).

Next we  define a flux four-vector:
\begin{equation}
N^{\nu} =
\int { d {\bf p} \over \gamma} p^{\nu} f( {\bf p}) ^{\rm inv} =
\{n({\bf r,t}), {\bf j}({\bf r,t}) \}
\label{qa11}
\end{equation}
An invariant density,  measured
 in a particular frame with the four velocity $U _{\nu}$
is then
\begin{equation}
n_0 = N^{\nu} U _{\nu}=
\int { d {\bf p} \over \gamma} ( p^{\nu} U _{\nu})  f( p) ^{\rm inv}
\label{qa12}
\end{equation}
 In particular, the
invariant density in the rest frame ( with $U _{\nu}^{o} =
\{ 1,0,0,0\}$) is $ n^{o} = N^{\nu} U _{\nu} ^{o}$.
We
normalize the distribution function (\ref{qa1}) to the
invariant
density of particles in the rest frame
\begin{equation}
n_0 =
\int  { d {\bf p} \over ( 2 \pi \hbar)^3 } \exp
\{-\beta_T \gamma \}
\label{qa2}
\end{equation}
Then, for a one dimensional distribution $ f({\bf p})^{\rm inv}
 = \delta( p_{\perp}^2)
f(p) ^{\rm inv} /\pi $ (below $p$ is a component of momentum along magnetic field)
\begin{equation}
f({\bf p}) ^{\rm inv} = { n_0 \over 2 K_1 (\beta_T)  }
\exp \left\{ - \beta_T p^{nu} U_{nu} \right\}
\label{qa3}
\end{equation}
where we introduced new variables $ \gamma = \cosh x$ and
$ \gamma_p = \cosh y $ and used a relation  (\cite{Gradstein}, (3.547.4))
\begin{equation}
\int _{- \inf} ^{\inf} dx \cosh x \exp \{ - \beta_T \cosh x \} =  2 K_1(\beta_T)
\label{qa31}
\end{equation}
The density in the frame moving with the four velocity
$U_{\nu}=\{ \gamma_p, {\bf v_p } \gamma_p \}$ ( here
$\gamma_p= 1/\sqrt{1-v_p^2}$)
is
\begin{equation}
n =N^{0} =
\int d {\bf p} f({\bf p}) ^{\rm inv} = \gamma_p n_0
\label{qa4}
\end{equation}

In this work we use the distribution function normalized to the
{\it laboratory } density $n$
\begin{equation}
f({\bf p}) = { \delta( p_{\perp}^2) \over  \pi } f(p),
\hskip .5 truein
f(p) =
 { n \over 2  K_1 (\beta_T) \gamma_p}
 \exp \left\{ - \beta_T p ^{\nu} U_{\nu} \right\}
\label{qa5}
\end{equation}

There is a natural simplification of the distribution
function (\ref{qa5}) in the case $\beta_T \gg 1 , \,
\gamma_p \gg 1$ (cold plasma streaming with large Lorentz factor).
In this case the distribution is strongly peaked at $\gamma= \gamma_p$ so
we can expand the distribution function, keeping  terms up to the second order
in $ \gamma -\gamma_p$:
\begin{equation}
f(p) = { n  \exp \{ - \beta_T \} \over 2  K_1 (\beta_T) \gamma_p}
\exp \left\{ -{ \beta_T ( \gamma - \gamma_p)^2 \over 2 ( \gamma_p^2 -1 ) }
\right\}
= { n \over \sqrt{ 2 \pi}  \Delta \gamma }
\exp \left\{ - { ( \gamma - \gamma_p)^2 \over 2 \Delta \gamma^2 } \right\}
\label{qa510}
\end{equation}
where we introduced $ \Delta \gamma = \sqrt{T} \gamma_p$ and used the fact that
$ \gamma_p \gg 1$.

In Table \ref{Maxwe} we give the estimates of the moments
of the  relativistic Maxwellian distribution.
$< {\bf ...} > $ implies $\int dp {\bf ...} f(p)/n $, where $n$ is the
noninvariant density in the laboratory frame.
The arguments of the Bessel functions in Table \ref{Maxwe} are $1/ T_p$.
When calculating moments we used  a relation
\begin{equation}
\int _0^{\infty}\, dx\, exp(-\beta_T \cosh(x)) \cosh(x)^n=(-1)^n\,
{d^n K_0(\beta_T)\over d \beta^{n} }
\label{ryzhik}
\end{equation}
and the asymptotic relations for the modified Bessel functions:
\begin{eqnarray} &
K_{\nu}(x)= \sqrt{{\pi\over  2 \, x}}\, e^{-x}\,\left(1 + {4\, \nu^2-1\over 8\,
x}
\right)& \mbox{ $x \rightarrow \infty$}\\
& K_0(x)=\,-\,\ln (x), \hskip .4 truein K_{\nu}(x)=
\, {1\over 2} \,\Gamma (\nu)\, \left( {x\over 2} \right) ^{-\nu}
& \mbox{ $x \rightarrow 0$}
\label{VVV}
\end{eqnarray}

\section{Cutoff and Cross-Over Points for Parallel Propagation}
\label{Cutoff}

Using streaming Maxwellian distribution it is not possible to find the
exact expressions for the two important frequencies:
cutoff frequency (a limit $k \rightarrow 0$ of the plasma wave dispersion)
and the  cross-over frequency (when the O mode has a
vacuum dispersion relation).

The cutoff frequency is
\begin{equation}
\omega_{\rm cutoff}^2 ={ 4 \pi e^2 \over  m_e} \int
{ d p \over \gamma^3} f(p) =
\label{qa15}
\end{equation}
and the cross-over frequency is
\begin{equation}
\omega_{\rm cross-over}^2 ={ 4 \pi e^2 \over  m_e} \int
{ d p \over \gamma^3}  { f(p) \over ( 1- v)^2 }
\label{qa6}
\end{equation}

For the case of relativistic Maxwellian distribution (Eq. \ref{flioj})
the expression for the cutoff frequency may be re written as
\begin{equation}
\omega_{\rm cutoff}^2 = {
{\omega_p^2 }  \over 2 K_1 (\beta_T) \gamma_p}
\int { d x \over \cosh^2 ( x+ y) } \exp \{ - \beta_T \cosh x \}
\label{qa51}
\end{equation}
where $\gamma_p =\cosh y$.
The corresponding integrations in the case $\beta_T \gg 1 $
(cold plasma) may be performed using the
steepest decent method.  For $\beta_T \gg \gamma_p $
the "sharply" peaked function under integral sign in (\ref{qa51}) is
$ \exp \{ - \beta_T \cosh x \}$ so we can use expand around the
point $x=0$ to obtain
\begin{equation}
\omega_{\rm cutoff}^2   =
\sqrt{ { \pi \over 2 \beta_T }}
{ \exp \{ -\beta_T \} \over 2\, \gamma_p^3 2\, K_1 (\beta_T)} =
{ \omega_p^ 2 \over \gamma_p^3 } \mbox { if $ \beta_T \gg \gamma_p $
 }
\label{qa7}
\end{equation}

For $\beta_T \ll 1 $
(hot plasma)
we can make the following approximation to  the exponential function:
\begin{equation}
 \exp \{ - \beta_T \cosh x \} \approx
\left\{ \begin{array}{ll}
1 & \mbox{ if $ \ln \beta_T < x < - \ln \beta_T $} \\
0 & \mbox{ otherwise}
\end{array} \right.
\label{qa8}
\end{equation}
We find then
\begin{equation}
\omega_{\rm cutoff}^2 \approx { n \over 2  K_1 (\beta_T) \gamma_p}
\int _{-\ln 2 \beta} ^ {\ln 2 \beta} { d x \over \cosh^2 ( x+ y) }
= \left\{ \begin{array}{ll}
\phantom{ {{{ {a\over b} \over {a\over b} } } \over {{ {a\over b}
\over {a\over b} } } } }
\mbox{ {\large $
{
\omega_p^ 2 T \over 4 \gamma_p ^3  } $}}
 & \mbox{ for $ \beta_T \gamma_p \gg 1 $ } \\
\phantom{ {{{ {a\over b} \over {a\over b} } } \over {{ {a\over b}
\over {a\over b} } } } }
\mbox{ {\large $
{ \omega_p^ 2 \over T   \gamma_p } $}}  & \mbox{ for $ \beta_T \gamma_p \ll 1 $ }
\end{array} \right.
\label{qa81}
\end{equation}
where we used
\begin{equation}
\int _{-\ln 2 \beta} ^ {\ln  2 \beta} { d x \over \cosh^2 ( x+ y) }
\approx { 2 \over 1+ 4 \beta^2 \gamma_p^2 } \,
\mbox{ for $ \beta_T  \ll  1$}
\label{qa9}
\end{equation}

An interesting consequence of Eqs. (\ref{qa7}) and (\ref{qa9}) is that
in the case of relativistically hot plasma streaming with very large
Lorentz factor, so that $ \gamma_p \gg T \gg 1$
thermal motion {\it increases } the cutoff frequency, while
for the lower streaming Lorentz factors thermal motion
decreases the the cutoff frequency.

The calculations of the cross-over frequency (\ref{qa6})
may be done exactly:
\begin{eqnarray}
&&
\omega_{\rm cross-over}^2 ={ \omega_p^2 \over 2 K_1 (\beta_T) \gamma_p}
\int_{-\infty} ^{\infty} { d x \over \cosh^2 x }
 { \exp \{ -\beta_T \cosh( x-y) \} \over
( 1 - \tanh x)^2 } = { \gamma_p ( 1+v_p)^2 K_2 (\beta_T) \omega_p^2 \over
2 K_1 (\beta_T) }
\mbox{} \nonumber \\ \mbox{}
&& \hskip 1 truein
\approx
\left\{ \begin{array} {ll}
{ 4 \gamma_p \omega_p^2 T } & \mbox{ $ T \gg 1$} \\
{ 2 \gamma_p \omega_p^2 } & \mbox{ $ T \ll 1$}
\end{array} \right.
\label{qa92}
\end{eqnarray}

\end{document}